\documentclass[sigconf]{acmart}

\usepackage{subcaption}
\usepackage{array}
\usepackage{tabularx}
\usepackage{booktabs}
\usepackage[normalem]{ulem}
\usepackage{float}
\usepackage{wrapfig}
\usepackage{etoolbox}
\usepackage{hyperref}
\usepackage{gensymb}
\usepackage{textgreek}
\usepackage{geometry}
\usepackage{amsmath}
\usepackage{siunitx}
\usepackage{textcomp}
\usepackage[utf8]{inputenc}
\usepackage{booktabs,tabularx,ragged2e}
\newcolumntype{Y}{>{\RaggedRight\arraybackslash}X}
\newcolumntype{L}[1]{>{\RaggedRight\arraybackslash}p{#1}}
\renewcommand{\arraystretch}{1.12} 
\usepackage{booktabs, adjustbox}
\usepackage{booktabs, tabularx}
\usepackage{multirow}
\usepackage{enumitem}
\usepackage{hyperref}

\newcommand{\mm}[1]{\textcolor{black}{#1}}

\AtBeginDocument{%
  }

\copyrightyear{2026}
\acmYear{2026}
\setcopyright{cc}
\setcctype{by}
\acmConference[CHI '26]{Proceedings of the 2026 CHI Conference on Human Factors in Computing Systems}{April 13--17, 2026}{Barcelona, Spain}
\acmBooktitle{Proceedings of the 2026 CHI Conference on Human Factors in Computing Systems (CHI '26), April 13--17, 2026, Barcelona, Spain}
\acmPrice{}
\acmDOI{10.1145/3772318.3790532}
\acmISBN{979-8-4007-2278-3/2026/04}

\begin{document}

\begin{teaserfigure}
    \centering
    \includegraphics[width=0.98\textwidth]{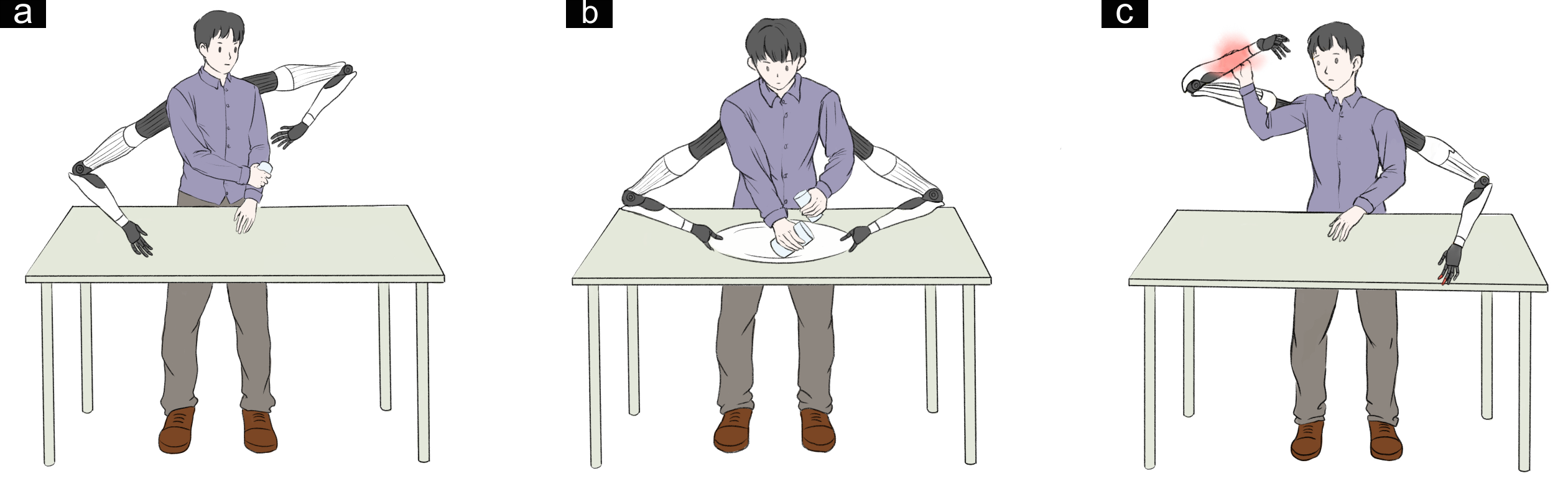}
    \caption{Study overview. (a) Cross-body handover. The SRL reaches across the participant’s midline to pass or receive an object. (b) Stabilize/assist. The SRLs position and hold while participants perform a two-handed manipulation. (c) Rule authoring. Users define spatial and autonomy rules for SRL behavior.}
    \label{fig:teaser}
    \Description{Panel (a): A user stands at a table while two back-mounted supernumerary arms perform a cross-body handover—one arm reaches across the torso to deliver or take an object from the user’s hand. Panel (b): The SRLs assist with a tabletop manipulation; both arms form a stable frame around a dish while the user works at the center, illustrating bracing and cooperative assistance. Panel (c): The user demonstrates or sets a preferred posture and motion for the SRLs (e.g., where to idle and how to approach), representing real-time rule authoring that the arms then follow.}
\end{teaserfigure}

\title{\mm{SRL Proxemics: Spatial Guidelines for Supernumerary Robotic Limbs in Near-Body Interactions}}
\author{Hongyu Zhou}
\orcid{0009-0007-3278-2122}
\affiliation{%
  \institution{The University of Sydney}
  \department{School of Computer Science}
  \country{Australia}
}
\email{hzho4130@uni.sydney.edu.au}

\author{Chia-An Fan}
\orcid{0009-0003-0023-9871}
\affiliation{%
  \institution{The University of Tokyo}
  \department{RCAST}
  \city{Tokyo}
  \country{Japan}
}
\email{chiaan.fan@star.rcast.u-tokyo.ac.jp}

\author{Yihao Dong}
\orcid{0009-0006-7979-2646}
\affiliation{%
  \institution{The University of Sydney}
  \department{School of Computer Science}
  \country{Australia}
}
\email{yihao.dong@sydney.edu.au}

\author{Shuto Takashita}
\orcid{0000-0001-6291-0673}
\affiliation{%
  \institution{The University of Tokyo}
  \department{RCAST}
  \city{Tokyo}
  \country{Japan}
}
\email{shuto.takashita@star.rcast.u-tokyo.ac.jp}

\author{Masahiko Inami}
\orcid{0000-0002-8652-0730}
\affiliation{%
  \institution{The University of Tokyo}
  \department{RCAST}
  \city{Tokyo}
  \country{Japan}
}
\email{drinami@star.rcast.u-tokyo.ac.jp}

\author{Zhanna Sarsenbayeva}
\orcid{0000-0002-1247-6036}
\affiliation{%
  \institution{The University of Sydney}
  \department{School of Computer Science}
  \country{Australia}
}
\email{zhanna.sarsenbayeva@sydney.edu.au}

\author{Anusha Withana}
\orcid{0000-0001-6587-1278}
\affiliation{%
  \institution{The University of Sydney}
  \department{School of Computer Science, Sydney Nano Institute}
  \country{Australia}
}
\email{anusha.withana@sydney.edu.au}

\renewcommand{\shortauthors}{Zhou et al.}

\begin{abstract}

Wearable supernumerary robotic limbs (SRLs) sit at the intersection of human augmentation and embodied AI, transforming into extensions of the human body. However, their movements within the intimate near-body space raise unresolved challenges for perceived safety, user control, and trust. In this paper, we present results from a Wizard-of-Oz study (n=18), where participants completed near-body collaboration tasks with SRLs to explore these challenges. We collected qualitative data through think-aloud protocols and semi-structured interviews, complemented by physiological signals and post-task ratings. Findings indicate that greater autonomy did not inherently enhance perceived safety or trust. Instead, participants identified near-body zones and paired them with clear coordination rules. They also expressed expectations for how different arm components should behave, shaping preferences around autonomy, perceived safety, and trust. Building on these insights, we introduce SRL Proxemics, zone- and segment-level design framework showing that autonomy is not monolithic: perceived safety hinges on spatially calibrated, legible behaviors, not higher autonomy.

\end{abstract}



\maketitle


\section{INTRODUCTION}

\mm{Supernumerary robotic limbs (SRLs) extend robotics from shared workspaces to shared bodies~\cite{prattichizzo2021human,muehlhaus2023need,inami2022cyborgs,arai2022embodiment}. By attaching directly to the user, they can be incorporated into the body schema as extensions~\cite{saraiji2018metaarms}, augmenting physical capabilities across domains from industrial assembly and surgical assistance to everyday support~\cite{parietti2016supernumerary,parietti2013dynamic,parietti2014supernumerary}.Prior systems have explored SRLs that operate in synchrony with the user’s natural limbs, such as wearable extra arms~\cite{vatsal2018design}, supporting tasks like stabilizing objects~\cite{bright2017supernumerary}, holding components in place, and executing parallel actions~\cite{cunningham2018supernumerary,lisini2025exploiting}.}

While prior work has established the mechanical feasibility of SRLs~\cite{parietti2013dynamic,parietti2014supernumerary,parietti2016supernumerary,hu2017hand,vatsal2018design,bright2017supernumerary,cunningham2018supernumerary,lisini2025exploiting}, realizing their full potential requires addressing the challenges of human–robot coexistence in shared physical spaces. 
Interactions in close proximity to the body, especially around sensitive regions such as the head and face, impose heightened demands for safety and predictability, as even small deviations can be perceived as intrusive~\cite{hall1966hidden,takayama2009influences}. However, limited attention has been paid to how these systems should behave to foster a user’s sense of assurance, a state of justified confidence that the system will act safely, predictably, and under the user’s ultimate control~\cite{rubagotti2022perceived,lyons2023explanations}.
\mm{In this work, we therefore treat functional physical safety as a provided baseline and focus instead on users’ sense of assurance, capturing perceived safety, predictability, and comfort during near-body interaction.}
Moreover, the autonomy that makes SRLs useful by reducing micromanagement further introduces uncertainty, since users cannot always anticipate where the system will move next~\cite{al2017challenges}. When such behavior occurs near vulnerable body regions, it poses not only functional concerns but also existential ones: the robot is no longer a passive tool, but a potential threat~\cite{lehmann2020should,lyons2023explanations}. To address these \mm{perceived safety} and predictability concerns, existing approaches often apply a uniform autonomy policy, either giving the system full control or requiring constant manual intervention~\cite{goodrich2008human,parasuraman2000model,bradshaw2017human}. However, the approaches overlook how user expectations differ across bodily regions, an aspect emphasized in prior work on proxemics and perceived safety~\cite{hall1966hidden,takayama2009influences}.

Building on these insights, there is a pressing need for a user-centered framework that captures how individuals negotiate autonomy and control in near-body contexts. Evidence suggests that users intuitively partition their bodies into zones of differentiated tolerance, adjusting their expectations for movement, signaling, and control accordingly~\cite{zhou2025juggling,al2017challenges}. Yet no systematic framework currently exists to elicit and formalize such rules for SRLs. Generative methods are needed to move beyond pre-set autonomy configurations and uncover how users themselves define boundaries, delegate control, and establish behavioral expectations for different zones and components.

To address this gap, we designed a formative, user-centered study that elicits user-defined rules for SRLs behavior in near-body contexts. We used a primed elicitation protocol that combines scenario-based rule authoring with a Wizard-of-Oz (WoZ) setup~\cite{buchenau2000experience,dahlback1993wizard}, in which a physical on-body SRLs prototype simulates participant-defined policies in real time, allowing experimenter-controlled movement, posture, and timing. Multi-sensory cues, such as motor hum, material presence, and real physical contact, elicit authentic responses to spatial intrusion~\cite{meehan2002physiological,walters2005influence,slater2009place}, enabling participants to iteratively refine SRLs behaviors during two tasks. This co-construction process supported zone- and component-level exploration and allowed systematic analysis of how comfort, trust, and embodiment evolve through spatial negotiation~\cite{hancock2011meta,devisser2018literature}.
The method grounds rule creation in user practice rather than current autonomy capabilities and reveals user-generated principles for designing future SRLs in intimate peripersonal space.

Our research is guided by the following questions: \textbf{RQ1}: \emph{What zone- and component-specific boundaries, motion strategies, and cueing requirements do users define to ensure assured interactions in peripersonal space?} \textbf{RQ2}: \emph{How do high-autonomy (fully autonomous baseline) vs. participant-defined rules affect physiological arousal during standardized near-body entries?} and 
\textbf{RQ3}: \emph{How do these autonomy arrangements shape subjective assurance (\mm{perceived safety}, trust) and embodiment in near-body contexts?}

In our study, we adopted a mixed-methods approach combining think-aloud protocols, semi-structured interviews, questionnaires on \mm{perceived safety}, trust, and embodiment, and event-locked skin conductance responses (SCR) to capture participants interacting with SRLs. Our findings revealed that users actively segment their body into sensitivity zones, adjusting SRLs autonomy and motion expectations accordingly. \textit{High-sensitivity zones} such as the head and torso elicited demands for \textit{tighter control}, while \textit{peripheral regions} allowed \textit{more freedom}.
The combined analysis uncovered three key design implications: the need for zone-specific autonomy gradients, component-level motion constraints, and adaptive cueing strategies. Together, these findings underscore the need for zone-responsive control mechanisms that orchestrate SRLs behavior in coordination with the user while preserving predictability, comfort, and agency, informing design guidelines for future SRLs systems operating within intimate peripersonal space. Taken together, these zone- and component-specific expectations form what we term \textbf{\textit{SRL Proxemics}}: the ways users negotiate where and how supernumerary limbs may move around their bodies as a function of bodily sensitivity and task context.

\section{RELATED WORK}

We organize related work across three areas central to our study: (1) the design and control of wearable SRLs, (2) spatial interaction and proxemics in human–robot interaction systems, and (3) autonomy configuration and role negotiation. Together, these domains inform how SRLs should behave near the body.

\subsection{Supernumerary Robotic Limbs}

Wearable Supernumerary Robotic Limbs (SRLs), such as extra arms~\cite{saraiji2018metaarms,vatsal2018design,Zhou2026OneBodyTwoMinds}, legs~\cite{parietti2015design}, or additional fingers~\cite{hu2017hand}, have been explored in human augmentation and embodied interaction research~\cite{prattichizzo2021human}. These systems integrate with the human body to enhance physical capability across varied domains. Early prototypes have demonstrated feasibility for wearable limbs and robotic fingers.

SRLs were initially developed for industrial use, aiding workers in construction and assembly~\cite{parietti2013dynamic,parietti2014supernumerary,bright2017supernumerary} (e.g., by bracing structures and holding parts while humans drill). Their potential has since expanded to medical support and rehabilitation~\cite{prattichizzo2021human,hussain2017hand}, as well as personal augmentation scenarios in everyday manipulation~\cite{saraiji2018metaarms,vatsal2018design,cunningham2018supernumerary}. 
In such applications, SRLs operate in concert with, or semi-independently from, the user’s limbs, blurring the boundary between tool and bodily extension~\cite{lisini2025exploiting,vatsal2018design}. To enable such coordination, various control strategies have been explored, including motion remapping (e.g., via foot or shoulder movements)~\cite{saraiji2018metaarms,sasaki2017metalimbs,shimobayashi2021independent} and brain-computer interfaces~\cite{penaloza2018towards}. However, most systems have been validated only in limited studies~\cite{saraiji2018metaarms,vatsal2017wearing,vatsal2018design}, with little evaluation of adaptability in real-world conditions.

Moreover, although prior work demonstrates mechanical feasibility, it offers limited guidance on how SRLs should behave near the body to sustain user assurance (perceived safety and trust) under varying autonomy. This motivates our focus on spatially adaptive behavior in intimate peripersonal space.

\subsection{\mm{Proxemics and Spatial Interaction in Robotic Systems}}

\mm{Proxemics, introduced in Edward T. Hall’s seminal work, describes how humans regulate space through zones---intimate, personal, social, and public---each reflecting expectations for perceived safety, comfort, and appropriate interaction. Building on this framework, recent robotics research has operationalized proxemic concepts in interaction design and motion planning~\cite{oechsner2025influence, chen2018planning}, not only at the level of social distance, but also by explicitly modelling human peripersonal space and perirobot space around the robot body~\cite{holthaus2012active,chen2018planning,rozlivek2023perirobot,oechsner2025influence}.}

Early HRI studies have shown that people are sensitive to how robots enter and occupy their proxemic zones: approach direction, robot behavior, and prior experience strongly modulate comfort and acceptance~\cite{takayama2009influences}. Studies using small humanoid platforms, such as the NAO robot, further demonstrate 
that people maintain human-like personal-space norms around robots, judging overly 
close or rapid approaches as unfriendly or intrusive~\cite{lehmann2020should}. These findings have informed rule-based proxemic controllers that guide social-robot navigation, approaching slowly, signaling intent, and adapting behavior to user pose or orientation~\cite{mead2013automated}.

More recent work has progressively refined these principles, expanding proxemics from fixed-distance rules to context-sensitive and socially aware models. This includes adaptive proxemic zones that vary with task demands, cultural norms, or social formations~\cite{gines2020defining}, as well as comprehensive proxemics taxonomies that parse the nuances and dynamics of user comfort across collaborative scenarios~\cite{mead2013automated}. Field studies with mixed-reality agents further show how combining subjective reports with objective tracking can reveal proxemic dynamics in situ~\cite{muller2025space}. However, these frameworks address robots that operate outside the human body, establishing expectations for external social navigation but leaving on-body or body-mounted systems unexamined.

\mm{Different from off-body robots, wearable robotic systems stay in close contact with the user. In this case, defining proximity by the distance between the robot and the human body become inadequate. In addition, since SRLs function as extensions of the human body~\cite{saraiji2018metaarms}, the proxemics involved in their interaction with users remain unexplored. Although prior work on wearable proxemics has explored how worn devices extend into interpersonal space~\cite{anacleto2015towards}, it typically treats them as static or single-point extensions. }

\mm{Body-mounted SRLs operate as articulated, (semi-) autonomous limbs that move within the wearer’s intimate peripersonal space, introducing spatial dynamics not accounted for in prior work. Despite this, there is no systematic framework describing how SRLs should manage distance, trajectories, or autonomy across different body zones relative to the wearer’s own limbs and vulnerable regions. We frame this gap as a distinct interaction paradigm, which we term \textit{SRL Proxemics}.}

\subsection{\mm{Safety and Trust in Wearable Robotic Systems}}

\mm{Wearable robotic systems operate within the user’s peripersonal space, making safety a central design consideration. To protect the wearer from physical harm, such devices must account for multiple sources of risk, including internal faults such as sensor failures or control errors~\cite{lee2022toward, naseri2022characterizing}, human–robot misalignment~\cite{rosenblatt2014active}, and unexpected external disturbances or impacts~\cite{olenvsek2021dynamic, major2020perturbation}. Addressing these risks is essential for ensuring safe operation in close proximity to the body.}

\mm{In addition to physical safety, perceived safety is essential for user acceptance. Its determinants, however, differ across wearable systems: exoskeletons, robotic prostheses, and SRLs each represent forms of cyborg-like human–machine augmentation and interact with the environment in distinct ways, leading to different sources of perceived risk or comfort. Exoskeletons aim to enhance human strength, precision, and performance~\cite{kirkwood2021s}. Given their close physical coupling with the user, their primary safety concerns are physical. In contrast, robotic prostheses and SRLs exhibit motions that are independent of the user's biological limbs~\cite{inami2022cyborgs}, which can introduce concerns about unintended impacts and unpredictable movements~\cite{lee2025e}, thereby affecting user trust and raising additional perceived safety issues. }

\mm{Beyond technical fault handling, work on wearable robotics has systematically examined how safety and comfort shape
acceptance and usability in practice. Meyer et al.\ review usability evaluation practices and contexts of use in
wearable robotics, highlighting how perceived safety, comfort, and effort are central to whether users accept such devices in
their daily lives~\cite{meyer2021analysis}. Shore et al.\ similarly show that older adults’ acceptance of robotic assistive devices
depends not only on functional benefit but also on how safe, predictable, and non-threatening the devices feel in close
proximity~\cite{shore2022technology}. In clinical contexts, feasibility and safety studies of cyborg exoskeletons such as the
Hybrid Assistive Limb (HAL) further demonstrate how detailed monitoring of adverse events, gait stability, and user
compensation strategies is required before deploying wearable systems with vulnerable populations~\cite{takahashi2023feasibility}.}

\mm{By exploring \textit{SRL Proxemics}, we provide a human-centered account of how individuals negotiate autonomy, movement, and signaling across different bodily zones, offering a foundation for designing SRLs that maintain predictable, comfortable, and trustworthy behavior around the body.}

\subsection{Autonomy Configuration and Role Negotiation in SRLs}

Research on wearable and SRLs has increasingly explored shared autonomy modes, ranging from fully manual to hybrid
shared control to fully autonomous for enhancing human performance and experience~\cite{goodrich2007human,dragan2013policy}.
Early studies in prosthetics and haptics emphasize the critical importance of embodiment and sense of control, showing
that systems must synchronize with user intent to be perceived as true extensions of the body~\cite{makin2017ownership,raspopovic2014restoring,zhou2025survey}.
\mm{More recent HRI work explicitly links robot autonomy levels to users’ experienced agency and trust: for example, changes in
autonomy can reduce people’s felt control even when objective performance improves~\cite{wozniak2024influence}, and shared
autonomy schemes in assistive robotics must carefully balance corrective intervention against preserving the user’s sense
of agency~\cite{collier2025sense}.} Beyond sensory integration, HRI research has also revealed that users expect robots to adhere
to spatial comfort and personal boundaries, using spatially-aware navigation or adaptive proxemic zones to regulate inter-agent
distance and orientation~\cite{hall1966hidden,takayama2009influences,mead2013automated}.

In the SRLs domain, WRLKit, a toolkit facilitating users’ location-based customization of wearable limb design, but it omits examination of real-time autonomy decisions or interaction nuances across bodily regions~\cite{viteckova2013wearable}. Similarly, wearable exoskeleton research emphasizes usability and user-centered design in the lower body~\cite{herrera2023qualitative}; however, the upper body SRLs space lacks investigation into how autonomy configurations (manual/shared/auto) interact with bodily proximity and user comfort.

Moreover, while several studies have explored global autonomy preferences (e.g., adjustable social autonomy in museum robots, none systematically address whether and how autonomy expectations vary depending on which part of the body a wearable limb operates in~\cite{newman2022helping}. That is, existing work treats autonomy as a monolithic property of the system, overlooking the potential for spatially adaptive autonomy, where configurations shift based on bodily region and proximity.

A clear gap lies at the intersection of three axes: prior work has not empirically mapped how user perceived safety, trust and embodiment varies with spatial context and consequence risk during near-body SRLs action. We address this by examining how autonomy should adapt to bodily region and proximity, and by deriving practical design principles for spatially adaptive autonomy that bridge theory and systems, grounded in user-driven measures of comfort, trust, and control.

\section{EXPERIMENTAL METHOD}
This study aims to elicit user-defined policies for the perceived safe operation of SRLs near the body, including where approaches are acceptable, how movements should be conducted, and when to delegate actions or require confirmation. To reduce fatigue and learning effects, we counterbalanced condition order and inserted breaks between tasks, using a primed elicitation protocol with a Wizard-of-Oz setup across two embodied tasks.

\mm{\subsection{Study Design and Research Goals}}

We employed a within-subjects, mixed-methods design to examine user perceptions of supernumerary robotic limbs (SRLs)~\cite{sasaki2017metalimbs}.
Given the nascent state of SRL proxemics, our study adopts an exploratory approach that combines two analytic layers: a generative elicitation of spatial policies (RQ1) and an assessment of assurance and arousal profiles under different autonomy configurations (RQ2–RQ3).
For RQ1, we adapted a User-Defined Elicitation (UDE) workflow~\cite{wobbrock2009user}, treating near-body interaction scenarios as \textit{referents} and participants’ preferred spatial boundaries and autonomy rules as \textit{symbols}, so that SRL policies emerge bottom-up from users’ own bodily comfort constraints.

To contextualise these user-defined policies (RQ2 and RQ3), we manipulated \textit{Autonomy Configuration} as a within-subjects factor with two levels: a \textbf{High-Autonomy Anchor} (a fixed, standardised behavior protocol specifying entry paths, segmented timing, and cueing, applied identically across participants~\cite{zhou2024coplayingvr,zhou2024pairplayvr}) and \textbf{Participant-Defined Rules (PDR)} (participant-authored, zone- and segment-specific constraints that the wizard enacted consistently during the subsequent task runs). While the study remains exploratory rather than confirmatory, prior work on agency guided our directional expectations for this comparison~\cite{wozniak2024influence,collier2025sense}: specifically, that \hypertarget{h1} {\textbf{(H1)}} high-autonomy entries would elicit higher event-locked SCR peaks due to reduced predictability, and that \hypertarget{h2}{\textbf{(H2)}} PDR would yield higher perceived safety and capacity-oriented trust by strengthening the user’s sense of control.

\subsubsection{Alignment of Research Goals and Measures}
\mm{To make the methodological alignment explicit, Table~\ref{mapping} maps each research question to the corresponding task segments, data sources, and dependent variables, linking qualitative spatial rules (RQ1), event-locked physiological signals (RQ2), and subjective assurance profiles (RQ3). Detailed construct definitions and justification for each measurement choice (e.g., capacity trust) are provided in Sec.~\ref{Data_Collection}.}

\begin{table*}[t]
  \centering
  \footnotesize
  \caption{Overview of Experimental Design: Alignment of research questions, task contexts, and measures.}
  \label{mapping}
  \renewcommand{\arraystretch}{1.3}
  \begin{tabularx}{\textwidth}{p{0.16\textwidth} p{0.22\textwidth} p{0.24\textwidth} X}
    \toprule
    \textbf{Research Goal (RQ)} & \textbf{Task Context} & \textbf{Data Sources} & \textbf{Dependent Variables \& Rationale} \\
    \midrule

    \textbf{RQ1: Spatial Policies} \newline (generative elicitation)
      & Comfort Zone Task \& \newline Control Handover Task
      & Think-aloud protocol; \newline
        Semi-structured interviews; \newline
        Participant-authored rules
      & \textbf{Outcomes:} zone- and component-specific boundaries, motion and cueing strategies, synthesised into the SRL Proxemics framework. \newline
        \textit{Rationale:} derives bottom-up guidelines grounded in users’ own near-body experience rather than designer assumptions. \\
    \midrule

    \textbf{RQ2: Physiological Arousal} \newline (exploratory assessment)
      & Interaction phases \newline
        (standardised entries across autonomy modes)
      & Skin conductance responses (SCR), analyzed per phase
      & \textbf{Metric:} baseline-corrected and Z-standardised SCR peak amplitudes. \newline
        \textit{Rationale:} objective, time-resolved arousal marker for the exploratory pattern expected in \hyperlink{h1}{H1}. \\
    \midrule

    \textbf{RQ3: Subjective Assurance} \newline (exploratory assessment)
      & Post-condition ratings \newline
        (after completing tasks in each mode)
      & Questionnaires: \newline
        Trust Scale (capacity trust); \newline
        Godspeed (safety); \newline
        AEQ (embodiment)
      & \textbf{Metrics:} within-subject scores for capacity trust, perceived safety, and embodiment indices, compared across autonomy configurations. \newline
        \textit{Rationale:} captures reflective assurance and bodily integration, informing the pattern anticipated in \hyperlink{h2}{H2}. \\
    \bottomrule
    \Description{The table summarizes how each research question is connected to the tasks, data sources, and dependent variables in the study. The first column lists three research questions: RQ1 on spatial policies (generative elicitation), RQ2 on physiological arousal (exploratory assessment), and RQ3 on subjective assurance (exploratory assessment). The second column describes the corresponding task contexts: RQ1 uses the Comfort Zone and Control Handover tasks; RQ2 focuses on standardized near-body entry phases across autonomy modes; RQ3 uses post-condition ratings collected after each autonomy configuration. The third column lists data sources: think-aloud, semi-structured interviews, and participant-authored rules for RQ1; phase-locked skin conductance responses (SCR) for RQ2; and three questionnaires (trust, Godspeed safety, and AEQ embodiment) for RQ3. The final column states the dependent variables and rationale: RQ1 produces zone- and component-specific policies aggregated into SRL Proxemics; RQ2 uses baseline-corrected, z-standardized SCR peaks as an objective arousal marker; and RQ3 compares within-subject scores for capacity trust, perceived safety, and embodiment across autonomy configurations to assess subjective assurance.}
  \end{tabularx}
\end{table*}

\mm{\subsection{Generative Elicitation}}\label{Generative_Elicitation}

\mm{To bridge user intent and SRLs behavior, we adopted a user-defined elicitation (UDE) method~\cite{wobbrock2009user,Ruiz2011CHI}. 
In traditional UDE, participants are shown a referent (desired effect) and asked to generate a symbol (input). 
In our adaptation, near-body interaction scenarios served as referents, while participants' preferred spatial boundaries and autonomy rules functioned as the elicited symbols. This treats users as experts on their own bodily comfort and sense of assurance, allowing candidate SRL policies to emerge bottom-up rather than from a priori assumptions.}

\subsection{Participants}

We recruited 18 participants (6 female, 12 male; $M=27.72$, $SD=9.91$; 16 right-handed, 2 left-handed), all with prior technology-interaction experience, a sample size consistent with mixed-methods elicitation norms~\cite{caine2016local,muehlhaus2023need}. We selected experienced participants because participants with little to no experience may be distracted by the novelty of technology-interaction and their near-body motions instead of the experience of defining and evaluating near-body SRLs policies. We recruited participants in the local area through an internal mailing list, word-of-mouth advertisement, and by snowball recruitment. Our participants' backgrounds spanned psychology, biology, physics, mathematics, computer science, art, mechanical engineering, and AI; three had off-body arm experience. All provided written informed consent prior participating in the study.

\subsection{Wizard-of-Oz Setup}
All sessions were conducted in a quiet 2.5\,m $\times$ 3.5\,m laboratory. Participants issued voice commands to the system; the operator listened and executed changes without verbal response. Interactions were video recorded, and a camera captured the participant’s activities for later analysis.

\subsubsection{Prototype}
We build on the \textit{MetaLimbs} wearable SRLs system~\cite{sasaki2017metalimbs}, which consists of two back-mounted, 7-DOF robotic arms with anthropomorphic hands, selected for safe, human-scale motion near the body (see Figure~\ref{tasks}). To support proxemics-focused testing, we made three changes: (i) We added a slim, detachable linkage at the end of the SRLs so a concealed operator could steer the limb from outside the participant’s personal space. The linkage transmits coarse pose commands while leaving the \mm{limb}’s joints free to follow smooth shoulder and wrist arcs, yielding natural approach/retreat motions; (ii) soft pads and shells at contact points to reduce perceived threat and enable safe close approaches;(iii) conservative software limits on joint ranges and velocities in the near-torso workspace, constraining maximum speed and reach to a predefined safe envelope. The operator was concealed from the participant’s main
field of view, so that participants interacted visually and physically with the SRLs rather than the operator. These changes were made to control the motion envelope presented to participants while remaining agnostic to any particular autonomy controller.

\subsubsection{Wizard-of-Oz Control}
Limbs were operated in a WoZ manner adapted from prior on-body SRLs elicitation~\cite{muehlhaus2023need}. The operator rehearsed a small library of trajectories (approach/entry, handover, support, withdrawal). To approximate imperfections typical of autonomous systems while maintaining experimental control, we used a scripted error-injection protocol during WoZ operation: short actuation delays, small transient angular perturbations on a single joint mid-movement, and brief joint freezes. Magnitudes were tuned in pilots to be noticeable yet safe, prompting strategy adaptation without physical risk. Safeguards included software joint limits, padded contact surfaces, and dual emergency stops. The protocol’s role is to keep stimuli realistic while preserving internal validity.

\noindent \textbf{\textit{\mm{Interface Design and Motion Dynamics.}}}
\mm{To generate consistent, human-like SRLs motions, the wizard controlled the \mm{limb} via a direct mechanical linkage attached near the end-effector, following prior work that similarly used body-driven control to elicit interaction strategies for wearable robotic limbs~\cite{muehlhaus2023need}. 
This ``master--slave'' rod allowed the operator to guide the SRLs by physically moving their hand in 3D space. 
We chose this mechanical interface to provide effectively zero control latency and high-fidelity, biologically plausible motion, which was important for maintaining the tempo of the high-paced Control Handover task (items released every 3--5\,s) without excessive operator workload. 
By physically coupling the operator's proprioception to the robot, the setup avoided the additional cognitive load and delay of GUI- or keyboard-based teleoperation. In practice, the wizard was able to follow this 3--5\,s release tempo throughout all sessions without skipped items or unplanned pauses, indicating that the interface was sufficient to handle the task pace reliably.}

\noindent \textbf{\textit{\mm{Operator Training and Safety Measures.}}}
\mm{A single experienced operator served as the wizard for all sessions to minimise inter-operator variability. 
Before the study, the wizard completed several hours of pilot training to standardise approach velocities and trajectories.}
\mm{During trials, the wizard followed a written protocol (Appendix~\ref{app:operator}) to keep behavior consistent and to avoid reacting to participants' social cues (e.g., gaze or body flinches) in the \textit{High-Autonomy Anchor} condition. }
\mm{We distinguished between scripted errors (intentional jitter specified in the protocol) and execution errors (unintended slips). Execution errors were rare and, when they occurred, were noted on a session log. If an execution error occurred (e.g., unexpected contact with the body or environment), the trial was immediately stopped and re-run from the beginning, and only the error-free repetition was retained for analysis. }
\mm{A stop button was placed on the table within the participant’s reach as a visible affordance to request an immediate interruption. In the event of activation, the operator must immediately cease motion and halt the trial, and the experimenter monitored for such requests throughout.}

\subsection{Procedure}
\mm{Each session followed a user-defined elicitation (UDE) protocol~\cite{wobbrock2009user,Ruiz2011CHI} and lasted approximately 60\,min. 
The two autonomy blocks (High-Autonomy Anchor and Participant-Defined Rules) were presented in counterbalanced order across participants.} The study received ethics clearance from the University of Tokyo (E2025ALS280).

\mm{\noindent \textbf{Setup and familiarisation.}
At the start, participants provided informed consent and demographics. The experimenter introduced the exploratory aim and fitted the SRLs and SCR sensors. To mitigate novelty effects, participants completed a 3--5\,min familiarisation period, freely exploring basic movements and practicing voice commands until comfortable. Data from this phase were not analysed.}

\mm{\noindent \textbf{High-Autonomy Anchor.}
In the \textit{High-Autonomy Anchor} block, participants performed the \emph{Comfort Zone} and \emph{Control Handover} tasks under a standardised semi - autonomous behavior profile adapted from prior work~\cite{zhou2024coplayingvr}. The concealed wizard followed a strict script (Appendix~\ref{app:operator}) specifying trajectories, timing, and occasional scripted perturbations to provide a consistent baseline experience. Participants were encouraged to think aloud throughout the interaction.}

\mm{\noindent \textbf{Participant-Defined Rules.}
In the \textit{PDR} block, structured to support generative design, participants first completed a brief think-aloud warm-up~\cite{ericsson1993protocol}, then entered the scenario-based elicitation stage. For each of the four interaction scenarios, they observed a baseline motion, critiqued it, and formulated alternative rules. The wizard implemented these proposals in real-time, allowing participants to iteratively refine the behavior until it matched their intended policy. To minimize legacy bias, participants were encouraged to propose multiple versions rather than settling on the first option; trigger conditions and cues were recorded for each finalised rule. Following this authoring phase, participants re-executed the \emph{Comfort Zone} and \emph{Control Handover} tasks using their own finalised PDR policies, again under continuous SCR recording.}

\noindent \textbf{Measures and debriefing.}
During all task runs, the experimenter time-stamped key interaction phases to align with continuous SCR recording. Subjective questionnaires (Trust, Godspeed, AEQ) were administered immediately after completing each autonomy block. Finally, a semi\allowbreak-\allowbreak structured interview invited participants to compare the configurations and reflect on boundary violations. The session concluded with debriefing.

\subsection{Interaction Scenarios and Task}
\mm{We designed our two tasks as controlled surrogates for specialised, high-demand SRL applications in which users are already bimanually loaded~\cite{parietti2016supernumerary,vatsal2018design}. 
To keep elicitation safe, we used generic, lightweight objects, while preserving the core spatial and cognitive structure of such scenarios. 
Across the four interaction phases, these tasks (i) elicit zone- and component-specific motion and cueing rules (RQ1) and (ii) provide repeated, comparable near-body entries for analysing arousal (RQ2) and subjective assurance profiles (RQ3).}

\begin{figure*}[t]
\centering
\includegraphics[width=\linewidth]{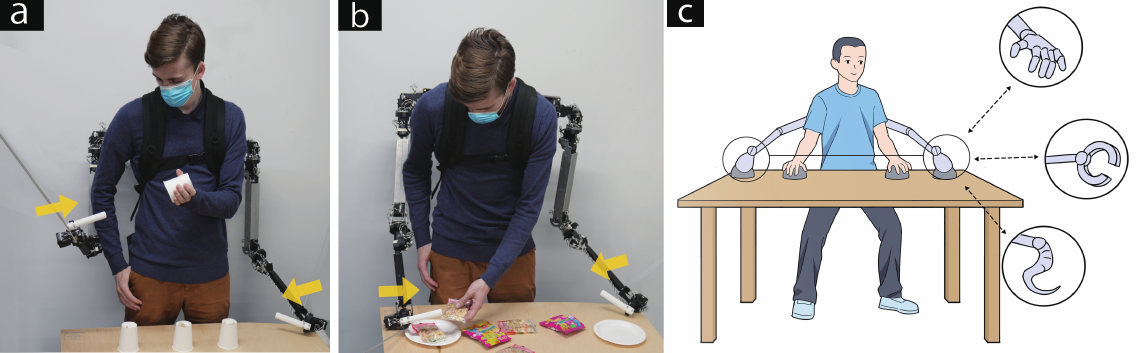}
\caption{Task layouts and wearable SRLs setup. (a) Comfort Zone Task handovers. (b) Control Handover Task: sorting deformable snack packs. (c) Wearable backpack-mounted SRLs with articulated joints and interchangeable end-effectors for near-body manipulation.}
\label{tasks}
\Description{A three-panel figure. (a) A participant wearing a backpack-mounted dual-arm SRLs stands at a table; yellow arrows indicate a handover toward the participant’s near-torso/hand region (Comfort Zone Task). (b) The same setup with deformable snack packages on the table; yellow arrows indicate the SRLs approaching for sorting/placement (Control Handover/Dynamic Collaborative Sorting). (c) A schematic showing a person with two back-mounted robotic \mm{limbs}; callouts depict example end-effectors (e.g., gripper/hand, hook-like tool, curved/tool tip), illustrating articulated joints and interchangeable end-effectors for near-body work.}
\end{figure*}

\subsubsection{Interaction Phases}

We employed four near-body interaction phases: approach/entry, handover, support/stabilize, and withdrawal/repositioning, to elicit participant-defined interaction policies rather than to test specific factor combinations. These phases were informed by prior work on wearable proxemics and interaction boundaries~\cite{profita2013don,zeagler2017wear}, spatial negotiation~\cite{sisbot2007human,rios2015proxemics}, safety control~\cite{kirschner2022iso}, and handover/co-manipulation~\cite{ortenzi2021object}. Each scenario was enacted through simple scripted sequences, with minor Wizard-of-Oz variations to guide reflection, not as part of a factorial design~\cite{riek2012wizard}.
For example, a typical sequence might involve: (1) approach, where the SRLs moved toward the participant either from a peripheral side or frontal trajectory; (2) entry, as the \mm{limb} entered reachable space or crossed the midline; (3) handover, where the object was transferred to or received from the participant’s hand; and (4) withdrawal, in which the SRLs retracted along a predefined path. These stages enabled situated reflection on bodily proximity, attentional demand, and acceptable autonomy behaviors.

\subsubsection{Comfort Zone Task}

The first task was a \emph{Comfort Zone Task}, adapted from the object handover paradigm in HRI~\cite{ortenzi2021object}, chosen for its ability to elicit coordination, trust, and peripersonal space negotiation (see Figure~\ref{tasks} (a)). To focus on interaction kinematics, a plain white cylindrical cup was used to minimize salience and orientation demands.

Each trial comprised 10 structured subtasks designed to probe autonomy preferences across spatial configurations. Subtasks varied along two key dimensions: directionality (inbound vs. outbound) and spatial configuration (ipsilateral vs. cross-body), forming four core interaction types. In inbound subtasks, participants received an object from the SRL; in outbound subtasks, they handed an object to the SRL. Interactions were either ipsilateral (on the same side of the body) or cross-body (requiring midline crossing). For instance, in one subtask, the right SRL handed a cup across the body to the participant’s left hand, entering the space in front of the torso. In another, the participant gave a cup from their right hand to the SRLs on the same side, engaging only the lateral elbow corridor.

We included two exemplars of each type, and selected two additional subtasks to balance lateral symmetry and enrich spatial coverage. For instance, by introducing frontal approaches toward the sternum and underarm deliveries near the waist.

These variations allowed us to systematically probe participant reactions across sensitive proximal areas (e.g., sternum, upper torso, neckline) and distal lateral zones (e.g., side waist, ribs). Within our design, the Comfort Zone Task primarily addresses RQ1 by eliciting fine-grained zone- and component-specific boundaries, motion strategies, and cueing requirements, while its standardised entry segments also supply the repeated near-body movements used for SCR (RQ2) and assurance comparisons (RQ3).

\subsubsection{Control Handover Task}

The second task was a \emph{Collaborative Sorting Task}, designed to probe how participants negotiated control and responsibility in a high-paced, semi-structured setting under human-led shared autonomy~\cite{pellegrinelli2016human,Parasuraman2000Automation} (see Figure~\ref{tasks} (b)). Building on prior work in robotic multitasking and asymmetrical coordination, this task mimicked a simplified industrial workflow where speed, spatial reasoning, and delegation were critical~\cite{Guiard1987Bimanual,Wickens2008MultipleResources}.
Participants collaborated with two shoulder-mounted SRLs to sort a total of 20 packages (10 red and 10 white) into two designated containers located to their left and right. The sorting rule was simple: red packages went into the left container, and white ones into the right. However, task dynamics introduced unpredictability and required moment-to-moment decision-making~\cite{Endsley1995SA,Wickens2008MultipleResources}. At the beginning of each trial, the workspace was empty. Packages were dropped one at a time into random locations within the shared workspace between the participant and the SRLs. These locations varied in lateral distance and elevation, with some landing near the participant and others nearer to one of the SRLs~\cite{Lasota2017HRC}. This intentional asymmetry in reachability and visibility elicited real-time delegation choices and revealed how participants allocate control under spatial pressure.
To introduce time pressure and encourage real-time strategy development, the items were released one at a time but at unpredictable intervals, occasionally resulting in multiple packages visible at once. \mm{The Control Handover Task complements the Comfort Zone Task by extending RQ1 into a high-paced, semi-structured setting where users negotiate turn-taking and shared workspace policies, while also supplying additional near-body entries and delegation episodes for the exploratory arousal and assurance comparisons (RQ2–RQ3).}


\subsection{Measures and Analysis}

\mm{We used established subjective and physiological measures to capture components of users’ assurance in near-body interaction, focusing on perceived safety, capacity trust, and event-locked arousal.}

\subsubsection{Data Collection}
\label{Data_Collection}
\mm{We collected three complementary data streams, selected to triangulate the qualitative, physiological, and subjective components of our research questions (Table~\ref{mapping}).}

\noindent \textbf{\textit{\mm{Qualitative Strategy (RQ1).}}} 
To elicit user-defined spatial policies, we recorded think-aloud sessions and post-task debriefs, then transcribed and theme-coded them to derive a taxonomy of spatial–autonomy rules.

\noindent \textbf{\textit{\mm{Physiological Arousal (RQ2).}}}
To complement subjective reports, we recorded skin conductance responses (SCR) using EmotiBit sensors (100 Hz, forearm electrodes)\footnote{\url{https://www.emotibit.com/product/all-in-one-emotibit-bundle/}}. SCR is a validated marker of autonomic arousal during near-body robot approaches~\cite{ogawa2020you}, offering a time-resolved index that complements self-report measures.

\noindent \textbf{\textit{\mm{Subjective Assurance and Embodiment (RQ3).}}}
Post-\hspace{0pt}condition questionnaires captured users’ assurance under each autonomy configuration:
\mm{
\begin{itemize}[leftmargin=*]
    \item \textit{Perceived safety and intelligence:} We used the relevant subscales of the Godspeed questionnaire~\cite{bartneck2009measurement}. Although newer instruments such as RoSAS have been proposed~\cite{carpinella2017rosas}, we retain Godspeed to ensure continuity with HRI and wearable-robotics studies on body-adjacent robots and augmentation~\cite{mandl2022embodied}. We use Godspeed mainly for comparability with prior work, not as standalone construct validation; future studies can further streamline to task-relevant subscales.
    \item \textit{Trust:} We used the 12-item Trust in Automated Systems scale by Jian et al.~\cite{jian2000foundations} on a 7-point Likert scale. In line with multi-dimensional trust frameworks~\cite{ullman2019mdmt}, we interpret responses as reflecting capacity trust (i.e., perceived competence, reliability, and safety of the SRL). Consistent with prior work, we computed scores for two distinct dimensions: \emph{Trust} and \emph{Distrust}, and administered items post-condition without rewording. Human–machine comparison prompts were omitted to focus on absolute assessments of the SRL’s behavior.
    \item \textit{Embodiment:} We used an adapted Avatar Embodiment Questionnaire (AEQ) focusing on body ownership, agency, and body image in near-body interaction~\cite{gonzalez2018avatar,umezawa2022bodily} (adaptation details and reliability in Appendix~\ref{Embodiment}). AEQ was administered post-conditions to capture condition-specific embodiment under different autonomy regimes, mirroring staged designs in prior enhanced-limb research~\cite{umezawa2022bodily}.
\end{itemize}
}

\subsubsection{Data Analysis}

Thematic analysis follows Braun and Clarke’s reflexive approach with Framework matrix charting; code frequencies indicate descriptive coverage, not saturation.
We adopt a split-resolution approach~\cite{mcduff2012affectaura} to accommodate the temporal mismatch between subjective and physiological measures. SCR captures rapid, local responses during entry, while session-level questionnaires (Godspeed, AEQ, Trust) assess reflective post-task judgments. SCR signals are baseline-corrected ($-$5 to 0\,s), peak-extracted (1--5\,s post-phase), log-transformed, and Z-standardized~\cite{dawson2007electrodermal}. Analyses contrast \textit{High-Autonomy Anchor} vs. \textit{PDR} entries, with descriptive summaries across spatial zones. Questionnaires are analyzed via paired Wilcoxon signed-rank tests with non-parametric effect sizes ($r$) and item-level medians reported. Between-phase contrasts are treated as exploratory.

\section{RESULTS}

In this section, we present our findings from mixed-methods analysis on near-body interaction with back-worn SRLs. To reflect the structure of our research questions, we report qualitative results first (RQ1) to frame the presentation of quantitative analyses (RQ2, RQ3). Unless otherwise noted, our quantitative results pool participant responses across the full set of scenarios.

\subsection{Qualitative Findings}

We conducted a reflexive thematic analysis of transcripts from participants' think-aloud sessions and post-task interviews, following Braun and Clarke's six-phase framework~\cite{braun2006using,braun2021one}. Consistent with an interpretivist epistemology, we treated coding as an analytic, interpretive act, rather than an objective classification~\cite{braun2019reflecting}. Coding treated the interaction phase (one of approach/entry, hand-over/manipulation, support/stabilize, or withdrawal/retreat) as the analytic unit, linking co-temporal think-aloud or interview utterances to that phase~\cite{miles2014qualitative,ericsson1993protocol}. Two researchers collaboratively developed initial open codes from a seed subset through joint familiarisation and reflexive memo-ing; these codes were iteratively refined into candidate themes via ongoing discussion, reorganisation, and clarification of inclusion/exclusion criteria~\cite{braun2021one,saldana2016coding}. We included only utterances with explicit spatial referents (body region, distance, trajectory, posture) tied to an interaction phase; generic remarks were memo-ed but not coded~\cite{saldana2016coding,miles2014qualitative}. To enable cross-participant comparison and design translation, we applied the Framework Method for matrix-based charting and alignment~\cite{gale2013using,ritchie2002qualitative}. In line with reflexive thematic analysis (reflexive TA), counts are reported as descriptive coverage to aid design communication, not as indicators of theme strength or for inference~\cite{braun2021one,braun2019reflecting}. Theme development was documented via merge/split decisions with an audit trail (memos, decision logs) and peer debriefs; we also ran a one-off ``cold'' co-analysis on a 15\% stratified subset to probe boundaries and negative cases. Consistent with reflexive TA, we did not compute inter-rater reliability, prioritising reflexivity over coder agreement metrics~\cite{braun2021one,nowell2017thematic,tracy2010qualitative,lincoln1985naturalistic}. Quotes are tagged by source as TA (think-aloud) or INT (interview) with participant IDs. Our analysis yielded three thematic layers: (1) embodied spatial boundaries, (2) component-specific motion and autonomy preferences, and (3) user-authored rules and delegation logics for SRLs control.

\subsubsection{\textbf{``My body has regions with rules'' — Eliciting a Body-Centric Trust Map} (RQ1)}\label{Movements}

\begin{figure*}[t]
\centering
\includegraphics[width=\linewidth]{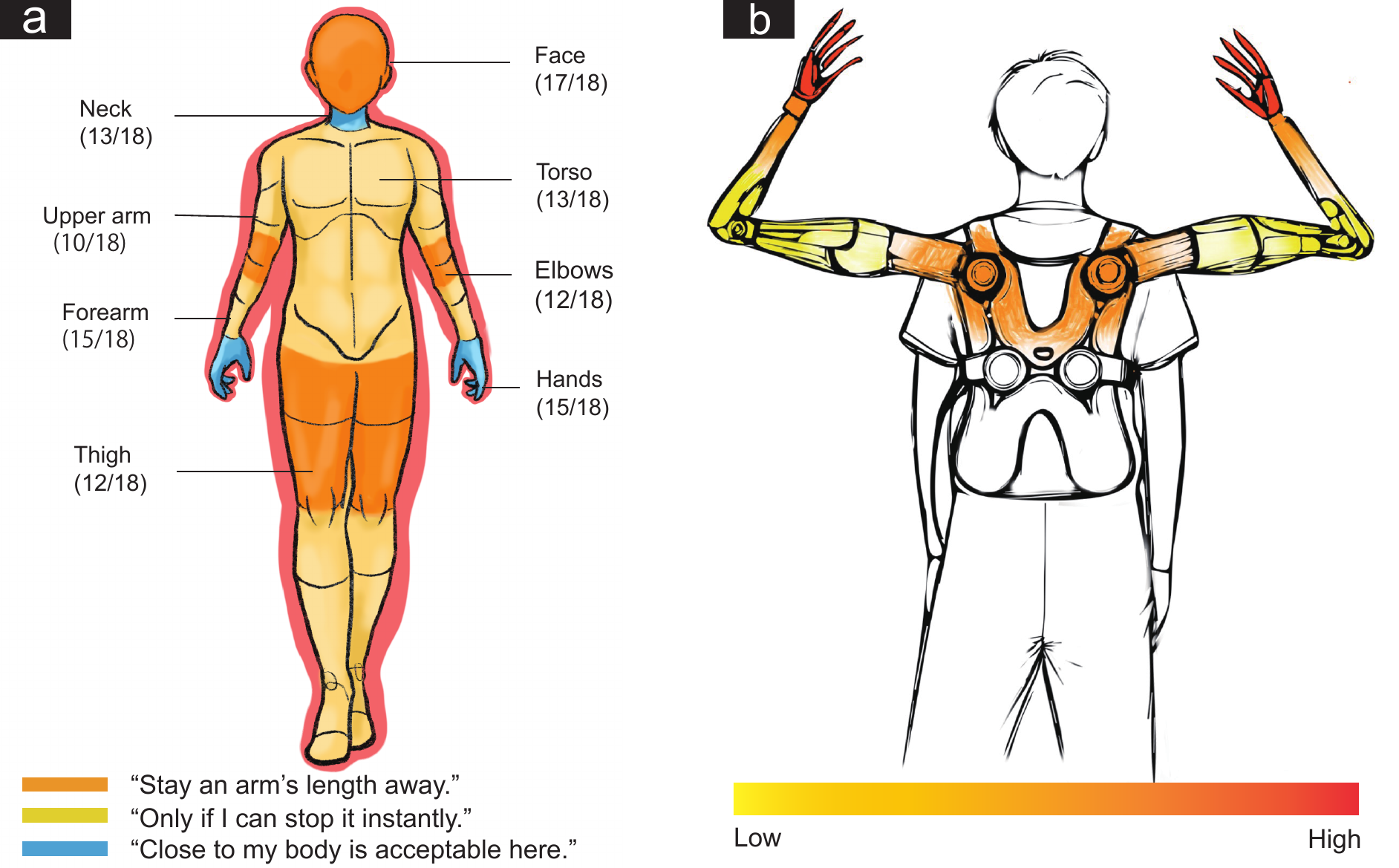}
\caption{Participant-defined sensitivity zones: orange = \emph{``stay an arm’s length away''}; yellow = \emph{``only if I can stop it instantly''}; blue = \emph{``close to my body is acceptable here.''} Labels reflect participants’ desired standoff distances for SRLs entry by body region. Numbers (e.g., 17/18) show how many participants assigned each zone that constraint.
(b) Participant-defined segment-level autonomy: Color gradient (yellow to red) reflects reduced comfort with autonomy: yellow for manual/confirm-only control, red for higher autonomy.}
\label{maps}
\Description{Two subfigures showing user preferences for SRLs proximity and autonomy. (a) A front-facing human figure with body parts colored in three levels: orange (e.g., face, torso, thighs) indicates ``keep at least one arm's length away''; yellow (e.g., upper arms) indicates ``close is acceptable if I can stop it''; and blue (e.g., hands, forearms) indicates ``can be worn close to the body.'' Labels show how many out of 18 participants agreed per body part (e.g., Face: 17/18). (b) A rear view of a person wearing back-mounted robotic limbs. The limbs are colored from yellow at the base to red at the hands, indicating increasing levels of caution or decreasing tolerance for unconfirmed autonomy. Red zones highlight regions where users prefer minimal or no autonomous motion.}
\end{figure*}

This subsection characterises how participants negotiated bodily boundaries with a wearable SRLs by combining a region-wise sensitivity map with user-defined safe-distance bands. Participants' descriptions converged on a clear taxonomy of sensitivity, as shown in Figure~\ref{maps} (a).

\noindent \textbf{\textit{Regions Treated as Out-of-Bounds.}} Participants commonly treated the head and face as the most sensitive and prohibited zones, typically maintaining a standoff distance of about an arm's length (17/18). Most participants (17/18) rejected any approach into this area regardless of speed or trajectory: ``\emph{I don't want the arm near my head}'' marked P03. However, majority (15/18) emphasized that predictability and user control influenced comfort near these regions. ``\emph{If I ask it to come closer just for the handover, its ok, because I'm expecting it and I can stop it}'' P08 said. Notably, many participants located their boundary within the frontal visual cone (i.e., the cone-shaped area directly visible in front of them), suggesting discomfort when the SRLs entered from the front, particularly when pointing directly at the face. These references highlight the importance of anticipation and user initiation, rather than proximity alone, in modulating perceived intrusiveness in the head/face region.

The elbow area was also widely considered a no-go zone (12/18), often described as prone to involuntary flinching. While the forearm was generally accepted, participants indicated that even brief contact at the elbow could trigger defensive reactions. As P04 noted, ``\emph{The forearm is fine, but the elbow is a no-go zone, because bump here made me hurt}''. Only a small number of participants (6/18) allowed brief proximity to the head or elbow, and only under strict conditions of user initiation or clear task framing.

\begin{figure*}[t]
\centering
\includegraphics[width=\linewidth]{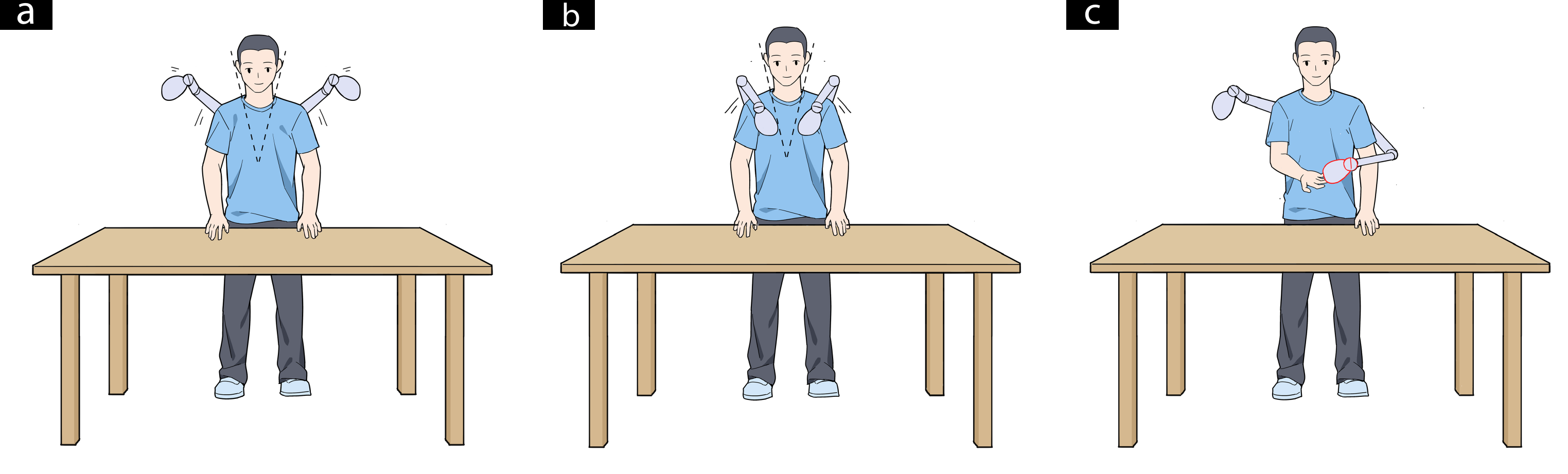}
\caption{Illustrations of participant-preferred SRLs behaviors. (\textbf{a--b}) Participants preferred SRLs to approach sensitive regions (e.g., head, face) using ``pause–move–pause'' styles with curved trajectories, rather than direct or unpredictable paths. (\textbf{c}) Autonomous actions were most often assigned to the distal segments (e.g., hands) of the non-dominant-side SRLs to support manipulation while preserving clear roles.}
\label{movement_prefs}
\Description{Three-panel diagram showing supernumerary robotic limbs (SRLs) mounted to a user's back. Panel (a) and (b) show the SRLs moving toward the user's head or shoulders in a deliberate, arced, and staged manner, emphasizing sensitivity near the face. Panel (c) shows the right-side SRLs (user’s non-dominant side) holding an object, highlighting participant preferences for autonomous behavior localized at the distal components (e.g., robotic hand).}
\end{figure*}

\noindent\parbox{\linewidth}{%
  \textbf{\textit{Supervisory Zone: Torso/Waist and Upper-Arm\allowbreak\textendash\allowbreak Buffered Proximity for Brief, Expected Moves.}}%
}\  The torso and waist were generally treated as conditionally accessible zones (13/18). Participants allowed brief entries but required clear purpose and user anticipation, especially for task-relevant handoffs. As P06 framed it: ``\emph{If you think of a person as a container ... if you touch my torso directly, it feels like too many boundaries have been crossed}''. The commonly accepted proximity was around a palm-width: ``\emph{About a hand away is comfortable; closer than feels like it's in my space}'', noted P02. Some participants (8/18) described relaxing this buffer when workload was high and rapid access preserved task flow: ``\emph{Normally I'd keep it away from my chest, but when I was holding the two parts and just needed the tool, I was fine with it coming closer}'', said P06.

Adjacent to the torso, the upper-arm region was also treated with caution (10/18). While not prohibited outright, participants requested spatial buffering to avoid tension or distraction. A fist-width of space was commonly mentioned: ``\emph{Along my upper arm, keep roughly a fist of space. When it brushes that area I tense up}'' (P15). These responses position the upper-arm as a caution zone, not off-limits, but movements there should be slower, indirect, and well signaled.

\begin{sloppypar}
\noindent\textbf{\textit{Utilitarian Zone:\allowbreak\ Hands, Forearms, and Periphery\allowbreak\ --\allowbreak\ Legible Near-Contact;\allowbreak\ Elbow as Caution Boundary.}}\ 
The hands and forearms were accepted as the SRL's active workspace (15/18). Near-contact and touch were permitted when aligned with task context, as P04 put it: ``\emph{Around my hands is its workspace},'' and P05 added, ``\emph{Even a brief brush during handover is okay if it leaves right away. A finger-width is fine for passing}.'' However, this tolerance held only for brief periods and when the \mm{limb} remained in view.
\end{sloppypar}

In contrast, acceptance dropped sharply at the elbow. A majority (14/18) identified this joint as a reflexive boundary, even within the otherwise accepted peripersonal zone. P14 described: ``\emph{Leave about a fist of space around my elbow}'', and it added ``\emph{if you graze it, I’ll flinch, so stay by the forearm, not across the elbow}''. The elbow thus marked a critical shift from utilitarian to protective perception.

The neck was generally considered a low-sensitivity area by most participants (13/18). They rarely flagged it as a concern, likely because it is less exposed to contact in daily activities and seldom involved in common gestures. As P13 put it, \emph{``It's not like anything to touch my neck in normal work anyway, just… not in the way}''. This suggests that people's sense of which body parts feel sensitive isn't just about where they are on the body, but about how often they're used or touched in daily life.

\noindent \textbf{\textit{Context-Dependent Zones: Behind-Body and Lower-Body Constraints.}} Two contexts reliably bent default distance expectations: behind-body interactions and lower-body movement while standing.

Interactions behind the body were often acceptable when paired with non-visual presence cues (e.g., soft hum), enabling participants to track the \mm{limb} without needing to turn (Figure~\ref{movement_prefs2} (c)). In these cases, most (14/18) preferred the SRL's end-effector to maintain roughly an arm’s-length standoff from the participant's back/torso. P10 explained, ``\emph{when it's behind me, about an arm's length feels right},'' and P11 added, ``\emph{With the soft hum I can track it}.''

In contrast, legs were treated as crucial to postural control, especially given that participants (12/18) were standing. As P05 explained, ``\emph{When I'm standing, my legs feel responsible because they're holding up my whole body, so I'm more sensitive about them}.'' Others reinforced the need for increased buffer in these zones: ``\emph{When I'm standing, keep a distance}'' (P08).

Taken together, these patterns constitute a body-centric trust map that specifies zone- and component-level boundaries, preferred approach paths, and cueing/confirmation requirements, directly answering RQ1. They also predict elevated arousal during approach/entry into sensitive fronts and midline crossings, motivating our SCR test in RQ2.

\subsubsection{\textbf{``Stop! Not that close.'' — Negotiating Distance and Autonomy Boundaries}}\label{Boundaries}

\begin{figure*}[t]
\centering
\includegraphics[width=\linewidth]{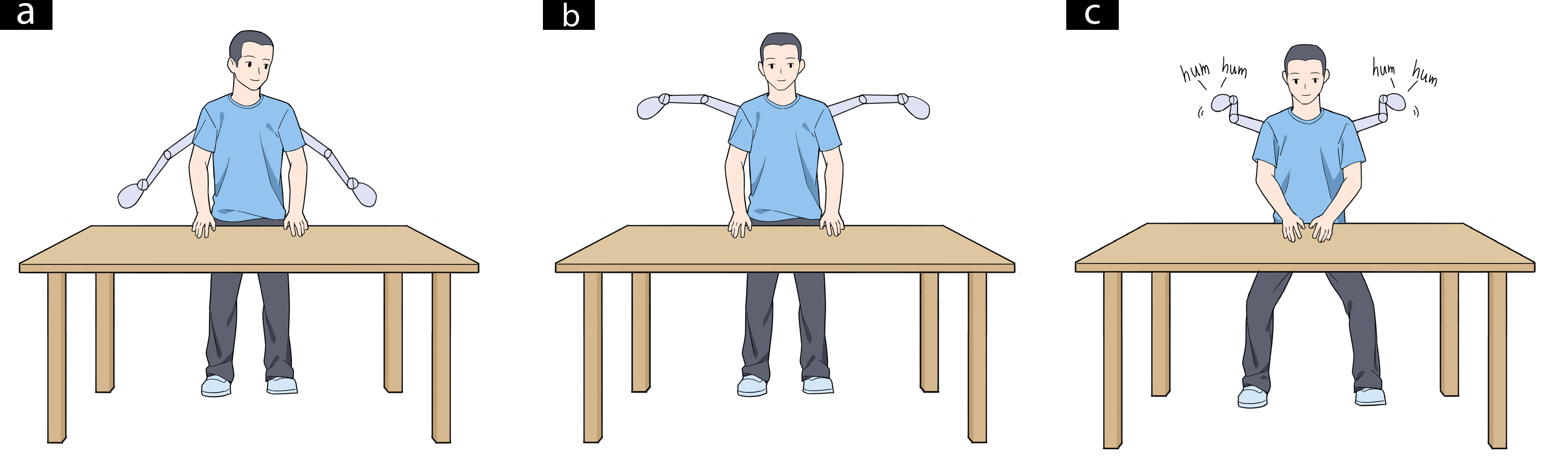}
\caption{Preferred SRLs motion strategies near the body. (a–b) Participants favored arc-shaped, whole-arm trajectories over abrupt vertical or frontal intrusions, especially near the face and torso. Lifting motions were more acceptable when the entire \mm{limb} rose cohesively rather than the hand popping up. (c) Auditory cues like subtle motor hums helped users anticipate SRLs movements occurring behind them.}
\label{movement_prefs2}
\Description{A three-panel diagram of robot \mm{limbs} interacting near a user's body. In (a) and (b), the \mm{limbs} approach in broad, curved arcs from the sides and below toward the torso. In (c), the \mm{limbs} move behind the user, with sound effects (``hum hum'') indicating motor noise.}
\end{figure*}

Participants expressed nuanced expectations for how supernumerary robotic limbs (SRLs) should behave near the body, especially under shared control. Preferences reflected a balance between trust in autonomy and a strong desire for predictability, spatial appropriateness, and minimal ambiguity. These expectations centered on two interconnected aspects: movement preferences in close proximity and autonomy allocation across \mm{limb} segments.

\noindent \textbf{\textit{Movement Preferences in Close Proximity}}
\textit{Segmented rhythm and timing:}
Participants widely preferred a segmented movement rhythm when SRLs entered sensitive areas like the facial zone or upper torso, for example a ``pause–move–pause'' style, as shown in Figure~\ref{movement_prefs} (a-b). This brief, deliberate pause at the edge of a critical boundary (e.g., before entering the frontal visual cone) provided time for users to mentally prepare and reduced surprise. P01 stated, `\emph{`too fast made me feel unpredictable''}. While P12 explained, \emph{``A short stop before coming in helps me brace''}. Rather than blanket slowness, participants emphasized purposeful pacing; P09 noted, \emph{``Creeping makes me nervous. Better to just do it and move away''}. Movements that appeared aimless were often perceived as intrusive or unsettling, such as drifting or hovering without a visible goal. As P05 said, \emph{``I don't know what it's waiting for when it hovers too long, it should either go or pull back''}.

\textit{Spatial Approach and Pathing:} Participants preferred arc-like approach paths over direct frontal or vertical intrusions, especially near the torso and face (Figure~\ref{movement_prefs2} (a-b)). From below, participants were particularly wary of sudden upward motions that brought the end-effector directly into sensitive regions. Many participants explicitly noted that a bottom-up lift should involve the entire \mm{limb}, like raising from the upper arm. \emph{``I saw the hand pops straight up at my face, it's startling. But if the whole arm arcs upward, I feel it coming''} (P03). P11 added, {``Don't sneak up from under my chin''}. Participants described this preference as a way to mitigate startle responses and preserve spatial awareness. \emph{``Don't lift the fingers straight at me even if it's coming to my face''} (P06). This preference mirrors those seen in interactions from the rear or side, where oblique arcs allow for more preview time and reduced perceived threat. \emph{``Diagonal or side entry gives me a warning''} (P10). Exceptions were granted when environmental or task constraints required a direct upward reach, participants insisted on modifying the end-effector orientation, such as angling the tool away from the eyes. \emph{``It's okay to come up if you're carrying something and not pointing it at me''} (P02).

\textit{Coordinating in Shared Workspaces:} Strategies for avoiding interference in shared workspaces emerged as a central concern for most participants (11 out of 18), especially when SRLs operated within overlapping hand zones. 11 out of 18 participants reported instinctively delegating supportive tasks to the SRLs on their non-dominant side. This spatial separation helped avoid collisions and streamline coordination, such as right-handed participants assigning complementary assistance to a left-positioned SRL. As P06 (right-handed) said, \emph{``the left arm takes care of things on its own, I don’t end up bumping into it or getting in the way''}.
To improve coordination and reduce interference, (6 out of 18) participants spontaneously defined informal spatial ``lanes'', specific paths where the robot could move without disrupting their actions. P02 described this as, \emph{``I stay front-right, it goes back-left''}. These self-organized protocols created a predictable rhythm of engagement. However, participants (3 out of 18)  struggled to maintain coordination in time-sensitive moments. \emph{``Sometimes we both reach, and both stop,''} said P11.

\noindent \textbf{\textit{Component-Specific Autonomy Expectations}}

Rather than endorsing global autonomy settings, participants expressed nuanced preferences tied to specific SRLs components, as shown in Figure~\ref{maps} (b). A common expectation was that the robotic hand could act more freely.

\textit{Hand / End-effector — High Autonomy with Guardrails: }
Participants (16/18) commonly treated the hand as the main locus of autonomous, expecting it to handle grasping, alignment, and handovers with minimal instruction, as shown in Figure~\ref{movement_prefs} (c). Often referred to as \emph{``the smart one''} or \emph{``my third teammate''}, (P16) the hand was delegated substantial autonomy. Many favored assigning more autonomous to the SRLs on their non-dominant side to support their dominant hand's precision work. As one right-handed participant (P14) noted, \textit{``My right hand is already doing the precise stuff. I need the left arm to be the smart one that helps without asking''}.
Even with high trust in SRLs, participants still preferred clear coordination rules. In particular, they favored taking turns in shared areas to avoid overlapping movements or spatial interference, one participant (P04) mentioned: \textit{``I want it to do its job, but don't crowd me''}.

\textit{Wrist — Supervisory Autonomy with Clear Boundaries}
Participants (15/18) envisioned the wrist as a mid-tier controller that supports the autonomous hand. While not expected to plan independently, the wrist was welcomed to perform micro-adjustments in orientation and roll, especially during handovers or in preparation for grasping. Participants emphasized that wrist autonomy should remain \emph{``in context,''} functioning like a connector rather than a decider. Majority participants (11/18) commented that a rotating wrist that lingered in their periphery felt unsettling: \textit{``Just roll and place''} (P02). The ideal autonomy model for the wrist was support autonomous distal action through localized adjustment. 9 out of 18 participants preferred fast, decisive wrist motions over slow or overly cautious ones. One participant noted, \textit{``Better to just do it and move away''} (P09). This suggests that participants tolerated very fast motion if purposeful.

\begin{figure*}[t]
    \centering
    \includegraphics[width=\linewidth]{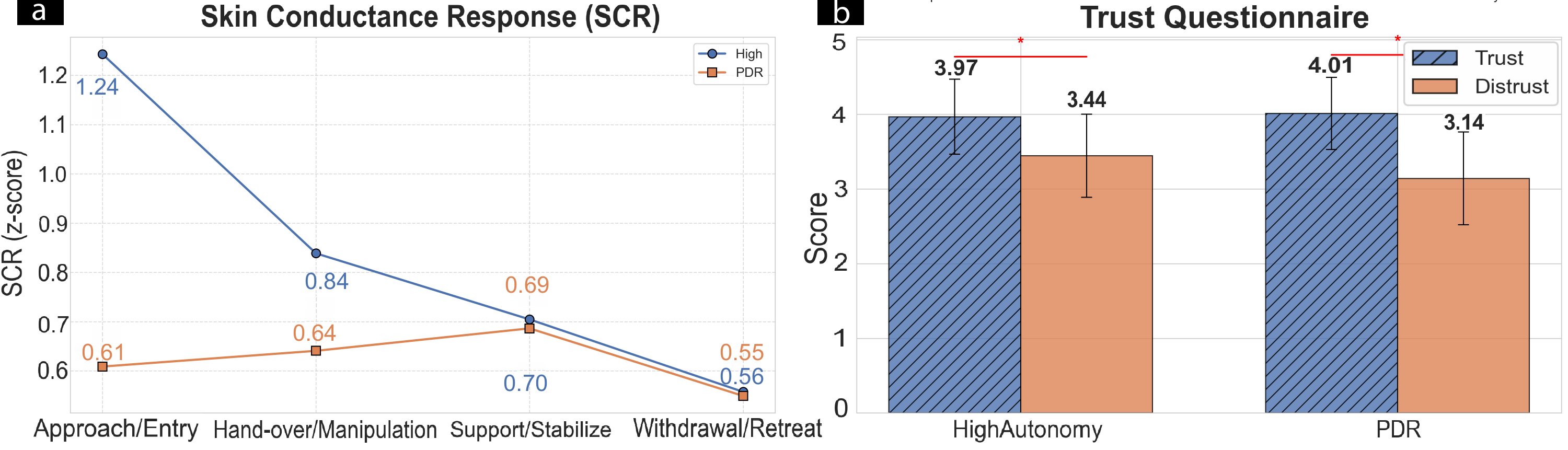} 
    \caption{Skin conductance responses across four interaction phases under two autonomy conditions (Participant-Defined Rules and High-Autonomy Anchor), and Trust in Automated Systems ratings comparing the same two conditions. (a) SCR across interaction stages. (b) Trust and distrust questionnaire ratings across conditions. * indicates $p < .05$.}
    \label{SCR}
    \Description{Two subfigures are shown. Subfigure (a) plots skin conductance response (SCR) z-scores across four SRLs action stages: Approach/Entry, Hand-over/Manipulation, Support/Stabilize, and Withdrawal/Retreat. The High-Autonomy condition shows a steep drop from 1.24 to 0.56 across stages, while the PDR condition remains flatter between 0.61 and 0.55. Subfigure (b) shows bar plots for Trust and Distrust questionnaire ratings under both High-Autonomy and PDR conditions. PDR shows higher trust (around 4.0) and lower distrust (around 3.1) compared to High-Autonomy. Asterisks mark statistically significant differences.}
\end{figure*}

\textit{Elbow — Reflex-Only Autonomy, Strong Spatial Boundaries: }
Participants (12/18) generally viewed the elbow as the sensitive and risky part of the SRLs when it moved near the body. They didn't want the elbow to make its own decisions, especially not to plan ahead. Instead, they preferred it to stay still unless it absolutely had to move in response to something unexpected, like avoiding a collision.
One participant (P11) explained, ``\emph{When I'm standing, I don't trust the elbow to move at all. It's close to my ribs}''. Others also described feeling uneasy when it passed too close to their sides. Several participants (4/18) mentioned that they had a mental \emph{``keep-out zone''} around the elbow, especially when it moved sideways or near their torso. ``\emph{It's okay if it self-corrects to avoid a crash}'' (P03), and P11 added, ``\emph{I usually don't want it planning with the elbow unless it's something small and far from me}''. Overall, participants didn't think of the elbow as something that should act on its own. They saw it more like a structural part that should stay predictable and only move if it had to for safety.

\textit{Upper arm and Base — Manual Control with Divergent Views on Repositioning: }
12 out of 18 participants recognized that the shoulder joint governs the entire \mm{limb}'s pose, and thus plays a pivotal role in facilitating workspace access. Specifically, participants (8/18) permitted autonomous repositioning to facilitate hand-level tasks, as long as such movements were preceded by clear audio or motion cues indicating intent. As P10 explained, \textit{``the hand's moving near my face, I actually feel safer when it's the shoulder not the elbow doing the moving''}. However, 9 out of 18 participants explicitly rejected midline-crossing or frontal repositioning, citing discomfort or surprise. P07 stated, \textit{``A small shoulder nudge is fine''} but emphasized that movements should stay predictable and within a limited range. Notably, 5 participants emphasized the shoulder's role in supporting the hand, rather than leading actions. \textit{``If it's guiding the hand somewhere, then yeah''}, said P05. In contrast to others who preferred stability, few participants  (6/18) accepted occasional repositioning if it helped reduce visual clutter. Still, these movements were only acceptable when the SRLs clearly signaled its intent and waited for user confirmation: \textit{``Just give me a heads-up first''}, said P12.

\noindent \textbf{\textit{Participant-Generated Design Ideas (Exploratory)}}
In the post-session debriefs, participants (11/18) proposed several coordination mechanisms they would like SRLs to support near the body. We report them here as design inspirations, not as tested effects.

\textit{Affect-Responsive Modulation: }Some participants (6/18) proposed SRLs that retreat upon sudden increases in physiological arousal (operationalized as skin conductance crossing a predefined threshold) and remain closer when signals indicate calm. As P07 noted, \textit{``If it detects that I suddenly tense up, it should stop all motion immediately''}.
\textit{Acoustic Proxemics: }Others (5/18) suggested low-salience audio to convey presence and distance. For example, use a soft hum for behind-the-back work, a rising tone near the torso, and silence near the face, so users need not visually track the \mm{limb}. \textit{``Its voice could ramps up when it gets closer to my torso, I can follow it without look''}, P11 noted. 3 out of 18 participants framed this as awareness support; annoyance and habituation trade-offs were not assessed.
\textit{Semantic Role Switching: } A subset (4/18) suggested assigning the \mm{limb} simple roles to switch behavior quickly. For example, \emph{``helper''} when it stays by the user's hand to hold or stabilize, and \emph{``fetcher''} when it works on its own, so they could move between collaboration and delegation without menus. As P02 said, \textit{``I told it ‘You're my third hand now', and it just follow me and help me hold stuff. But after I said ‘Gofer mode', and it should start doing fetching or sorting on its own''}.

These movement and autonomy allocations operationalize user expectations for assured near-body action (RQ1). They further suggest that user-initiated and well-cued entries, particularly those that avoid sensitive frontal zones or midline crossings, may mitigate physiological arousal compared to unannounced approaches, motivating RQ2. These policies also point to higher perceived safety and trust, as well as preserved agency and embodiment, when participants retain rule-setting control over SRLs behavior, as explored in RQ3. \mm{Across all sessions, no stop-button activations occurred. In post-session interviews, several participants noted that while some frontal or near-face approaches felt ``too close'' or ``a bit scary'', they did not perceive the robot as mechanically ``out of control'', and therefore did not feel compelled to hit the stop button.}

\subsection{Quantitative Findings}

\subsubsection{Physiological Data: Skin Conductance Response (RQ2)}\label{SCRs}

\begin{figure*}[t]
    \centering
    \includegraphics[width=\linewidth]{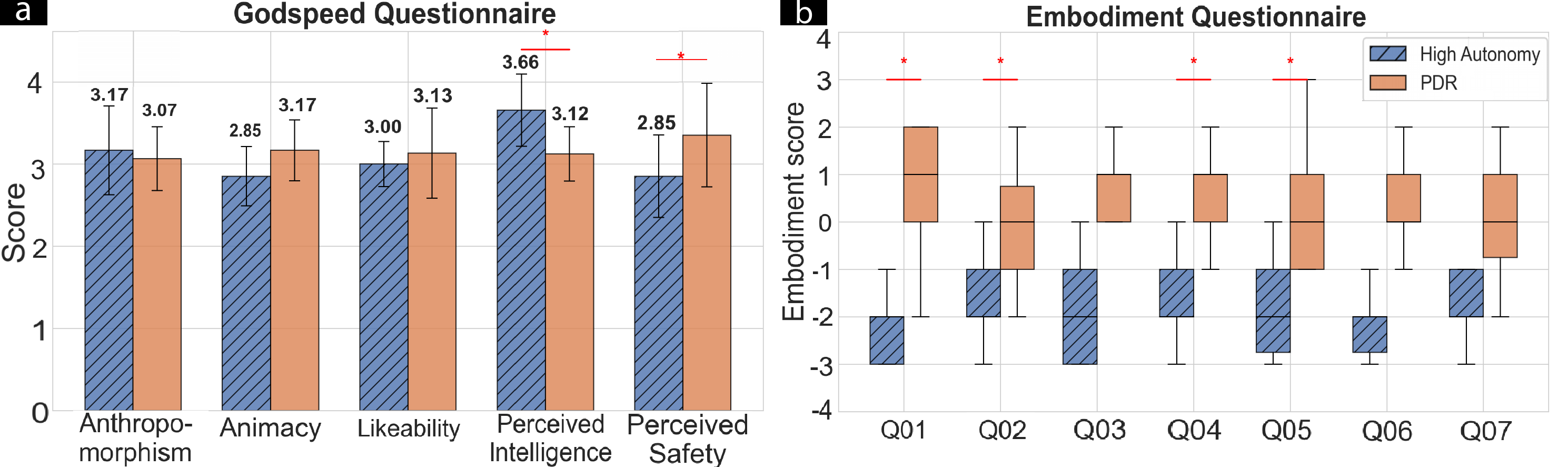} 
    \caption{The results for Participant-Defined Rules and High-Autonomy Anchor in (a) the Godspeed Questionnaire subscales and (b) the individual items (Q1–Q7) of the Embodiment Questionnaire. * indicates $p < .05$.}
    \label{EmbodimentResult}
    \Description{Two subfigures comparing user ratings between high-autonomy and low-autonomy (PDR) conditions. (a) A bar chart shows mean scores across five Godspeed dimensions: Anthropomorphism, Animacy, Likeability, Perceived Intelligence, and Perceived Safety. PDR condition has higher scores in Perceived Intelligence and Perceived Safety, marked with * ($p < .05$). (b) A grouped boxplot shows user responses on seven embodiment questions (Q1–Q7). PDR scores are consistently higher across questions, with significant differences (marked with *) on Q01, Q02, Q03, Q05.}
\end{figure*}

\begin{sloppypar}
Data were analyzed across four SRLs action categories (Figure~\ref{SCR}\allowbreak{} (a)) under two execution logics: the \emph{High-Autonomy Anchor} and participant-authored \emph{PDR}. Trials were baseline-corrected using a $-5$ to $0$\,s pre-action window; peak responses within 1--5\,s post-onset were extracted, log-transformed ($\log(\mathrm{SCR}+1)$), and Z-standardized. Standardized values are unitless. Trials without a detectable peak were assigned 0.
\end{sloppypar}

\begin{sloppypar}
A Friedman test revealed a significant main effect of action category ($\chi^2(3)=11.53,\allowbreak\,p=.009,\allowbreak\,W=.19$). Post-hoc Wilcoxon tests (Bonferroni-corrected) showed that \emph{approach/\allowbreak entry} (Mdn = 0.55, IQR = 0.60) elicited stronger standardized responses than \emph{hand-over/\allowbreak manipulation}, \emph{support/\allowbreak stabilize}, and \emph{withdrawal/\allowbreak retreat} (all $p < .01$); no differences were observed among the latter three (all $p > .10$).
\end{sloppypar}

To probe the Execution Logic $\times$ Action Category relationship, Wilcoxon signed-rank tests compared the \emph{High-Autonomy Anchor} and \emph{PDR} within each category. A significant difference emerged only for \emph{Proxemic Approach} (Z = $-3.15$, $p = .002$, $r = .74$), with Anchor showing higher standardized responses (Mdn = 0.85, IQR = 0.50) than PDR (Mdn = 0.25, IQR = 0.45). No Anchor–PDR differences were found for \emph{Handover}, \emph{Support}, or \emph{Withdrawal} (all $p > .10$, ns).

These findings answer RQ2 by showing that physiological arousal concentrates in the approach/entry phase, especially when SRLs enter sensitive frontal or midline zones identified in our qualitative trust maps. Elevated arousal during unannounced or direct approaches in the High-Autonomy condition aligns with participants' rejection of abrupt or ambiguous trajectories near the head, chest, and elbows. In contrast, participant-defined rules (PDR) were associated with significantly lower arousal. 

\subsubsection{Trust Questionnaire Results (RQ3)}\label{Trust}

Scores from an adapted trust-in-automation instrument~\cite{jian2000foundations} were combined into two dimensions: \emph{Trust} (mean of 6 items) and \emph{Distrust} (mean of 6 items). Because items are Likert and $N$ is modest, we analyzed per-participant scale means with paired non-parametric tests (Wilcoxon signed-rank), reporting medians and $z$–statistics. Internal consistency was acceptable for both subscales (High-Autonomy: Trust α = .70, Distrust α = .72; PDR: Trust α = .79, Distrust α = .81).

Wilcoxon signed-rank tests revealed a significant increase  (Figure~\ref{SCR} (b)) in \emph{Trust} for the PDR condition compared to High-Autonomy (median$_{\mathrm{PDR}}=5.42$ $>$ median$_{\mathrm{High}}=4.18$, $z=3.74$, $p < .05$). Conversely, \emph{Distrust} was significantly lower in PDR than in High-Autonomy (median$_{\mathrm{PDR}}=2.06$ $<$ median$_{\mathrm{High}}=2.96$, $z=-2.85$, $p < .05$). Normality checks on paired differences (Shapiro–Wilk) indicated approximate normality, but we report Wilcoxon statistics given the ordinal response format; complementary permutation checks yielded consistent inferences.

Taken together, these results indicate that participant-defined rules (PDR) increase trust while lowering distrust.

\subsubsection{Embodiment Questionnaire Results (RQ3)}\label{Embodiment}

Results from the embodiment questionnaire, administered after the High-Autonomy Anchor and PDR Application, revealed statistically significant differences on several key items  (see Figure~\ref{EmbodimentResult} (b)), analyzed using a Wilcoxon signed-rank test.
Within-subject paired comparisons showed significantly higher ratings under
\textit{PDR} than under the \textit{High-Autonomy Anchor} on items indexing
\textbf{ownership (Q4)}, \textbf{body image (Q2)}, and \textbf{agency (Q1, Q5)}
(all $p<.05$).

For RQ3, these item-level gains indicate that participant-defined rules increase felt embodiment of the SRL.

\subsubsection{Godspeed Questionnaire Results (RQ3)}\label{Godspeed}

Scores from the Godspeed Questionnaire Series~\cite{bartneck2009measurement} were analyzed to compare participants' perceptions of the SRLs between the High-Autonomy and PDR conditions, as shown in Figure~\ref{EmbodimentResult} (a). Across all subscales, internal consistency was acceptable (Cronbach's $\alpha$: Anthropomorphism = .80–.86; Animacy = .81–.88; Likeability = .84–.89; Perceived Intelligence = .77–.84; Perceived Safety = .74–.79).
Wilcoxon signed-rank tests revealed a significant increase in Perceived Intelligence for the High-Autonomy condition compared to PDR (median$\text{High}=3.70$ > median$\text{PDR}=3.20$, $z=-3.111$, $p < .05$). Conversely, Perceived Safety was significantly higher in the PDR condition than in High-Autonomy (median$\text{PDR}=3.33$ > median$\text{High}=3.00$, $z=2.282$, $p < .05$). No significant differences were found for Anthropomorphism ($z=-1.300$, $p=0.193$), Animacy ($z=2.310$, $p=0.0609$), or Likeability ($z=0.475$, $p=0.635$).

These results answer RQ3: autonomy arrangement shapes subjective assurance in near-body contexts. Participants perceived the SRLs as more `intelligent' under the High-Autonomy, but felt safer and more assured under PDR.

{\paragraph{Exploratory check for condition order.}
We conducted a post-hoc check for carry-over effects (Anchor--first vs.\ PDR--first) on questionnaire and SCR outcomes. No significant \textit{Condition}~$\times$~\textit{Order} interactions were found across primary measures (all $p>.05$). Given the limited power for between-subjects effects ($N=18$), we report pooled results and treat condition comparisons as exploratory.



\section{DISCUSSION}

Our study revealed that participants define spatial and autonomy expectations using a body-centric trust gradient (RQ1). Rather than applying uniform policies, they articulated region-specific rules, favouring higher autonomy at distal joints while requiring manual or confirmation-based control near sensitive zones. These participant-defined rules reduced physiological arousal during near-body entries compared to the high-autonomy baseline (RQ2) and fostered greater subjective assurance and embodiment (RQ3). Synthesising these results, we now discuss their implications for SRLs Proxemics and zone-responsive control. For clarity, we organize the discussion around our hypotheses (H1--H3), revisiting each in light of the results and prior work.

\subsection{Zone- and Component-Specific Strategies for Assured Near-Body Interaction (RQ1)}\label{RQ1}

\subsubsection{Entry Strategies and Motion Cues for Reducing Discomfort}
Our qualitative analysis identified four movement features that consistently helped reduce participants’ discomfort during near-body interactions with SRLs.
First, a segmented temporal structure characterized by pause\allowbreak--\allowbreak move\allowbreak--\allowbreak pause rhythms made the motion more predictable and less intrusive. Participants often disliked ambiguous drifting motions and instead preferred brief, deliberate actions, consistent with prior findings in legible robot motion~\cite{dragan2013legibility}.
Second, non-frontal approach directions (such as from the side or behind) were favored over direct frontal movements, highlighting the importance of entry geometry in proxemic design~\cite{mumm2011human}.
Third, clear action intent played a key role: participants were more comfortable when the robot’s posture (e.g., outward-facing hands), visible payloads, or preparatory movements clearly signaled its goal. This aligns with previous work advocating for pre-motion cues~\cite{knepper2017implicit}.
Fourth, participants preferred turn-taking over simultaneous action when working in shared spaces, as it reduced both physical interference and mental workload. Importantly, these were not just ``preferences,'' but formed the basis of embodied risk management strategies, with significant implications for SRLs autonomy.

\subsubsection{Allocating Autonomy by Limb Component}

Participants expressed consistent preferences regarding the appropriate level of autonomy across different segments of the robotic limb. The hand was generally trusted to operate with a high degree of autonomy, particularly when its movements were segmented and easily interpretable. Participants welcomed actions such as autonomous grasping and item delivery, which aligns with prior work framing SRLs as collaborative tools~\cite{parietti2016supernumerary,zhou2024coplayingvr}. In contrast, the elbow was viewed as suitable only for reactive or reflex-like behavior. Movements initiated at the elbow, especially near sensitive regions like the ribs, were frequently described as uncomfortable or intrusive. This challenges prior robotic planning frameworks that assume equal autonomy across all joints~\cite{lavalle2006planning}. The shoulder and base were perceived as acceptable for autonomous repositioning, but only under specific conditions: when following constrained trajectories (e.g., from the rear or side), using clear cues (e.g., auditory or haptic), and moving at moderate speeds. This segment-specific autonomy model offers an alternative to traditional full-body planning approaches, and complements recent perspectives in embodied interaction that advocate context-dependent autonomy allocation~\cite{zhou2025juggling}.

\subsection{Entry is the Risk Locus: Physiological and Subjective Convergence (RQ2, RQ3)}\label{RQ2, RQ3}

Across both physiological and subjective measures, we found that participants perceived the greatest risk not during task execution, but during entry when SRLs moved into near-body spaces. This phase consistently triggered elevated skin conductance responses, and was uniquely sensitive to execution logic: \emph{PDR} modes attenuated arousal more effectively than \emph{high-autonomy anchor}. Subjective measures mirrored this trend. Participants rated PDR as safer and more trustworthy, even though its perceived intelligence (per Godspeed) was slightly lower.
These findings suggest that spatial thresholds, rather than task complexity, are the primary locus of safety negotiation. This aligns with prior work in human–robot proxemics~\cite{mead2016proxemic,takayama2009influences}, and approach behavior~\cite{walters2005influence,mumm2011human}, where perceived safety is shaped not by what the robot does, but how it enters shared space. More broadly, boundary negotiation and trust calibration have been studied across other interactive AI/wearable systems, highlighting that users’ acceptance often depends on legibility and perceived control rather than raw capability~\cite{jayasiriwardene2023adaptive,jayasiriwardene2021interactive,jayasiriwardene2021architectural,jayasiriwardene2022knowledge,jayasiriwardene2025more,jayasiriwardene2026fixed,fan2021high,ferianc2021improving}.

\subsection{A Three-Tiered Policy for Zone-Responsive Control}

\mm{Synthesizing the motion preferences (Section~\ref{RQ1}) and physiological and experiential sensitivities (Section~\ref{RQ2, RQ3}), our findings coalesce into an initial framework for \textit{SRL Proxemics}: how body-mounted SRLs should distribute distance, motion, and autonomy across the user's body. To make this initial framework actionable, we cast it as a three-tiered, zone-responsive policy (Table~\ref{tab:zone_segment_synthesis_revised}).}

We define three adjacent body zones: Critical (e.g., face, neck, torso midline), Supervisory (chest, shoulders, upper arms), and Utilitarian (hands, periphery, non-dominant side). For each zone, we specify acceptable SRLs actions (e.g., approach, support, reposition), tolerated limb components, expected cueing strategies, and preferred control modes. For example, subtle wrist movements were acceptable in Supervisory zones if the intent was clear, while frontal base motion often required explicit user consent. \mm{Applying these guidelines to professional use cases clarifies their practical value. In industrial assembly, the Critical Zone guideline implies that an SRL holding a hazardous tool 
(e.g., a soldering iron) should not move autonomously through the face region~\cite{bright2017supernumerary,prattichizzo2021human}. In surgical assistance, the Supervisory Zone guideline suggests that a robotic retractor should require explicit surgeon confirmation before repositioning near the torso~\cite{amadeo2022soft,van2024capacitive}.}

These insights are consolidated into a structured table linking thematic actions (e.g., entry, handover) to user - derived rules and concrete design suggestions. These zone - responsive policies are intended to complement, not replace, established machinery and robot safety requirements by adding a user-centred proxemic layer for near-body SRLs behavior. Taken together, the three zones and their associated rules form an initial, empirically grounded account of SRLs Proxemics. We deliberately treat SRLs Proxemics as an emerging conceptual space rather than a finished theory, inviting future work to test, refine, and extend this framework across SRLs morphologies, tasks, and user populations.

\begin{table*}[t]
\centering
\renewcommand{\thesubtable}{\Alph{subtable}}
\caption{\mm{Summary of SRLs Proxemics guidelines: (A) body-centric zones and their sensitivity profiles; (B) segment-level policies for distance, motion, and autonomy within each zone. Consensus strength ($\bullet\bullet\bullet$, $\bullet\bullet\circ$, $\bullet\circ\circ$) denotes strong, moderate, and emerging participant agreement, respectively.}}
\label{tab:zone_segment_synthesis_revised}
\small

\begin{subtable}{\textwidth}
\centering
\caption{Body-Centric Zones}
\label{tab:zones}
\begin{adjustbox}{width=\textwidth}
\begin{tabular}{p{3.1cm}|p{4.6cm}p{4.6cm}p{4.6cm}p{1.6cm}}
\toprule
\textbf{Policy Aspect} & \textbf{Critical Zone} (Face, Neck, Midline) & \textbf{Supervisory Zone} (Upper Torso, Shoulders) & \textbf{Utilitarian Zone} (Hands, Forearms, Periphery) & \textbf{Consensus} \\
\midrule
\textbf{Safe Distance \& Clearance} 
& Default to arm's-length distance. Treat frontal cone of vision as the primary boundary with low tolerance for entry.
& Maintain a palm- or fist-width buffer. Buffer can be briefly reduced under high workload if action is user-cancellable.
& Permit close proximity when movement is predictable and task-relevant. Periphery has the highest tolerance.
& $\bullet\bullet\bullet$
\\ \midrule

\textbf{Entry Approach (Path \& Timing)} 
& \textbf{Confirmation Required}: Require clear pre-motion cues. Use curved, non-frontal paths. Angle tools away from user; avoid hovering.
& \textbf{Bounded Entry}: Use side/rear paths with a brief, deliberate pause at the zone boundary. Avoid crossing midline without cues.
& \textbf{Permitted if Legible}: Use decisive, curved paths. Avoid aimless drift or hesitation.
& $\bullet\bullet\bullet$
\\ \midrule

\textbf{Handover} 
& User-initiated or single user confirmation required. Use brief cues (e.g., audio) and tilt wrist away. Enforce turn-taking.
& Employ turn-taking. Use brief, deliberate pauses; do not linger near the chest.
& Delegate for quick exchanges. Clear the area immediately after completion.
& $\bullet\bullet\circ$
\\ \midrule

\textbf{Support / Stabilization} 
& Single user confirmation required. Hold a fixed pose; prohibit autonomous micro-adjustments near face/neck.
& Bounded autonomy; engage slowly and stiffly. Avoid sweeping motions near the ribs.
& Delegated autonomy; can re-engage to offer support after providing a clear cue.
& $\bullet\bullet\circ$
\\ \midrule

\textbf{Crossing Midline \& Repositioning} 
& User consent required. Use a curved, non-frontal path. Prohibit direct frontal crossing.
& Single user confirmation required, accompanied by a clear cue (e.g., audio ping). Use a single, smooth path.
& Allowed if intent is announced. Execute as a single pass; prohibit back-and-forth movement.
& $\bullet\bullet\bullet$
\\ \midrule

\textbf{Idle / Hovering} 
& Retract to a safe distance or hold a static pose. Prohibit any hovering near the face.
& Limit idling to zero velocity. Signal intent if pausing for an extended period.
& Permitted if motion is predictable and intent is clear (e.g., holding an object). Avoid empty-handed drift.
& $\bullet\bullet\bullet$
\\ \midrule

\textbf{Withdrawal / Retreat} 
& Reactive, but user-cancellable. Use a swift, clean, non-frontal exit. Do not cut across the face.
& Reactive. Use a side or back path. Avoid sweeping near the ribs.
& Reactive. Take the shortest safe path away from the user.
& $\bullet\bullet\circ$
\\ \midrule

\textbf{Coordination in Shared Hand Space} 
& Enforce strict turn-taking. Prohibit simultaneous entry by user and SRL.
& Alternate actions. Encourage use of implicit ``lanes'' to reduce collisions.
& Support delegation to the arm on the user's non-dominant side.
& $\bullet\bullet\circ$
\\
\bottomrule
\end{tabular}
\end{adjustbox}
\end{subtable}

\vspace{6pt}

\begin{subtable}{\textwidth}
\centering
\caption{Supernumerary limb Segments}
\label{tab:segments}
\begin{adjustbox}{width=\textwidth}
\begin{tabular}{p{3.1cm}|p{4.0cm}p{4.0cm}p{4.0cm}p{4.0cm}p{1.8cm}}
\toprule
\textbf{Policy Aspect} & \textbf{Hand / End-Effector} & \textbf{Wrist} & \textbf{Elbow} & \textbf{Shoulder–Base} & \textbf{Consensus} \\
\midrule
\textbf{Safe Distance \& Clearance} 
& Permit delegated autonomy for tasks near the hands; ensure user can cancel proximity at any time.
& Adjust orientation without reducing safe distance. Limit acceleration when near the torso.
& Maintain a strict ``no-go zone'' near the ribs/waist. Treat as a simple structural connector.
& Use rear/side paths for clearance; avoid entering the user's primary field of view without permission.
& $\bullet\bullet\circ$
\\ \midrule

\textbf{Entry Approach (Path \& Timing)} 
& Employ a ``pause–present–pause'' motion. Present held objects to clearly signal intent; avoid pointing at the face.
& Perform minor, predictable adjustments to support hand's intent. Lock position quickly; avoid ambiguous, lingering rotations.
& Use for reactive, corrective movements only; never lead the approach.
& Reposition from rear/side to set up an entry. Use a continuous, low-salience audio cue for presence awareness.
& $\bullet\bullet\circ$
\\ \midrule

\textbf{Handover} 
& Permit autonomy with safety guardrails. Present object clearly; grasp must be user-cancellable.
& Rotate to align for handover, then hold position. Should not make independent decisions.
& Prohibit sweeping motions across the torso during a handover.
& Adopt a posture that supports the handover, preferably using side or rear paths.
& $\bullet\bullet\circ$
\\ \midrule

\textbf{Support / Stabilization} 
& Maintain a stable grasp. Provide a brief warning signal before making re-contact.
& Use purposeful orientation changes only; avoid jittery movements.
& Use for reactive smoothing movements only. Prohibit any planned actions.
& Limit posture changes to rear or side paths.
& $\bullet\bullet\circ$
\\ \midrule

\textbf{Crossing Midline \& Repositioning} 
& Announce intent before movement. Prepare for a pass rather than cutting directly across.
& Provide clear warning before entering the user's frontal cone of vision.
& Prohibit sideways sweeps. If repositioning is necessary, only retreat backward.
& Use slow, autonomous paths on rear/side. Maintain an audio cue for spatial awareness.
& $\bullet\bullet\circ$
\\ \midrule

\textbf{Idle / Hovering} 
& Hold objects visibly to maintain intent legibility. Avoid ambiguous, empty-handed hovering.
& Freeze orientation quickly. Avoid slow, lingering rotations that create ambiguity.
& Must remain clear of the user's personal space. Do not use elbow for repositioning.
& Hold in a quiet, background position outside the user's primary field of view.
& $\bullet\bullet\circ$
\\ \midrule

\textbf{Withdrawal / Retreat} 
& Release object and clear the zone decisively and predictably.
& Snap to a neutral, non-threatening orientation, then stop.
& Employ ``reflex-only'' retreat. Prohibit any planned forward movements.
& Use a non-frontal exit path. Maintain a soft audio cue during retreat.
& $\bullet\bullet\circ$
\\ \midrule

\textbf{Coordination in Shared Hand Space} 
& Lead with legible movement signals. Time actions to avoid overlapping with the user's hands.
& Support the hand's intended action; never compete for the same space.
& Maintain designated ``no-go zones'' to guarantee clearance for user.
& Adjust whole-arm pose to set up actions. Require confirmation for movements that cross the midline.
& $\bullet\circ\circ$
\\
\bottomrule
\end{tabular}
\Description{Two-part table summarizing SRL Proxemics guidelines. Subtable (A) organizes policies by body-centric zones: a Critical Zone (face, neck, midline), a Supervisory Zone (upper torso, shoulders), and a Utilitarian Zone (hands, forearms, periphery). Each row corresponds to a policy aspect (safe distance and clearance, entry approach and timing, handover, support and stabilization, crossing the midline and repositioning, idling or hovering, withdrawal or retreat, and coordination in shared hand space) and gives recommended SRL behavior for each zone. A final column shows how strongly participants agreed with each guideline using three, two, or one filled bullets, indicating strong, moderate, or emerging consensus.
Subtable (B) organizes policies by SRL segment: Hand or End-Effector, Wrist, Elbow, and Shoulder–Base. For the same set of policy aspects, it describes how much autonomy each segment should have, where it may move, and how it should enter, idle, and withdraw near the body. The consensus column again uses bullet symbols to indicate the strength of participant agreement. Together, the two subtables translate qualitative themes and SCR findings into zone- and segment-specific control guidelines for safe, legible SRL behavior around the body.}
\end{adjustbox}
\end{subtable}
\end{table*}

\subsection{Future Directions: State-Aware and Signal-First Control Overlays}
Participants independently proposed adaptive modifiers to the zone-based design guidelines, pointing to future extensions. A state-aware overlay would adjust autonomy thresholds in response to user workload or arousal (e.g., increasing standoff distance during high SCR interaction phases), consistent with biosignal-mediated shared control paradigms~\cite{abiri2019brain}. A signal-first overlay would enhance anticipatory motion, including gaze-mimetic sequences and multimodal previews (e.g., acoustic tones), echoing findings in anticipatory intent signalling~\cite{strabala2013toward}. Although only a minority mentioned role-based control (e.g., ``third teammate'' vs. ``tool handler''), such framing aligns with interface metaphors explored in cooperative telepresence systems~\cite{Saraiji2018Fusion}. These overlays should be viewed not as replacements for the core policy, but as compositional additions that further personalise and modulate SRLs behavior. Importantly, each remains grounded in user expectations rather than designer heuristics, positioning them as testable hypotheses for future implementation.


\section{Limitations and Future Work}

Our study has several limitations that qualify the scope of the findings and point to concrete next steps. \mm{At the same time, we position these findings as exploratory design guidance grounded in 
a single-session WoZ study with expert users and a focused set of measures, rather than as precise population-level estimates.}

\paragraph{Wizard-of-Oz implementation only} All SRLs behaviors were enacted using a Wizard-of-Oz (WoZ) arrangement, with a hidden operator reproducing the requested motion grammar (e.g., non-frontal arcs, segmented timing). This allowed us to probe potentially intrusive, near-body interactions without physical hazard, but WoZ cannot fully capture closed-loop perception-control couplings, recovery from misdetections, or distribution of autonomous errors in the wild~\cite{dahlback1993wizard}. Future work should implement the zone policies as runtime constraints in an actual controller and compare WoZ versus autonomous execution with matched tasks and systematic ablations (e.g., removing pre-cues or arc entries) to identify which policy elements drive \mm{perceived safety} and embodiment gains. Implementation-wise, recent toolkits for rapidly prototyping wearable sensing and on-body feedback~\cite{yu2023drivingvibe,fan2024spinshot,wu2025headturner} make it increasingly feasible to embed proximity-aware sensing and pre-cue signaling into SRL mounts or garments~\cite{yu2024fabricating,gough2023design,tong2023fully,yu2024irontex,perera2024integrating,Perera2025eTactileKit,yu2025designing,chau2021composite,dong2025just,Dong_2026_TactDeform}.

\paragraph{Task and apparatus specificity} Findings are grounded in a back-worn, multi-joint SRLs configuration operating in two safe surrogate tasks designed to elicit near-body rules (proximal handover and shared sorting). This design maximizes elicitation value while minimizing physical hazard, but may under-represent settings with heavy payloads, sharp tools, or constrained workcells. Replication with alternative mount points (waist/chest), different end-effectors, and higher-inertia arms, as well as in-situ studies (e.g., bench assembly, assistive care), are needed to assess robustness and transfer.

\paragraph{Single-session evaluation.} The study was single-session (50–60 minutes). Participants authored rules and then immediately experienced their integrated effect. \mm{This captures first-use judgements, but does not provide a prolonged familiarisation phase that could stabilise behavior or psychophysiological baselines, even for expert users. As a result, some of the quantitative contrasts should be interpreted as exploratory.} Longitudinal deployments will be necessary to track how distance thresholds, approach preferences, and delegation patterns shift over days or weeks.

\paragraph{Lightweight measurement coverage.} Physiological arousal was indexed with skin conductance responses; we did not triangulate startle or attentional capture with eye tracking, pupillometry, EMG, or head pose, and questionnaire ordering could interact with condition effects. Future studies may employ multi-channel sensing and preregister approach-phase analyses to strengthen inference.

\paragraph{\mm{Scale choice and construct coverage.}}
\mm{We selected established HRI instruments (Godspeed subscales for perceived safety and intelligence~\cite{bartneck2009measurement}, Jian et al.'s trust scale for capacity-oriented trust~\cite{jian2000foundations}, and an adapted AEQ for embodiment~\cite{gonzalez2018avatar,umezawa2022bodily}) to align with prior SRLs and wearable-robotics work and to keep the questionnaire burden manageable. However, we did not include newer multi-dimensional measures such as RoSAS~\cite{carpinella2017rosas} or MDMT~\cite{ullman2019mdmt}, which could disentangle additional facets of social perception and trust. Future studies might combine our spatial policies with richer measurement batteries to capture, for example, moral versus capacity trust or social attributes beyond those probed here.}

\paragraph{\mm{Sample Bias, Expert Users, and Order Effects}}
\mm{This study was designed for generative insight rather than population inference. 
Our $N\!=\!18$ robotics- and technology-familiar sample meets common targets for elicitation in HCI~\cite{caine2016local,muehlhaus2023need}, but it limits statistical power and generalisability. 
Expert users may articulate richer spatial policies than novices, yet their prior experience can also compress variance in questionnaire scores and trust judgements, making our quantitative findings best read as exploratory patterns and, if anything, an optimistic boundary condition for lay users’ perceived risk. 
Although autonomy-condition order was counterbalanced, residual ordering and carry-over effects (e.g., priming from the High-Autonomy Anchor) cannot be ruled out given the sample size. 
SCR analyses on small paired samples are therefore reported as relative contrasts rather than absolute physiological claims. 
Cultural and anthropometric factors, as well as handedness, may further moderate preferred zones; future work should use larger~\cite{liu2025understanding,liu2025effects,wang2025vr}, stratified, cross-cultural samples to test the stability and external validity of the Three-Tiered Zone Policies.}

\paragraph{Untested overlay features.} The proposed affect-responsive and anticipatory-signaling overlays surfaced during debrief as user suggestions, not tested features. Prototyping these overlays raises practical questions (sensor reliability, individual calibration, privacy, and false positives). Future work should implement conservative, user-cancellable versions, evaluate them against the baseline policies, and examine ethical guardrails for state-adaptive autonomy.

\paragraph{Single-human, single-workspace setting.} Our policies were derived for one human collaborating with two SRLs in a single workspace. Multi-actor settings introduce contention (simultaneous entries), etiquette (“turn-taking by zone”) and arbitration between competing policies. Extending the framework to multi-human/multi-robot contexts, formalizing the policies as verifiable runtime constraints with monitoring for cue delivery, consent logging, and safe degradation under sensor loss are important steps toward deployment.

\paragraph{\mm{Scope of SRLs Proxemics.}} \mm{Our account of SRLs Proxemics is necessarily partial. It is grounded in a single WoZ study with one SRLs configuration and cultural context, and should therefore be read as an initial conceptual space to be further explored and refined in future on-body robot research, rather than as a prescriptive or closed set of rules.}


\section{CONCLUSION}

In this work, we investigated how users define spatial boundaries and autonomy preferences for Supernumerary Robotic Limbs (SRLs) during close-proximity interaction. Using a Wizard-of-Oz setup, participants completed two embodied tasks: Comfort Zone and Control Handover, under two modes: a high-autonomy baseline and Participant-Defined Rules (PDR). Our findings show that greater autonomy did not necessarily promote \mm{perceived safety}y or trust. Instead, participants segmented the body into functional zones and expressed context-dependent expectations for movement and control. They reported greater \mm{perceived safety}, trust, and embodiment under the PDR condition, while the high-autonomy mode, though more capable, was perceived as less predictable. These insights support SRLs systems that go beyond functional efficiency to reflect users' intuitive expectations for \mm{perceived safety}, legible, and trustworthy near-body interaction. This underscores the value of zone-responsive policies tailored to body region and task stage.

\begin{acks}
\mm{This project was supported by the Australian Research Council Discovery Early Career Award (DECRA) - DE200100479. Dr. Anusha Withana is the recipient of a DECRA fellowship funded by the Australian Government. We are grateful for the support provided by the Neurodisability Assist Trust and Cerebral Palsy Alliance, Australia - PRG04219. Finally, we would like to express our sincere gratitude to the members of the Information Somatics Lab for their generous support and assistance throughout this project. It has been a privilege to conduct this research alongside such an inspiring team.}
\end{acks}

\bibliographystyle{ACM-Reference-Format}
\bibliography{Safety.bib}


\begin{thebibliography}{140}


\ifx \showCODEN    \undefined \def \showCODEN     #1{\unskip}     \fi
\ifx \showISBNx    \undefined \def \showISBNx     #1{\unskip}     \fi
\ifx \showISBNxiii \undefined \def \showISBNxiii  #1{\unskip}     \fi
\ifx \showISSN     \undefined \def \showISSN      #1{\unskip}     \fi
\ifx \showLCCN     \undefined \def \showLCCN      #1{\unskip}     \fi
\ifx \shownote     \undefined \def \shownote      #1{#1}          \fi
\ifx \showarticletitle \undefined \def \showarticletitle #1{#1}   \fi
\ifx \showURL      \undefined \def \showURL       {\relax}        \fi
\providecommand\bibfield[2]{#2}
\providecommand\bibinfo[2]{#2}
\providecommand\natexlab[1]{#1}
\providecommand\showeprint[2][]{arXiv:#2}

\bibitem[Abiri et~al\mbox{.}(2019)]%
        {abiri2019brain}
\bibfield{author}{\bibinfo{person}{Reza Abiri}, \bibinfo{person}{Saeid Borhani}, \bibinfo{person}{Eric~W. Sellers}, \bibinfo{person}{Yang Jiang}, {and} \bibinfo{person}{Xueliang Zhao}.} \bibinfo{year}{2019}\natexlab{}.
\newblock \showarticletitle{A comprehensive review of {EEG}-based brain–computer interface paradigms}.
\newblock \bibinfo{journal}{\emph{Journal of Neural Engineering}} \bibinfo{volume}{16}, \bibinfo{number}{1} (\bibinfo{year}{2019}), \bibinfo{pages}{011001}.
\newblock
\href{https://doi.org/10.1088/1741-2552/aaf12e}{doi:\nolinkurl{10.1088/1741-2552/aaf12e}}


\bibitem[Al~Sada et~al\mbox{.}(2017)]%
        {al2017challenges}
\bibfield{author}{\bibinfo{person}{Mohammed Al~Sada}, \bibinfo{person}{Mohamed Khamis}, \bibinfo{person}{Akira Kato}, \bibinfo{person}{Shigeki Sugano}, \bibinfo{person}{Tatsuo Nakajima}, {and} \bibinfo{person}{Florian Alt}.} \bibinfo{year}{2017}\natexlab{}.
\newblock \showarticletitle{Challenges and opportunities of supernumerary robotic limbs}.
\newblock  (\bibinfo{year}{2017}).
\newblock


\bibitem[Amadeo et~al\mbox{.}(2022)]%
        {amadeo2022soft}
\bibfield{author}{\bibinfo{person}{Tomas Amadeo}, \bibinfo{person}{Daniel Van~Lewen}, \bibinfo{person}{Taylor Janke}, \bibinfo{person}{Tommaso Ranzani}, \bibinfo{person}{Anand Devaiah}, \bibinfo{person}{Urvashi Upadhyay}, {and} \bibinfo{person}{Sheila Russo}.} \bibinfo{year}{2022}\natexlab{}.
\newblock \showarticletitle{Soft robotic deployable origami actuators for neurosurgical brain retraction}.
\newblock \bibinfo{journal}{\emph{Frontiers in Robotics and AI}}  \bibinfo{volume}{8} (\bibinfo{year}{2022}), \bibinfo{pages}{731010}.
\newblock


\bibitem[Anacleto and Fels(2015)]%
        {anacleto2015towards}
\bibfield{author}{\bibinfo{person}{Junia Anacleto} {and} \bibinfo{person}{Sidney Fels}.} \bibinfo{year}{2015}\natexlab{}.
\newblock \showarticletitle{Towards a model of virtual proxemics for wearables}. In \bibinfo{booktitle}{\emph{IFIP Conference on Human-Computer Interaction}}. Springer, \bibinfo{pages}{433--447}.
\newblock


\bibitem[Arai et~al\mbox{.}(2022)]%
        {arai2022embodiment}
\bibfield{author}{\bibinfo{person}{Ken Arai}, \bibinfo{person}{Hiroto Saito}, \bibinfo{person}{Masaaki Fukuoka}, \bibinfo{person}{Sachiyo Ueda}, \bibinfo{person}{Maki Sugimoto}, \bibinfo{person}{Michiteru Kitazaki}, {and} \bibinfo{person}{Masahiko Inami}.} \bibinfo{year}{2022}\natexlab{}.
\newblock \showarticletitle{Embodiment of supernumerary robotic limbs in virtual reality}.
\newblock \bibinfo{journal}{\emph{Scientific reports}} \bibinfo{volume}{12}, \bibinfo{number}{1} (\bibinfo{year}{2022}), \bibinfo{pages}{9769}.
\newblock


\bibitem[Bartneck et~al\mbox{.}(2009)]%
        {bartneck2009measurement}
\bibfield{author}{\bibinfo{person}{Christoph Bartneck}, \bibinfo{person}{Dana Kuli{\'c}}, \bibinfo{person}{Elizabeth Croft}, {and} \bibinfo{person}{Susana Zoghbi}.} \bibinfo{year}{2009}\natexlab{}.
\newblock \showarticletitle{Measurement instruments for the anthropomorphism, animacy, likeability, perceived intelligence, and perceived safety of robots}.
\newblock \bibinfo{journal}{\emph{International journal of social robotics}} \bibinfo{volume}{1}, \bibinfo{number}{1} (\bibinfo{year}{2009}), \bibinfo{pages}{71--81}.
\newblock


\bibitem[Bradshaw et~al\mbox{.}(2017)]%
        {bradshaw2017human}
\bibfield{author}{\bibinfo{person}{Jeffrey~M Bradshaw}, \bibinfo{person}{Paul~J Feltovich}, {and} \bibinfo{person}{Matthew Johnson}.} \bibinfo{year}{2017}\natexlab{}.
\newblock \showarticletitle{Human--agent interaction}.
\newblock In \bibinfo{booktitle}{\emph{The handbook of human-machine interaction}}. \bibinfo{publisher}{CRC Press}, \bibinfo{pages}{283--300}.
\newblock


\bibitem[Braun and Clarke(2006)]%
        {braun2006using}
\bibfield{author}{\bibinfo{person}{Virginia Braun} {and} \bibinfo{person}{Victoria Clarke}.} \bibinfo{year}{2006}\natexlab{}.
\newblock \showarticletitle{Using thematic analysis in psychology}.
\newblock \bibinfo{journal}{\emph{Qualitative Research in Psychology}} \bibinfo{volume}{3}, \bibinfo{number}{2} (\bibinfo{year}{2006}), \bibinfo{pages}{77--101}.
\newblock
\href{https://doi.org/10.1191/1478088706qp063oa}{doi:\nolinkurl{10.1191/1478088706qp063oa}}


\bibitem[Braun and Clarke(2019)]%
        {braun2019reflecting}
\bibfield{author}{\bibinfo{person}{Virginia Braun} {and} \bibinfo{person}{Victoria Clarke}.} \bibinfo{year}{2019}\natexlab{}.
\newblock \showarticletitle{Reflecting on reflexive thematic analysis}.
\newblock \bibinfo{journal}{\emph{Qualitative research in sport, exercise and health}} \bibinfo{volume}{11}, \bibinfo{number}{4} (\bibinfo{year}{2019}), \bibinfo{pages}{589--597}.
\newblock


\bibitem[Braun and Clarke(2021)]%
        {braun2021one}
\bibfield{author}{\bibinfo{person}{Virginia Braun} {and} \bibinfo{person}{Victoria Clarke}.} \bibinfo{year}{2021}\natexlab{}.
\newblock \showarticletitle{One size fits all? What counts as quality practice in (reflexive) thematic analysis?}
\newblock \bibinfo{journal}{\emph{Qualitative research in psychology}} \bibinfo{volume}{18}, \bibinfo{number}{3} (\bibinfo{year}{2021}), \bibinfo{pages}{328--352}.
\newblock


\bibitem[Bright(2017)]%
        {bright2017supernumerary}
\bibfield{author}{\bibinfo{person}{Lawrence Lawrence~Zack Bright}.} \bibinfo{year}{2017}\natexlab{}.
\newblock \emph{\bibinfo{title}{Supernumerary robotic limbs for human augmentation in overhead assembly tasks}}.
\newblock \bibinfo{thesistype}{Ph.\,D. Dissertation}. \bibinfo{school}{Massachusetts Institute of Technology}.
\newblock


\bibitem[Buchenau and Suri(2000)]%
        {buchenau2000experience}
\bibfield{author}{\bibinfo{person}{Marion Buchenau} {and} \bibinfo{person}{Jane~Fulton Suri}.} \bibinfo{year}{2000}\natexlab{}.
\newblock \showarticletitle{Experience Prototyping}. In \bibinfo{booktitle}{\emph{Proceedings of the 3rd Conference on Designing Interactive Systems (DIS)}}. \bibinfo{pages}{424--433}.
\newblock
\href{https://doi.org/10.1145/347642.347802}{doi:\nolinkurl{10.1145/347642.347802}}


\bibitem[Caine(2016)]%
        {caine2016local}
\bibfield{author}{\bibinfo{person}{Kelly Caine}.} \bibinfo{year}{2016}\natexlab{}.
\newblock \showarticletitle{Local standards for sample size at CHI}. In \bibinfo{booktitle}{\emph{Proceedings of the 2016 CHI conference on human factors in computing systems}}. \bibinfo{pages}{981--992}.
\newblock


\bibitem[Carpinella et~al\mbox{.}(2017)]%
        {carpinella2017rosas}
\bibfield{author}{\bibinfo{person}{Colleen~M Carpinella}, \bibinfo{person}{Alisa~B Wyman}, \bibinfo{person}{Michael~A Perez}, {and} \bibinfo{person}{Steven~J Stroessner}.} \bibinfo{year}{2017}\natexlab{}.
\newblock \showarticletitle{The robotic social attributes scale (RoSAS) development and validation}. In \bibinfo{booktitle}{\emph{Proceedings of the 2017 ACM/IEEE International Conference on Human-Robot Interaction (HRI)}}. ACM, \bibinfo{pages}{254--262}.
\newblock
\href{https://doi.org/10.1145/2909824.3020208}{doi:\nolinkurl{10.1145/2909824.3020208}}


\bibitem[Chau et~al\mbox{.}(2021)]%
        {chau2021composite}
\bibfield{author}{\bibinfo{person}{Edwin Chau}, \bibinfo{person}{Jiakun Yu}, \bibinfo{person}{Cagatay Goncu}, {and} \bibinfo{person}{Anusha Withana}.} \bibinfo{year}{2021}\natexlab{}.
\newblock \showarticletitle{Composite line designs and accuracy measurements for tactile line tracing on touch surfaces}.
\newblock \bibinfo{journal}{\emph{Proceedings of the ACM on Human-Computer Interaction}} \bibinfo{volume}{5}, \bibinfo{number}{ISS} (\bibinfo{year}{2021}), \bibinfo{pages}{1--17}.
\newblock


\bibitem[Chen et~al\mbox{.}(2018)]%
        {chen2018planning}
\bibfield{author}{\bibinfo{person}{Lipeng Chen}, \bibinfo{person}{Luis~FC Figueredo}, {and} \bibinfo{person}{Mehmet~R Dogar}.} \bibinfo{year}{2018}\natexlab{}.
\newblock \showarticletitle{Planning for muscular and peripersonal-space comfort during human-robot forceful collaboration}. In \bibinfo{booktitle}{\emph{2018 IEEE-RAS 18th International Conference on Humanoid Robots (Humanoids)}}. IEEE, \bibinfo{pages}{1--8}.
\newblock


\bibitem[Collier et~al\mbox{.}(2025)]%
        {collier2025sense}
\bibfield{author}{\bibinfo{person}{Maggie~A Collier}, \bibinfo{person}{Rithika Narayan}, {and} \bibinfo{person}{Henny Admoni}.} \bibinfo{year}{2025}\natexlab{}.
\newblock \showarticletitle{The sense of agency in assistive robotics using shared autonomy}. In \bibinfo{booktitle}{\emph{2025 20th ACM/IEEE International Conference on Human-Robot Interaction (HRI)}}. IEEE, \bibinfo{pages}{880--888}.
\newblock


\bibitem[Cunningham et~al\mbox{.}(2018)]%
        {cunningham2018supernumerary}
\bibfield{author}{\bibinfo{person}{James Cunningham}, \bibinfo{person}{Anita Hapsari}, \bibinfo{person}{Pierre Guilleminot}, \bibinfo{person}{Ali Shafti}, {and} \bibinfo{person}{A~Aldo Faisal}.} \bibinfo{year}{2018}\natexlab{}.
\newblock \showarticletitle{The supernumerary robotic 3 rd thumb for skilled music tasks}. In \bibinfo{booktitle}{\emph{2018 7th IEEE International Conference on Biomedical Robotics and Biomechatronics (Biorob)}}. IEEE, \bibinfo{pages}{665--670}.
\newblock


\bibitem[Dahlb{\"a}ck et~al\mbox{.}(1993)]%
        {dahlback1993wizard}
\bibfield{author}{\bibinfo{person}{Nils Dahlb{\"a}ck}, \bibinfo{person}{Arne J{\"o}nsson}, {and} \bibinfo{person}{Lars Ahrenberg}.} \bibinfo{year}{1993}\natexlab{}.
\newblock \showarticletitle{Wizard of Oz Studies: Why and How}.
\newblock \bibinfo{journal}{\emph{Knowledge-Based Systems}} \bibinfo{volume}{6}, \bibinfo{number}{4} (\bibinfo{year}{1993}), \bibinfo{pages}{258--266}.
\newblock
\href{https://doi.org/10.1016/0950-7051(93)90017-N}{doi:\nolinkurl{10.1016/0950-7051(93)90017-N}}


\bibitem[Dawson et~al\mbox{.}(2007)]%
        {dawson2007electrodermal}
\bibfield{author}{\bibinfo{person}{Michael~E Dawson}, \bibinfo{person}{Anne~M Schell}, \bibinfo{person}{Diane~L Filion}, {et~al\mbox{.}}} \bibinfo{year}{2007}\natexlab{}.
\newblock \showarticletitle{The electrodermal system}.
\newblock \bibinfo{journal}{\emph{Handbook of psychophysiology}}  \bibinfo{volume}{2} (\bibinfo{year}{2007}), \bibinfo{pages}{200--223}.
\newblock


\bibitem[de~Visser et~al\mbox{.}(2018)]%
        {devisser2018literature}
\bibfield{author}{\bibinfo{person}{Ewart~J. de Visser}, \bibinfo{person}{Richard Pak}, {and} \bibinfo{person}{Tyler~H. Shaw}.} \bibinfo{year}{2018}\natexlab{}.
\newblock \showarticletitle{From Automation to Autonomy: The Trust Continuum in Human–Machine Interaction}.
\newblock \bibinfo{journal}{\emph{IEEE Transactions on Human–Machine Systems}} \bibinfo{volume}{48}, \bibinfo{number}{4} (\bibinfo{year}{2018}), \bibinfo{pages}{283--290}.
\newblock
\href{https://doi.org/10.1109/THMS.2017.2777404}{doi:\nolinkurl{10.1109/THMS.2017.2777404}}


\bibitem[Dong et~al\mbox{.}(2026)]%
        {Dong_2026_TactDeform}
\bibfield{author}{\bibinfo{person}{Yihao Dong}, \bibinfo{person}{Praneeth~Bimsara Perera}, \bibinfo{person}{Chin-Teng Lin}, \bibinfo{person}{Craig~T. Jin}, {and} \bibinfo{person}{Anusha Withana}.} \bibinfo{year}{2026}\natexlab{}.
\newblock \showarticletitle{{TactDeform}: Finger Pad Deformation Inspired Spatial Tactile Feedback for Virtual Geometry Exploration}. In \bibinfo{booktitle}{\emph{Proceedings of the 2026 {CHI} Conference on Human Factors in Computing Systems ({CHI '26})}} (Barcelona, Spain). \bibinfo{publisher}{Association for Computing Machinery}, \bibinfo{address}{New York, NY, USA}.
\newblock
\showISBNx{979-8-4007-2278-3/2026/04}
\href{https://doi.org/10.1145/3772318.3791699}{doi:\nolinkurl{10.1145/3772318.3791699}}


\bibitem[Dong et~al\mbox{.}(2025)]%
        {dong2025just}
\bibfield{author}{\bibinfo{person}{Yihao Dong}, \bibinfo{person}{Pamuditha Somarathne}, \bibinfo{person}{Craig~T Jin}, \bibinfo{person}{Juno Kim}, \bibinfo{person}{Andrea Bianchi}, {and} \bibinfo{person}{Anusha Withana}.} \bibinfo{year}{2025}\natexlab{}.
\newblock \showarticletitle{Just Before Touch: Manipulating Perceived Haptic Sensations through Proactive Vibrotactile Cues in Virtual Reality}. In \bibinfo{booktitle}{\emph{Proceedings of the Augmented Humans International Conference 2025}}. \bibinfo{pages}{79--91}.
\newblock


\bibitem[Dragan et~al\mbox{.}(2013)]%
        {dragan2013legibility}
\bibfield{author}{\bibinfo{person}{Anca~D. Dragan}, \bibinfo{person}{Kenton C.~T. Lee}, {and} \bibinfo{person}{Siddhartha~S. Srinivasa}.} \bibinfo{year}{2013}\natexlab{}.
\newblock \showarticletitle{Legibility and Predictability of Robot Motion}. In \bibinfo{booktitle}{\emph{Proceedings of the 8th ACM/IEEE International Conference on Human-Robot Interaction (HRI)}}. \bibinfo{publisher}{IEEE}, \bibinfo{pages}{301--308}.
\newblock


\bibitem[Dragan and Srinivasa(2013)]%
        {dragan2013policy}
\bibfield{author}{\bibinfo{person}{Anca~D. Dragan} {and} \bibinfo{person}{Siddhartha~S. Srinivasa}.} \bibinfo{year}{2013}\natexlab{}.
\newblock \showarticletitle{A Policy-Blending Formalism for Shared Control}. In \bibinfo{booktitle}{\emph{Proceedings of the ACM/IEEE International Conference on Human-Robot Interaction (HRI)}}. \bibinfo{pages}{1--8}.
\newblock


\bibitem[Eisinga et~al\mbox{.}(2013)]%
        {eisinga2013reliability}
\bibfield{author}{\bibinfo{person}{Rob Eisinga}, \bibinfo{person}{Manfred~te Grotenhuis}, {and} \bibinfo{person}{Ben Pelzer}.} \bibinfo{year}{2013}\natexlab{}.
\newblock \showarticletitle{The reliability of a two-item scale: Pearson, Cronbach, or Spearman-Brown?}
\newblock \bibinfo{journal}{\emph{International journal of public health}} \bibinfo{volume}{58}, \bibinfo{number}{4} (\bibinfo{year}{2013}), \bibinfo{pages}{637--642}.
\newblock


\bibitem[Endsley(1995)]%
        {Endsley1995SA}
\bibfield{author}{\bibinfo{person}{Mica~R. Endsley}.} \bibinfo{year}{1995}\natexlab{}.
\newblock \showarticletitle{Toward a Theory of Situation Awareness in Dynamic Systems}.
\newblock \bibinfo{journal}{\emph{Human Factors}} \bibinfo{volume}{37}, \bibinfo{number}{1} (\bibinfo{year}{1995}), \bibinfo{pages}{32--64}.
\newblock


\bibitem[Ericsson and Simon(1993)]%
        {ericsson1993protocol}
\bibfield{author}{\bibinfo{person}{K.~Anders Ericsson} {and} \bibinfo{person}{Herbert~A. Simon}.} \bibinfo{year}{1993}\natexlab{}.
\newblock \bibinfo{booktitle}{\emph{Protocol Analysis: Verbal Reports as Data} (\bibinfo{edition}{revised edition} ed.)}.
\newblock \bibinfo{publisher}{MIT Press}, \bibinfo{address}{Cambridge, MA}.
\newblock


\bibitem[Fan et~al\mbox{.}(2024)]%
        {fan2024spinshot}
\bibfield{author}{\bibinfo{person}{Chia-An Fan}, \bibinfo{person}{En-Huei Wu}, \bibinfo{person}{Chia-Yu Cheng}, \bibinfo{person}{Yu-Cheng Chang}, \bibinfo{person}{Alvaro Lopez}, \bibinfo{person}{Yu Chen}, \bibinfo{person}{Chia-Chen Chi}, \bibinfo{person}{Yi-Sheng Chan}, \bibinfo{person}{Ching-Yi Tsai}, {and} \bibinfo{person}{Mike~Y Chen}.} \bibinfo{year}{2024}\natexlab{}.
\newblock \showarticletitle{SpinShot: Optimizing both physical and perceived force feedback of flywheel-based, directional impact handheld devices}. In \bibinfo{booktitle}{\emph{Proceedings of the 37th Annual ACM Symposium on User Interface Software and Technology}}. \bibinfo{pages}{1--15}.
\newblock


\bibitem[Fan et~al\mbox{.}(2021)]%
        {fan2021high}
\bibfield{author}{\bibinfo{person}{Hongxiang Fan}, \bibinfo{person}{Martin Ferianc}, \bibinfo{person}{Miguel Rodrigues}, \bibinfo{person}{Hongyu Zhou}, \bibinfo{person}{Xinyu Niu}, {and} \bibinfo{person}{Wayne Luk}.} \bibinfo{year}{2021}\natexlab{}.
\newblock \showarticletitle{High-performance FPGA-based accelerator for Bayesian neural networks}. In \bibinfo{booktitle}{\emph{2021 58th ACM/IEEE Design Automation Conference (DAC)}}. IEEE, \bibinfo{pages}{1063--1068}.
\newblock


\bibitem[Ferianc et~al\mbox{.}(2021)]%
        {ferianc2021improving}
\bibfield{author}{\bibinfo{person}{Martin Ferianc}, \bibinfo{person}{Hongxiang Fan}, \bibinfo{person}{Divyansh Manocha}, \bibinfo{person}{Hongyu Zhou}, \bibinfo{person}{Shuanglong Liu}, \bibinfo{person}{Xinyu Niu}, {and} \bibinfo{person}{Wayne Luk}.} \bibinfo{year}{2021}\natexlab{}.
\newblock \showarticletitle{Improving performance estimation for design space exploration for convolutional neural network accelerators}.
\newblock \bibinfo{journal}{\emph{Electronics}} \bibinfo{volume}{10}, \bibinfo{number}{4} (\bibinfo{year}{2021}), \bibinfo{pages}{520}.
\newblock


\bibitem[Gale et~al\mbox{.}(2013)]%
        {gale2013using}
\bibfield{author}{\bibinfo{person}{Nicola~K Gale}, \bibinfo{person}{Gemma Heath}, \bibinfo{person}{Elaine Cameron}, \bibinfo{person}{Sabina Rashid}, {and} \bibinfo{person}{Sabi Redwood}.} \bibinfo{year}{2013}\natexlab{}.
\newblock \showarticletitle{Using the framework method for the analysis of qualitative data in multi-disciplinary health research}.
\newblock \bibinfo{journal}{\emph{BMC medical research methodology}} \bibinfo{volume}{13}, \bibinfo{number}{1} (\bibinfo{year}{2013}), \bibinfo{pages}{117}.
\newblock


\bibitem[Gin{\'e}s~Clavero et~al\mbox{.}(2020)]%
        {gines2020defining}
\bibfield{author}{\bibinfo{person}{Jonatan Gin{\'e}s~Clavero}, \bibinfo{person}{Francisco Mart{\'\i}n~Rico}, \bibinfo{person}{Francisco~J Rodr{\'\i}guez-Lera}, \bibinfo{person}{Jos{\'e}~Miguel Guerrero~Hern{\'a}ndez}, {and} \bibinfo{person}{Vicente Matell{\'a}n~Olivera}.} \bibinfo{year}{2020}\natexlab{}.
\newblock \showarticletitle{Defining adaptive proxemic zones for activity-aware navigation}. In \bibinfo{booktitle}{\emph{Workshop of Physical Agents}}. Springer, \bibinfo{pages}{3--17}.
\newblock


\bibitem[Gonzalez-Franco and Peck(2018)]%
        {gonzalez2018avatar}
\bibfield{author}{\bibinfo{person}{Mar Gonzalez-Franco} {and} \bibinfo{person}{Tabitha~C Peck}.} \bibinfo{year}{2018}\natexlab{}.
\newblock \showarticletitle{Avatar embodiment. towards a standardized questionnaire}.
\newblock \bibinfo{journal}{\emph{Frontiers in Robotics and AI}}  \bibinfo{volume}{5} (\bibinfo{year}{2018}), \bibinfo{pages}{74}.
\newblock


\bibitem[Goodrich and Schultz(2007)]%
        {goodrich2007human}
\bibfield{author}{\bibinfo{person}{Michael~A. Goodrich} {and} \bibinfo{person}{Alan~C. Schultz}.} \bibinfo{year}{2007}\natexlab{}.
\newblock \showarticletitle{Human--Robot Interaction: A Survey}.
\newblock \bibinfo{journal}{\emph{Foundations and Trends in Human--Computer Interaction}} \bibinfo{volume}{1}, \bibinfo{number}{3} (\bibinfo{year}{2007}), \bibinfo{pages}{203--275}.
\newblock
\href{https://doi.org/10.1561/1100000005}{doi:\nolinkurl{10.1561/1100000005}}


\bibitem[Goodrich et~al\mbox{.}(2008)]%
        {goodrich2008human}
\bibfield{author}{\bibinfo{person}{Michael~A Goodrich}, \bibinfo{person}{Alan~C Schultz}, {et~al\mbox{.}}} \bibinfo{year}{2008}\natexlab{}.
\newblock \showarticletitle{Human--robot interaction: a survey}.
\newblock \bibinfo{journal}{\emph{Foundations and trends{\textregistered} in human--computer interaction}} \bibinfo{volume}{1}, \bibinfo{number}{3} (\bibinfo{year}{2008}), \bibinfo{pages}{203--275}.
\newblock


\bibitem[Gough et~al\mbox{.}(2023)]%
        {gough2023design}
\bibfield{author}{\bibinfo{person}{Phillip Gough}, \bibinfo{person}{Praneeth~Bimsara Perera}, \bibinfo{person}{Michael~A Kertesz}, {and} \bibinfo{person}{Anusha Withana}.} \bibinfo{year}{2023}\natexlab{}.
\newblock \showarticletitle{Design, mould, grow!: A fabrication pipeline for growing 3d designs using myco-materials}. In \bibinfo{booktitle}{\emph{Proceedings of the 2023 CHI conference on human factors in computing systems}}. \bibinfo{pages}{1--15}.
\newblock


\bibitem[Guiard(1987)]%
        {Guiard1987Bimanual}
\bibfield{author}{\bibinfo{person}{Yves Guiard}.} \bibinfo{year}{1987}\natexlab{}.
\newblock \showarticletitle{Asymmetric Division of Labor in Human Skilled Bimanual Action: The Kinematic Chain as a Model}.
\newblock \bibinfo{journal}{\emph{Journal of Motor Behavior}} \bibinfo{volume}{19}, \bibinfo{number}{4} (\bibinfo{year}{1987}), \bibinfo{pages}{486--517}.
\newblock


\bibitem[Hall(1966)]%
        {hall1966hidden}
\bibfield{author}{\bibinfo{person}{Edward~T. Hall}.} \bibinfo{year}{1966}\natexlab{}.
\newblock \bibinfo{booktitle}{\emph{The Hidden Dimension}}.
\newblock \bibinfo{publisher}{Doubleday}.
\newblock


\bibitem[Hancock et~al\mbox{.}(2011)]%
        {hancock2011meta}
\bibfield{author}{\bibinfo{person}{Peter~A. Hancock}, \bibinfo{person}{Deborah~R. Billings}, \bibinfo{person}{Kristin~E. Schaefer}, \bibinfo{person}{Jessie Y.~C. Chen}, \bibinfo{person}{Ewart~J. de Visser}, {and} \bibinfo{person}{Raja Parasuraman}.} \bibinfo{year}{2011}\natexlab{}.
\newblock \showarticletitle{A Meta-Analysis of Factors Affecting Trust in Human–Robot Interaction}. In \bibinfo{booktitle}{\emph{Proceedings of the 6th ACM/IEEE International Conference on Human–Robot Interaction (HRI)}}. \bibinfo{pages}{517--524}.
\newblock
\href{https://doi.org/10.1145/1957656.1957786}{doi:\nolinkurl{10.1145/1957656.1957786}}


\bibitem[Hauke and Kossowski(2011)]%
        {hauke2011comparison}
\bibfield{author}{\bibinfo{person}{Jan Hauke} {and} \bibinfo{person}{Tomasz Kossowski}.} \bibinfo{year}{2011}\natexlab{}.
\newblock \showarticletitle{Comparison of values of Pearson's and Spearman's correlation coefficients on the same sets of data}.
\newblock \bibinfo{journal}{\emph{Quaestiones geographicae}} \bibinfo{volume}{30}, \bibinfo{number}{2} (\bibinfo{year}{2011}), \bibinfo{pages}{87--93}.
\newblock


\bibitem[Herrera-Valenzuela et~al\mbox{.}(2023)]%
        {herrera2023qualitative}
\bibfield{author}{\bibinfo{person}{Diana Herrera-Valenzuela}, \bibinfo{person}{Laura D{\'\i}az-Pe{\~n}a}, \bibinfo{person}{Carolina Redondo-Gal{\'a}n}, \bibinfo{person}{Mar{\'\i}a~Jos{\'e} Arroyo}, \bibinfo{person}{L{\'\i}a Cascante-Guti{\'e}rrez}, \bibinfo{person}{{\'A}ngel Gil-Agudo}, \bibinfo{person}{Juan~C Moreno}, {and} \bibinfo{person}{Antonio~J Del-Ama}.} \bibinfo{year}{2023}\natexlab{}.
\newblock \showarticletitle{A qualitative study to elicit user requirements for lower limb wearable exoskeletons for gait rehabilitation in spinal cord injury}.
\newblock \bibinfo{journal}{\emph{Journal of NeuroEngineering and Rehabilitation}} \bibinfo{volume}{20}, \bibinfo{number}{1} (\bibinfo{year}{2023}), \bibinfo{pages}{138}.
\newblock


\bibitem[Holthaus and Wachsmuth(2012)]%
        {holthaus2012active}
\bibfield{author}{\bibinfo{person}{Patrick Holthaus} {and} \bibinfo{person}{Sven Wachsmuth}.} \bibinfo{year}{2012}\natexlab{}.
\newblock \showarticletitle{Active peripersonal space for more intuitive HRI}. In \bibinfo{booktitle}{\emph{2012 12th IEEE-RAS International Conference on Humanoid Robots (Humanoids 2012)}}. IEEE, \bibinfo{pages}{508--513}.
\newblock


\bibitem[Hu et~al\mbox{.}(2017)]%
        {hu2017hand}
\bibfield{author}{\bibinfo{person}{Yuhan Hu}, \bibinfo{person}{Sang-won Leigh}, {and} \bibinfo{person}{Pattie Maes}.} \bibinfo{year}{2017}\natexlab{}.
\newblock \showarticletitle{Hand development kit: Soft robotic fingers as prosthetic augmentation of the hand}. In \bibinfo{booktitle}{\emph{Adjunct Proceedings of the 30th Annual ACM Symposium on User Interface Software and Technology}}. \bibinfo{pages}{27--29}.
\newblock


\bibitem[Hussain et~al\mbox{.}(2017)]%
        {hussain2017hand}
\bibfield{author}{\bibinfo{person}{Irfan Hussain}, \bibinfo{person}{Gionata Salvietti}, \bibinfo{person}{Giovanni Spagnoletti}, \bibinfo{person}{Monica Malvezzi}, \bibinfo{person}{David Cioncoloni}, \bibinfo{person}{Simone Rossi}, {and} \bibinfo{person}{Domenico Prattichizzo}.} \bibinfo{year}{2017}\natexlab{}.
\newblock \showarticletitle{A Soft Supernumerary Robotic Finger and Mobile Arm Support for Grasping Compensation and Hemiparetic Upper Limb Rehabilitation}.
\newblock \bibinfo{journal}{\emph{Robotics and Autonomous Systems}}  \bibinfo{volume}{93} (\bibinfo{year}{2017}), \bibinfo{pages}{1--12}.
\newblock
\href{https://doi.org/10.1016/j.robot.2017.03.015}{doi:\nolinkurl{10.1016/j.robot.2017.03.015}}


\bibitem[Inami et~al\mbox{.}(2022)]%
        {inami2022cyborgs}
\bibfield{author}{\bibinfo{person}{Masahiko Inami}, \bibinfo{person}{Daisuke Uriu}, \bibinfo{person}{Zendai Kashino}, \bibinfo{person}{Shigeo Yoshida}, \bibinfo{person}{Hiroto Saito}, \bibinfo{person}{Azumi Maekawa}, {and} \bibinfo{person}{Michiteru Kitazaki}.} \bibinfo{year}{2022}\natexlab{}.
\newblock \showarticletitle{Cyborgs, human augmentation, cybernetics, and JIZAI body}. In \bibinfo{booktitle}{\emph{Proceedings of the Augmented Humans International Conference 2022}}. \bibinfo{pages}{230--242}.
\newblock


\bibitem[Jayasiriwardene and Meedeniya(2021a)]%
        {jayasiriwardene2021architectural}
\bibfield{author}{\bibinfo{person}{Shakyani Jayasiriwardene} {and} \bibinfo{person}{Dulani Meedeniya}.} \bibinfo{year}{2021}\natexlab{a}.
\newblock \showarticletitle{Architectural framework for an interactive learning toolkit}. In \bibinfo{booktitle}{\emph{2021 International Research Conference on Smart Computing and Systems Engineering (SCSE)}}, Vol.~\bibinfo{volume}{4}. IEEE, \bibinfo{pages}{14--21}.
\newblock


\bibitem[Jayasiriwardene and Meedeniya(2021b)]%
        {jayasiriwardene2021interactive}
\bibfield{author}{\bibinfo{person}{Shakyani Jayasiriwardene} {and} \bibinfo{person}{Dulani Meedeniya}.} \bibinfo{year}{2021}\natexlab{b}.
\newblock \showarticletitle{Interactive and adaptive learning content authoring framework for an m-learning toolkit}. In \bibinfo{booktitle}{\emph{2021 1st Conference on Online Teaching for Mobile Education (OT4ME)}}. IEEE, \bibinfo{pages}{153--160}.
\newblock


\bibitem[Jayasiriwardene and Meedeniya(2022)]%
        {jayasiriwardene2022knowledge}
\bibfield{author}{\bibinfo{person}{Shakyani Jayasiriwardene} {and} \bibinfo{person}{Dulani Meedeniya}.} \bibinfo{year}{2022}\natexlab{}.
\newblock \showarticletitle{A knowledge-based adaptive algorithm to recommend interactive learning assessments}. In \bibinfo{booktitle}{\emph{2022 2nd International Conference on Advanced Research in Computing (ICARC)}}. IEEE, \bibinfo{pages}{379--384}.
\newblock


\bibitem[Jayasiriwardene and Meedeniya(2023)]%
        {jayasiriwardene2023adaptive}
\bibfield{author}{\bibinfo{person}{Shakyani Jayasiriwardene} {and} \bibinfo{person}{Dulani Meedeniya}.} \bibinfo{year}{2023}\natexlab{}.
\newblock \showarticletitle{An adaptive and interactive learning toolkit (iLearn)}.
\newblock \bibinfo{journal}{\emph{Software Impacts}}  \bibinfo{volume}{15} (\bibinfo{year}{2023}), \bibinfo{pages}{100471}.
\newblock


\bibitem[Jayasiriwardene et~al\mbox{.}(2025)]%
        {jayasiriwardene2025more}
\bibfield{author}{\bibinfo{person}{Shakyani Jayasiriwardene}, \bibinfo{person}{Benjamin Tag}, \bibinfo{person}{Anusha Withana}, {and} \bibinfo{person}{Zhanna Sarsenbayeva}.} \bibinfo{year}{2025}\natexlab{}.
\newblock \showarticletitle{More Than Words: The Impact of Voice Assistant Personality Traits on Failure Mitigation}.
\newblock \bibinfo{journal}{\emph{Proceedings of the ACM on Interactive, Mobile, Wearable and Ubiquitous Technologies}} \bibinfo{volume}{9}, \bibinfo{number}{3} (\bibinfo{year}{2025}), \bibinfo{pages}{1--33}.
\newblock


\bibitem[Jayasiriwardene et~al\mbox{.}(2026)]%
        {jayasiriwardene2026fixed}
\bibfield{author}{\bibinfo{person}{Shakyani Jayasiriwardene}, \bibinfo{person}{Hongyu Zhou}, \bibinfo{person}{Weiwei Jiang}, \bibinfo{person}{Benjamin Tag}, \bibinfo{person}{Emmanuel Stamatakis}, \bibinfo{person}{Anusha Withana}, {and} \bibinfo{person}{Zhanna Sarsenbayeva}.} \bibinfo{year}{2026}\natexlab{}.
\newblock \showarticletitle{From Fixed to Flexible: Shaping AI Personality in Context-Sensitive Interaction}.
\newblock \bibinfo{journal}{\emph{arXiv preprint arXiv:2601.08194}} (\bibinfo{year}{2026}).
\newblock


\bibitem[Jian et~al\mbox{.}(2000)]%
        {jian2000foundations}
\bibfield{author}{\bibinfo{person}{Jiun-Yin Jian}, \bibinfo{person}{Ann~M Bisantz}, {and} \bibinfo{person}{Colin~G Drury}.} \bibinfo{year}{2000}\natexlab{}.
\newblock \showarticletitle{Foundations for an empirically determined scale of trust in automated systems}.
\newblock \bibinfo{journal}{\emph{International journal of cognitive ergonomics}} \bibinfo{volume}{4}, \bibinfo{number}{1} (\bibinfo{year}{2000}), \bibinfo{pages}{53--71}.
\newblock


\bibitem[Kirkwood et~al\mbox{.}(2021)]%
        {kirkwood2021s}
\bibfield{author}{\bibinfo{person}{Gavin~Lawrence Kirkwood}, \bibinfo{person}{Christopher~D Otmar}, {and} \bibinfo{person}{Mohemmad Hansia}.} \bibinfo{year}{2021}\natexlab{}.
\newblock \showarticletitle{Who's leading this dance?: Theorizing automatic and strategic synchrony in human-exoskeleton interactions}.
\newblock \bibinfo{journal}{\emph{Frontiers in Psychology}}  \bibinfo{volume}{12} (\bibinfo{year}{2021}), \bibinfo{pages}{624108}.
\newblock


\bibitem[Kirschner et~al\mbox{.}(2022)]%
        {kirschner2022iso}
\bibfield{author}{\bibinfo{person}{Robin~Jeanne Kirschner}, \bibinfo{person}{Nico Mansfeld}, \bibinfo{person}{Saeed Abdolshah}, {and} \bibinfo{person}{Sami Haddadin}.} \bibinfo{year}{2022}\natexlab{}.
\newblock \showarticletitle{ISO/TS 15066: How different interpretations affect risk assessment}.
\newblock \bibinfo{journal}{\emph{arXiv preprint arXiv:2203.02706}} (\bibinfo{year}{2022}).
\newblock


\bibitem[Knepper et~al\mbox{.}(2017)]%
        {knepper2017implicit}
\bibfield{author}{\bibinfo{person}{Ross~A. Knepper}, \bibinfo{person}{Siddhartha~S. Srinivasa}, \bibinfo{person}{Johan~E. Bergstr{\"o}m}, {and} \bibinfo{person}{G{\"u}nter Niemeyer}.} \bibinfo{year}{2017}\natexlab{}.
\newblock \showarticletitle{Implicit Communication in a Joint Action}. In \bibinfo{booktitle}{\emph{Proceedings of the 2017 ACM/IEEE International Conference on Human-Robot Interaction (HRI)}}. \bibinfo{publisher}{ACM}, \bibinfo{pages}{283--292}.
\newblock


\bibitem[Lasota et~al\mbox{.}(2017)]%
        {Lasota2017HRC}
\bibfield{author}{\bibinfo{person}{Przemyslaw~A. Lasota}, \bibinfo{person}{Terrence Fong}, {and} \bibinfo{person}{Julie~A. Shah}.} \bibinfo{year}{2017}\natexlab{}.
\newblock \showarticletitle{A Survey of Methods for Safe Human–Robot Interaction}.
\newblock \bibinfo{journal}{\emph{Foundations and Trends in Robotics}} \bibinfo{volume}{5}, \bibinfo{number}{4} (\bibinfo{year}{2017}), \bibinfo{pages}{261--349}.
\newblock


\bibitem[LaValle(2006)]%
        {lavalle2006planning}
\bibfield{author}{\bibinfo{person}{Steven~M. LaValle}.} \bibinfo{year}{2006}\natexlab{}.
\newblock \bibinfo{booktitle}{\emph{Planning Algorithms}}.
\newblock \bibinfo{publisher}{Cambridge University Press}.
\newblock


\bibitem[Lee et~al\mbox{.}(2025)]%
        {lee2025e}
\bibfield{author}{\bibinfo{person}{I-Chieh Lee}, \bibinfo{person}{Ming Liu}, {and} \bibinfo{person}{He Huang}.} \bibinfo{year}{2025}\natexlab{}.
\newblock \showarticletitle{Enhancing Robot Transparency in Human–Robot Prosthesis Interaction to Mitigate Terrain Misrecognition Error}.
\newblock \bibinfo{journal}{\emph{IEEE Transactions on Medical Robotics and Bionics}} \bibinfo{volume}{7}, \bibinfo{number}{2} (\bibinfo{year}{2025}), \bibinfo{pages}{734--742}.
\newblock
\href{https://doi.org/10.1109/TMRB.2025.3552924}{doi:\nolinkurl{10.1109/TMRB.2025.3552924}}


\bibitem[Lee et~al\mbox{.}(2022)]%
        {lee2022toward}
\bibfield{author}{\bibinfo{person}{I-Chieh Lee}, \bibinfo{person}{Ming Liu}, \bibinfo{person}{Michael~D. Lewek}, \bibinfo{person}{Xiaogang Hu}, \bibinfo{person}{William~G. Filer}, {and} \bibinfo{person}{He Huang}.} \bibinfo{year}{2022}\natexlab{}.
\newblock \showarticletitle{Toward Safe Wearer-Prosthesis Interaction: Evaluation of Gait Stability and Human Compensation Strategy Under Faults in Robotic Transfemoral Prostheses}.
\newblock \bibinfo{journal}{\emph{IEEE Transactions on Neural Systems and Rehabilitation Engineering}}  \bibinfo{volume}{30} (\bibinfo{year}{2022}), \bibinfo{pages}{2773--2782}.
\newblock
\href{https://doi.org/10.1109/TNSRE.2022.3208778}{doi:\nolinkurl{10.1109/TNSRE.2022.3208778}}


\bibitem[Lehmann et~al\mbox{.}(2020)]%
        {lehmann2020should}
\bibfield{author}{\bibinfo{person}{Hagen Lehmann}, \bibinfo{person}{Adam Rojik}, {and} \bibinfo{person}{Matej Hoffmann}.} \bibinfo{year}{2020}\natexlab{}.
\newblock \showarticletitle{Should a small robot have a small personal space? Investigating personal spatial zones and proxemic behavior in human-robot interaction}.
\newblock \bibinfo{journal}{\emph{arXiv preprint arXiv:2009.01818}} (\bibinfo{year}{2020}).
\newblock


\bibitem[Lincoln and Guba(1985)]%
        {lincoln1985naturalistic}
\bibfield{author}{\bibinfo{person}{Yvonna~S. Lincoln} {and} \bibinfo{person}{Egon~G. Guba}.} \bibinfo{year}{1985}\natexlab{}.
\newblock \bibinfo{booktitle}{\emph{Naturalistic Inquiry}}.
\newblock \bibinfo{publisher}{SAGE Publications}, \bibinfo{address}{Newbury Park, CA}.
\newblock


\bibitem[Lisini~Baldi et~al\mbox{.}(2025)]%
        {lisini2025exploiting}
\bibfield{author}{\bibinfo{person}{Tommaso Lisini~Baldi}, \bibinfo{person}{Nicole D’Aurizio}, \bibinfo{person}{Chiara Gaudeni}, \bibinfo{person}{Sergio Gurgone}, \bibinfo{person}{Daniele Borzelli}, \bibinfo{person}{Andrea d’Avella}, {and} \bibinfo{person}{Domenico Prattichizzo}.} \bibinfo{year}{2025}\natexlab{}.
\newblock \showarticletitle{Exploiting body redundancy to control supernumerary robotic limbs in human augmentation}.
\newblock \bibinfo{journal}{\emph{The International Journal of Robotics Research}} \bibinfo{volume}{44}, \bibinfo{number}{2} (\bibinfo{year}{2025}), \bibinfo{pages}{291--316}.
\newblock


\bibitem[Liu et~al\mbox{.}(2025a)]%
        {liu2025effects}
\bibfield{author}{\bibinfo{person}{Ruilin Liu}, \bibinfo{person}{Tinghui Li}, {and} \bibinfo{person}{Zhanna Sarsenbayeva}.} \bibinfo{year}{2025}\natexlab{a}.
\newblock \showarticletitle{Effects of Ambient Illumination and Screen Luminance on Mixed Reality Interaction}.
\newblock \bibinfo{journal}{\emph{IEEE Access}}  \bibinfo{volume}{13} (\bibinfo{year}{2025}), \bibinfo{pages}{192837--192855}.
\newblock


\bibitem[Liu et~al\mbox{.}(2025b)]%
        {liu2025understanding}
\bibfield{author}{\bibinfo{person}{Ruilin Liu}, \bibinfo{person}{Tinghui Li}, {and} \bibinfo{person}{Zhanna Sarsenbayeva}.} \bibinfo{year}{2025}\natexlab{b}.
\newblock \showarticletitle{Understanding and addressing ambient illumination as a situational impairment for digital devices}. In \bibinfo{booktitle}{\emph{Proceedings of the 37th Australian Conference on Human-Computer Interaction}}. \bibinfo{pages}{556--566}.
\newblock


\bibitem[Lyons et~al\mbox{.}(2023)]%
        {lyons2023explanations}
\bibfield{author}{\bibinfo{person}{Joseph~B Lyons}, \bibinfo{person}{Izz aldin Hamdan}, {and} \bibinfo{person}{Thy~Q Vo}.} \bibinfo{year}{2023}\natexlab{}.
\newblock \showarticletitle{Explanations and trust: What happens to trust when a robot partner does something unexpected?}
\newblock \bibinfo{journal}{\emph{Computers in Human Behavior}}  \bibinfo{volume}{138} (\bibinfo{year}{2023}), \bibinfo{pages}{107473}.
\newblock


\bibitem[Major et~al\mbox{.}(2020)]%
        {major2020perturbation}
\bibfield{author}{\bibinfo{person}{Matthew~J Major}, \bibinfo{person}{Chelsi~K Serba}, {and} \bibinfo{person}{Keith~E Gordon}.} \bibinfo{year}{2020}\natexlab{}.
\newblock \showarticletitle{Perturbation recovery during walking is impacted by knowledge of perturbation timing in below-knee prosthesis users and non-impaired participants}.
\newblock \bibinfo{journal}{\emph{Plos one}} \bibinfo{volume}{15}, \bibinfo{number}{7} (\bibinfo{year}{2020}), \bibinfo{pages}{e0235686}.
\newblock


\bibitem[Makin et~al\mbox{.}(2017)]%
        {makin2017ownership}
\bibfield{author}{\bibinfo{person}{Tamar~R. Makin}, \bibinfo{person}{Fr{\'e}d{\'e}rique de Vignemont}, {and} \bibinfo{person}{A.~Aldo Faisal}.} \bibinfo{year}{2017}\natexlab{}.
\newblock \showarticletitle{Neurocognitive Considerations of Body Ownership and Agency}.
\newblock \bibinfo{journal}{\emph{Trends in Cognitive Sciences}} \bibinfo{volume}{21}, \bibinfo{number}{7} (\bibinfo{year}{2017}), \bibinfo{pages}{542--553}.
\newblock
\href{https://doi.org/10.1016/j.tics.2017.04.005}{doi:\nolinkurl{10.1016/j.tics.2017.04.005}}


\bibitem[Mandl et~al\mbox{.}(2022)]%
        {mandl2022embodied}
\bibfield{author}{\bibinfo{person}{Sarah Mandl}, \bibinfo{person}{Maximilian Bretschneider}, \bibinfo{person}{Stefanie Meyer}, {and} \bibinfo{person}{Anja Strobel}.} \bibinfo{year}{2022}\natexlab{}.
\newblock \showarticletitle{Embodied Digital Technologies: First Insights in the Social and Legal Perception of Robots and Users of Prostheses}.
\newblock \bibinfo{journal}{\emph{Frontiers in Robotics and AI}}  \bibinfo{volume}{9} (\bibinfo{year}{2022}), \bibinfo{pages}{867649}.
\newblock


\bibitem[McDuff et~al\mbox{.}(2012)]%
        {mcduff2012affectaura}
\bibfield{author}{\bibinfo{person}{Daniel McDuff}, \bibinfo{person}{Amy Karlson}, \bibinfo{person}{Ashish Kapoor}, \bibinfo{person}{Asta Roseway}, {and} \bibinfo{person}{Mary Czerwinski}.} \bibinfo{year}{2012}\natexlab{}.
\newblock \showarticletitle{AffectAura: an intelligent system for emotional memory}. In \bibinfo{booktitle}{\emph{Proceedings of the SIGCHI conference on human factors in computing systems}}. \bibinfo{pages}{849--858}.
\newblock


\bibitem[Mead et~al\mbox{.}(2013)]%
        {mead2013automated}
\bibfield{author}{\bibinfo{person}{Ross Mead}, \bibinfo{person}{Amin Atrash}, {and} \bibinfo{person}{Maja~J Matari{\'c}}.} \bibinfo{year}{2013}\natexlab{}.
\newblock \showarticletitle{Automated proxemic feature extraction and behavior recognition: Applications in human-robot interaction}.
\newblock \bibinfo{journal}{\emph{International Journal of Social Robotics}} \bibinfo{volume}{5}, \bibinfo{number}{3} (\bibinfo{year}{2013}), \bibinfo{pages}{367--378}.
\newblock


\bibitem[Mead and Matari{\'c}(2016)]%
        {mead2016proxemic}
\bibfield{author}{\bibinfo{person}{Ross Mead} {and} \bibinfo{person}{Maja~J. Matari{\'c}}.} \bibinfo{year}{2016}\natexlab{}.
\newblock \showarticletitle{Perceptual Models of Human--Robot Proxemics}.
\newblock In \bibinfo{booktitle}{\emph{Experimental Robotics}}. \bibinfo{series}{Springer Tracts in Advanced Robotics}, Vol.~\bibinfo{volume}{109}. \bibinfo{publisher}{Springer}, \bibinfo{pages}{261--276}.
\newblock


\bibitem[Meehan et~al\mbox{.}(2002)]%
        {meehan2002physiological}
\bibfield{author}{\bibinfo{person}{Michael Meehan}, \bibinfo{person}{Brent Insko}, \bibinfo{person}{Mary~C. Whitton}, {and} \bibinfo{person}{Frederick~P. Brooks}.} \bibinfo{year}{2002}\natexlab{}.
\newblock \showarticletitle{Physiological Measures of Presence in Stressful Virtual Environments}.
\newblock \bibinfo{journal}{\emph{ACM Transactions on Graphics}} \bibinfo{volume}{21}, \bibinfo{number}{3} (\bibinfo{year}{2002}), \bibinfo{pages}{645--652}.
\newblock
\href{https://doi.org/10.1145/566654.566630}{doi:\nolinkurl{10.1145/566654.566630}}


\bibitem[Meyer et~al\mbox{.}(2021)]%
        {meyer2021analysis}
\bibfield{author}{\bibinfo{person}{Jan~Thomas Meyer}, \bibinfo{person}{Roger Gassert}, {and} \bibinfo{person}{Olivier Lambercy}.} \bibinfo{year}{2021}\natexlab{}.
\newblock \showarticletitle{An analysis of usability evaluation practices and contexts of use in wearable robotics}.
\newblock \bibinfo{journal}{\emph{Journal of neuroengineering and rehabilitation}} \bibinfo{volume}{18}, \bibinfo{number}{1} (\bibinfo{year}{2021}), \bibinfo{pages}{170}.
\newblock


\bibitem[Miles et~al\mbox{.}(2014)]%
        {miles2014qualitative}
\bibfield{author}{\bibinfo{person}{Matthew~B. Miles}, \bibinfo{person}{A.~Michael Huberman}, {and} \bibinfo{person}{Johnny Salda{\~n}a}.} \bibinfo{year}{2014}\natexlab{}.
\newblock \bibinfo{booktitle}{\emph{Qualitative Data Analysis: A Methods Sourcebook} (\bibinfo{edition}{3} ed.)}.
\newblock \bibinfo{publisher}{SAGE Publications}, \bibinfo{address}{Thousand Oaks, CA}.
\newblock


\bibitem[Muehlhaus et~al\mbox{.}(2023)]%
        {muehlhaus2023need}
\bibfield{author}{\bibinfo{person}{Marie Muehlhaus}, \bibinfo{person}{Marion Koelle}, \bibinfo{person}{Artin Saberpour}, {and} \bibinfo{person}{J{\"u}rgen Steimle}.} \bibinfo{year}{2023}\natexlab{}.
\newblock \showarticletitle{I need a third arm! eliciting body-based interactions with a wearable robotic arm}. In \bibinfo{booktitle}{\emph{Proceedings of the 2023 CHI Conference on Human Factors in Computing Systems}}. \bibinfo{pages}{1--15}.
\newblock


\bibitem[M{\"u}ller and Richert(2025)]%
        {muller2025space}
\bibfield{author}{\bibinfo{person}{Ana M{\"u}ller} {and} \bibinfo{person}{Anja Richert}.} \bibinfo{year}{2025}\natexlab{}.
\newblock \showarticletitle{The Space Between Us: A Methodological Framework for Researching Bonding and Proxemics in Situated Group-Agent Interactions}.
\newblock \bibinfo{journal}{\emph{arXiv preprint arXiv:2506.11829}} (\bibinfo{year}{2025}).
\newblock


\bibitem[Mumm and Mutlu(2011)]%
        {mumm2011human}
\bibfield{author}{\bibinfo{person}{Jonathan Mumm} {and} \bibinfo{person}{Bilge Mutlu}.} \bibinfo{year}{2011}\natexlab{}.
\newblock \showarticletitle{Human-robot proxemics: physical and psychological distancing in human-robot interaction}. In \bibinfo{booktitle}{\emph{Proceedings of the 6th international conference on Human-robot interaction}}. \bibinfo{pages}{331--338}.
\newblock


\bibitem[Naseri et~al\mbox{.}(2022)]%
        {naseri2022characterizing}
\bibfield{author}{\bibinfo{person}{Amirreza Naseri}, \bibinfo{person}{Ming Liu}, \bibinfo{person}{I-Chieh Lee}, \bibinfo{person}{Wentao Liu}, {and} \bibinfo{person}{He Huang}.} \bibinfo{year}{2022}\natexlab{}.
\newblock \showarticletitle{Characterizing Prosthesis Control Fault During Human-Prosthesis Interactive Walking Using Intrinsic Sensors}.
\newblock \bibinfo{journal}{\emph{IEEE Robotics and Automation Letters}} \bibinfo{volume}{7}, \bibinfo{number}{3} (\bibinfo{year}{2022}), \bibinfo{pages}{8307--8314}.
\newblock
\href{https://doi.org/10.1109/LRA.2022.3186503}{doi:\nolinkurl{10.1109/LRA.2022.3186503}}


\bibitem[Newman et~al\mbox{.}(2022)]%
        {newman2022helping}
\bibfield{author}{\bibinfo{person}{Benjamin~A Newman}, \bibinfo{person}{Reuben~M Aronson}, \bibinfo{person}{Kris Kitani}, {and} \bibinfo{person}{Henny Admoni}.} \bibinfo{year}{2022}\natexlab{}.
\newblock \showarticletitle{Helping people through space and time: Assistance as a perspective on human-robot interaction}.
\newblock \bibinfo{journal}{\emph{Frontiers in Robotics and AI}}  \bibinfo{volume}{8} (\bibinfo{year}{2022}), \bibinfo{pages}{720319}.
\newblock


\bibitem[Nowell et~al\mbox{.}(2017)]%
        {nowell2017thematic}
\bibfield{author}{\bibinfo{person}{Lorelli~S. Nowell}, \bibinfo{person}{Jill~M. Norris}, \bibinfo{person}{Deborah~E. White}, {and} \bibinfo{person}{Nancy~J. Moules}.} \bibinfo{year}{2017}\natexlab{}.
\newblock \showarticletitle{Thematic analysis: Striving to meet the trustworthiness criteria}.
\newblock \bibinfo{journal}{\emph{The Qualitative Report}} \bibinfo{volume}{22}, \bibinfo{number}{3} (\bibinfo{year}{2017}), \bibinfo{pages}{160--177}.
\newblock


\bibitem[Oechsner et~al\mbox{.}(2025)]%
        {oechsner2025influence}
\bibfield{author}{\bibinfo{person}{Carl Oechsner}, \bibinfo{person}{Jan Leusmann}, \bibinfo{person}{Robin Welsch}, \bibinfo{person}{Andreas~Martin Butz}, {and} \bibinfo{person}{Sven Mayer}.} \bibinfo{year}{2025}\natexlab{}.
\newblock \showarticletitle{Influence of Perceived Danger on Proxemics in Human-Robot Object Handovers}. In \bibinfo{booktitle}{\emph{Proceedings of the Mensch und Computer 2025}}. \bibinfo{publisher}{Association for Computing Machinery}, \bibinfo{address}{New York, NY, USA}, \bibinfo{pages}{111--120}.
\newblock
\showISBNx{9798400715822}
\href{https://doi.org/10.1145/3743049.3743064}{doi:\nolinkurl{10.1145/3743049.3743064}}


\bibitem[Ogawa et~al\mbox{.}(2020)]%
        {ogawa2020you}
\bibfield{author}{\bibinfo{person}{Nami Ogawa}, \bibinfo{person}{Takuji Narumi}, \bibinfo{person}{Hideaki Kuzuoka}, {and} \bibinfo{person}{Michitaka Hirose}.} \bibinfo{year}{2020}\natexlab{}.
\newblock \showarticletitle{Do you feel like passing through walls?: Effect of self-avatar appearance on facilitating realistic behavior in virtual environments}. In \bibinfo{booktitle}{\emph{Proceedings of the 2020 CHI conference on human factors in computing systems}}. \bibinfo{pages}{1--14}.
\newblock


\bibitem[Olen{\v{s}}ek et~al\mbox{.}(2021)]%
        {olenvsek2021dynamic}
\bibfield{author}{\bibinfo{person}{Andrej Olen{\v{s}}ek}, \bibinfo{person}{Matja{\v{z}} Zadravec}, \bibinfo{person}{Helena Burger}, {and} \bibinfo{person}{Zlatko Matja{\v{c}}i{\'c}}.} \bibinfo{year}{2021}\natexlab{}.
\newblock \showarticletitle{Dynamic balancing responses in unilateral transtibial amputees following outward-directed perturbations during slow treadmill walking differ considerably for amputated and non-amputated side}.
\newblock \bibinfo{journal}{\emph{Journal of neuroengineering and rehabilitation}} \bibinfo{volume}{18}, \bibinfo{number}{1} (\bibinfo{year}{2021}), \bibinfo{pages}{123}.
\newblock


\bibitem[Ortenzi et~al\mbox{.}(2021)]%
        {ortenzi2021object}
\bibfield{author}{\bibinfo{person}{Valerio Ortenzi}, \bibinfo{person}{Akansel Cosgun}, \bibinfo{person}{Tommaso Pardi}, \bibinfo{person}{Wesley~P Chan}, \bibinfo{person}{Elizabeth Croft}, {and} \bibinfo{person}{Dana Kuli{\'c}}.} \bibinfo{year}{2021}\natexlab{}.
\newblock \showarticletitle{Object handovers: a review for robotics}.
\newblock \bibinfo{journal}{\emph{IEEE Transactions on Robotics}} \bibinfo{volume}{37}, \bibinfo{number}{6} (\bibinfo{year}{2021}), \bibinfo{pages}{1855--1873}.
\newblock


\bibitem[Parasuraman et~al\mbox{.}(2000a)]%
        {parasuraman2000model}
\bibfield{author}{\bibinfo{person}{Raja Parasuraman}, \bibinfo{person}{Thomas~B Sheridan}, {and} \bibinfo{person}{Christopher~D Wickens}.} \bibinfo{year}{2000}\natexlab{a}.
\newblock \showarticletitle{A model for types and levels of human interaction with automation}.
\newblock \bibinfo{journal}{\emph{IEEE Transactions on systems, man, and cybernetics-Part A: Systems and Humans}} \bibinfo{volume}{30}, \bibinfo{number}{3} (\bibinfo{year}{2000}), \bibinfo{pages}{286--297}.
\newblock


\bibitem[Parasuraman et~al\mbox{.}(2000b)]%
        {Parasuraman2000Automation}
\bibfield{author}{\bibinfo{person}{Raja Parasuraman}, \bibinfo{person}{Thomas~B. Sheridan}, {and} \bibinfo{person}{Christopher~D. Wickens}.} \bibinfo{year}{2000}\natexlab{b}.
\newblock \showarticletitle{A Model for Types and Levels of Human Interaction with Automation}.
\newblock \bibinfo{journal}{\emph{IEEE Transactions on Systems, Man, and Cybernetics - Part A}} \bibinfo{volume}{30}, \bibinfo{number}{3} (\bibinfo{year}{2000}), \bibinfo{pages}{286--297}.
\newblock


\bibitem[Parietti and Asada(2016)]%
        {parietti2016supernumerary}
\bibfield{author}{\bibinfo{person}{Federico Parietti} {and} \bibinfo{person}{Harry Asada}.} \bibinfo{year}{2016}\natexlab{}.
\newblock \showarticletitle{Supernumerary robotic limbs for human body support}.
\newblock \bibinfo{journal}{\emph{IEEE Transactions on Robotics}} \bibinfo{volume}{32}, \bibinfo{number}{2} (\bibinfo{year}{2016}), \bibinfo{pages}{301--311}.
\newblock


\bibitem[Parietti and Asada(2013)]%
        {parietti2013dynamic}
\bibfield{author}{\bibinfo{person}{Federico Parietti} {and} \bibinfo{person}{Harry~H Asada}.} \bibinfo{year}{2013}\natexlab{}.
\newblock \showarticletitle{Dynamic analysis and state estimation for wearable robotic limbs subject to human-induced disturbances}. In \bibinfo{booktitle}{\emph{2013 IEEE International Conference on Robotics and Automation}}. IEEE, \bibinfo{pages}{3880--3887}.
\newblock


\bibitem[Parietti and Asada(2014)]%
        {parietti2014supernumerary}
\bibfield{author}{\bibinfo{person}{Federico Parietti} {and} \bibinfo{person}{H~Harry Asada}.} \bibinfo{year}{2014}\natexlab{}.
\newblock \showarticletitle{Supernumerary robotic limbs for aircraft fuselage assembly: body stabilization and guidance by bracing}. In \bibinfo{booktitle}{\emph{2014 IEEE International Conference on Robotics and Automation (ICRA)}}. IEEE, \bibinfo{pages}{1176--1183}.
\newblock


\bibitem[Parietti et~al\mbox{.}(2015)]%
        {parietti2015design}
\bibfield{author}{\bibinfo{person}{Federico Parietti}, \bibinfo{person}{Kameron~C Chan}, \bibinfo{person}{Banks Hunter}, {and} \bibinfo{person}{H~Harry Asada}.} \bibinfo{year}{2015}\natexlab{}.
\newblock \showarticletitle{Design and control of supernumerary robotic limbs for balance augmentation}. In \bibinfo{booktitle}{\emph{2015 IEEE International Conference on Robotics and Automation (ICRA)}}. IEEE, \bibinfo{pages}{5010--5017}.
\newblock


\bibitem[Pellegrinelli et~al\mbox{.}(2016)]%
        {pellegrinelli2016human}
\bibfield{author}{\bibinfo{person}{Stefania Pellegrinelli}, \bibinfo{person}{Henny Admoni}, \bibinfo{person}{Shervin Javdani}, {and} \bibinfo{person}{Siddhartha Srinivasa}.} \bibinfo{year}{2016}\natexlab{}.
\newblock \showarticletitle{Human-robot shared workspace collaboration via hindsight optimization}. In \bibinfo{booktitle}{\emph{2016 IEEE/RSJ International Conference on Intelligent Robots and Systems (IROS)}}. IEEE, \bibinfo{pages}{831--838}.
\newblock


\bibitem[Penaloza et~al\mbox{.}(2018)]%
        {penaloza2018towards}
\bibfield{author}{\bibinfo{person}{Christian Penaloza}, \bibinfo{person}{David Hernandez-Carmona}, {and} \bibinfo{person}{Shuichi Nishio}.} \bibinfo{year}{2018}\natexlab{}.
\newblock \showarticletitle{Towards intelligent brain-controlled body augmentation robotic limbs}. In \bibinfo{booktitle}{\emph{2018 IEEE International Conference on Systems, Man, and Cybernetics (SMC)}}. IEEE, \bibinfo{pages}{1011--1015}.
\newblock


\bibitem[Perera et~al\mbox{.}(2024)]%
        {perera2024integrating}
\bibfield{author}{\bibinfo{person}{Praneeth~Bimsara Perera}, \bibinfo{person}{Hansa Marasinghe}, \bibinfo{person}{Taiki Takami}, \bibinfo{person}{Hiroyuki Kajimoto}, {and} \bibinfo{person}{Anusha Withana}.} \bibinfo{year}{2024}\natexlab{}.
\newblock \showarticletitle{Integrating Force Sensing with Electro-Tactile Feedback in 3D Printed Haptic Interfaces}. In \bibinfo{booktitle}{\emph{Proceedings of the 2024 ACM International Symposium on Wearable Computers}}. \bibinfo{pages}{48--54}.
\newblock


\bibitem[Perera et~al\mbox{.}(2025)]%
        {Perera2025eTactileKit}
\bibfield{author}{\bibinfo{person}{Praneeth~Bimsara Perera}, \bibinfo{person}{Ravindu~Madhushan Pushpakumara}, \bibinfo{person}{Hiroyuki Kajimoto}, \bibinfo{person}{Arata Jingu}, \bibinfo{person}{J{\"u}rgen Steimle}, {and} \bibinfo{person}{Anusha Withana}.} \bibinfo{year}{2025}\natexlab{}.
\newblock \showarticletitle{eTactileKit: A Toolkit for Design Exploration and Rapid Prototyping of Electro-Tactile Interfaces}. In \bibinfo{booktitle}{\emph{Proceedings of the 38th Annual ACM Symposium on User Interface Software and Technology (UIST ’25)}}. \bibinfo{pages}{1--17}.
\newblock
\href{https://doi.org/10.1145/3746059.3747796}{doi:\nolinkurl{10.1145/3746059.3747796}}


\bibitem[Prattichizzo et~al\mbox{.}(2021)]%
        {prattichizzo2021human}
\bibfield{author}{\bibinfo{person}{Domenico Prattichizzo}, \bibinfo{person}{Maria Pozzi}, \bibinfo{person}{Tommaso~Lisini Baldi}, \bibinfo{person}{Monica Malvezzi}, \bibinfo{person}{Irfan Hussain}, \bibinfo{person}{Simone Rossi}, {and} \bibinfo{person}{Gionata Salvietti}.} \bibinfo{year}{2021}\natexlab{}.
\newblock \showarticletitle{Human augmentation by wearable supernumerary robotic limbs: review and perspectives}.
\newblock \bibinfo{journal}{\emph{Progress in Biomedical Engineering}} \bibinfo{volume}{3}, \bibinfo{number}{4} (\bibinfo{year}{2021}), \bibinfo{pages}{042005}.
\newblock


\bibitem[Profita et~al\mbox{.}(2013)]%
        {profita2013don}
\bibfield{author}{\bibinfo{person}{Halley~P Profita}, \bibinfo{person}{James Clawson}, \bibinfo{person}{Scott Gilliland}, \bibinfo{person}{Clint Zeagler}, \bibinfo{person}{Thad Starner}, \bibinfo{person}{Jim Budd}, {and} \bibinfo{person}{Ellen Yi-Luen Do}.} \bibinfo{year}{2013}\natexlab{}.
\newblock \showarticletitle{Don't mind me touching my wrist: a case study of interacting with on-body technology in public}. In \bibinfo{booktitle}{\emph{Proceedings of the 2013 international symposium on wearable computers}}. \bibinfo{pages}{89--96}.
\newblock


\bibitem[Raspopovic et~al\mbox{.}(2014)]%
        {raspopovic2014restoring}
\bibfield{author}{\bibinfo{person}{Stanisa Raspopovic}, \bibinfo{person}{Marco Capogrosso}, \bibinfo{person}{Francesco~Maria Petrini}, {and} \bibinfo{person}{et al.}} \bibinfo{year}{2014}\natexlab{}.
\newblock \showarticletitle{Restoring Natural Sensory Feedback in Hand Prostheses via Intraneural Stimulation}.
\newblock \bibinfo{journal}{\emph{Science Translational Medicine}} \bibinfo{volume}{6}, \bibinfo{number}{222} (\bibinfo{year}{2014}), \bibinfo{pages}{222ra19}.
\newblock
\href{https://doi.org/10.1126/scitranslmed.3006820}{doi:\nolinkurl{10.1126/scitranslmed.3006820}}


\bibitem[Riek(2012)]%
        {riek2012wizard}
\bibfield{author}{\bibinfo{person}{Laurel~D Riek}.} \bibinfo{year}{2012}\natexlab{}.
\newblock \showarticletitle{Wizard of oz studies in hri: a systematic review and new reporting guidelines}.
\newblock \bibinfo{journal}{\emph{Journal of Human-Robot Interaction}} \bibinfo{volume}{1}, \bibinfo{number}{1} (\bibinfo{year}{2012}), \bibinfo{pages}{119--136}.
\newblock


\bibitem[Rios-Martinez et~al\mbox{.}(2015)]%
        {rios2015proxemics}
\bibfield{author}{\bibinfo{person}{Jorge Rios-Martinez}, \bibinfo{person}{Anne Spalanzani}, {and} \bibinfo{person}{Christian Laugier}.} \bibinfo{year}{2015}\natexlab{}.
\newblock \showarticletitle{From proxemics theory to socially-aware navigation: A survey}.
\newblock \bibinfo{journal}{\emph{International Journal of Social Robotics}} \bibinfo{volume}{7}, \bibinfo{number}{2} (\bibinfo{year}{2015}), \bibinfo{pages}{137--153}.
\newblock


\bibitem[Ritchie and Spencer(2002)]%
        {ritchie2002qualitative}
\bibfield{author}{\bibinfo{person}{Jane Ritchie} {and} \bibinfo{person}{Liz Spencer}.} \bibinfo{year}{2002}\natexlab{}.
\newblock \showarticletitle{Qualitative data analysis for applied policy research}.
\newblock In \bibinfo{booktitle}{\emph{Analyzing qualitative data}}. \bibinfo{publisher}{routledge}, \bibinfo{pages}{173--194}.
\newblock


\bibitem[Rosenblatt et~al\mbox{.}(2014)]%
        {rosenblatt2014active}
\bibfield{author}{\bibinfo{person}{Noah~J Rosenblatt}, \bibinfo{person}{ACPO Bauer}, \bibinfo{person}{DCPO Rotter}, \bibinfo{person}{Mark~D Grabiner}, {et~al\mbox{.}}} \bibinfo{year}{2014}\natexlab{}.
\newblock \showarticletitle{Active dorsiflexing prostheses may reduce trip-related fall risk in people with transtibial amputation}.
\newblock \bibinfo{journal}{\emph{J Rehabil Res Dev}} \bibinfo{volume}{51}, \bibinfo{number}{8} (\bibinfo{year}{2014}), \bibinfo{pages}{1229--1242}.
\newblock


\bibitem[Rozlivek et~al\mbox{.}(2023)]%
        {rozlivek2023perirobot}
\bibfield{author}{\bibinfo{person}{Jakub Rozlivek}, \bibinfo{person}{Petr Svarny}, {and} \bibinfo{person}{Matej Hoffmann}.} \bibinfo{year}{2023}\natexlab{}.
\newblock \showarticletitle{Perirobot space representation for HRI: measuring and designing collaborative workspace coverage by diverse sensors}. In \bibinfo{booktitle}{\emph{2023 IEEE/RSJ International Conference on Intelligent Robots and Systems (IROS)}}. IEEE, \bibinfo{pages}{5958--5965}.
\newblock


\bibitem[Rubagotti et~al\mbox{.}(2022)]%
        {rubagotti2022perceived}
\bibfield{author}{\bibinfo{person}{Matteo Rubagotti}, \bibinfo{person}{Inara Tusseyeva}, \bibinfo{person}{Sara Baltabayeva}, \bibinfo{person}{Danna Summers}, {and} \bibinfo{person}{Anara Sandygulova}.} \bibinfo{year}{2022}\natexlab{}.
\newblock \showarticletitle{Perceived safety in physical human--robot interaction—A survey}.
\newblock \bibinfo{journal}{\emph{Robotics and Autonomous Systems}}  \bibinfo{volume}{151} (\bibinfo{year}{2022}), \bibinfo{pages}{104047}.
\newblock


\bibitem[Ruiz et~al\mbox{.}(2011)]%
        {Ruiz2011CHI}
\bibfield{author}{\bibinfo{person}{Jaime Ruiz}, \bibinfo{person}{Yang Li}, {and} \bibinfo{person}{Edward Lank}.} \bibinfo{year}{2011}\natexlab{}.
\newblock \showarticletitle{User-Defined Motion Gestures for Mobile Interaction}. In \bibinfo{booktitle}{\emph{Proceedings of the SIGCHI Conference on Human Factors in Computing Systems (CHI '11)}}. \bibinfo{publisher}{ACM}, \bibinfo{pages}{197--206}.
\newblock
\href{https://doi.org/10.1145/1978942.1978971}{doi:\nolinkurl{10.1145/1978942.1978971}}


\bibitem[Salda{\~n}a(2016)]%
        {saldana2016coding}
\bibfield{author}{\bibinfo{person}{Johnny Salda{\~n}a}.} \bibinfo{year}{2016}\natexlab{}.
\newblock \bibinfo{booktitle}{\emph{The Coding Manual for Qualitative Researchers} (\bibinfo{edition}{3} ed.)}.
\newblock \bibinfo{publisher}{SAGE Publications}, \bibinfo{address}{London}.
\newblock


\bibitem[Saraiji et~al\mbox{.}(2018a)]%
        {saraiji2018metaarms}
\bibfield{author}{\bibinfo{person}{MHD~Yamen Saraiji}, \bibinfo{person}{Tomoya Sasaki}, \bibinfo{person}{Kai Kunze}, \bibinfo{person}{Kouta Minamizawa}, {and} \bibinfo{person}{Masahiko Inami}.} \bibinfo{year}{2018}\natexlab{a}.
\newblock \showarticletitle{Metaarms: Body remapping using feet-controlled artificial arms}. In \bibinfo{booktitle}{\emph{Proceedings of the 31st Annual ACM Symposium on User Interface Software and Technology}}. \bibinfo{pages}{65--74}.
\newblock


\bibitem[Saraiji et~al\mbox{.}(2018b)]%
        {Saraiji2018Fusion}
\bibfield{author}{\bibinfo{person}{MHD~Yamen Saraiji}, \bibinfo{person}{Tomoya Sasaki}, \bibinfo{person}{Reo Matsumura}, \bibinfo{person}{Kouta Minamizawa}, {and} \bibinfo{person}{Masahiko Inami}.} \bibinfo{year}{2018}\natexlab{b}.
\newblock \showarticletitle{Fusion: full body surrogacy for collaborative communication}.
\newblock In \bibinfo{booktitle}{\emph{ACM SIGGRAPH 2018 Emerging Technologies}}. \bibinfo{pages}{1--2}.
\newblock


\bibitem[Sasaki et~al\mbox{.}(2017)]%
        {sasaki2017metalimbs}
\bibfield{author}{\bibinfo{person}{Tomoya Sasaki}, \bibinfo{person}{MHD~Yamen Saraiji}, \bibinfo{person}{Charith~Lasantha Fernando}, \bibinfo{person}{Kouta Minamizawa}, {and} \bibinfo{person}{Masahiko Inami}.} \bibinfo{year}{2017}\natexlab{}.
\newblock \showarticletitle{MetaLimbs: Multiple arms interaction metamorphism}.
\newblock In \bibinfo{booktitle}{\emph{ACM SIGGRAPH 2017 emerging technologies}}. \bibinfo{pages}{1--2}.
\newblock


\bibitem[Shimobayashi et~al\mbox{.}(2021)]%
        {shimobayashi2021independent}
\bibfield{author}{\bibinfo{person}{Hideki Shimobayashi}, \bibinfo{person}{Tomoya Sasaki}, \bibinfo{person}{Arata Horie}, \bibinfo{person}{Riku Arakawa}, \bibinfo{person}{Zendai Kashino}, {and} \bibinfo{person}{Masahiko Inami}.} \bibinfo{year}{2021}\natexlab{}.
\newblock \showarticletitle{Independent control of supernumerary appendages exploiting upper limb redundancy}. In \bibinfo{booktitle}{\emph{Proceedings of the Augmented Humans International Conference 2021}}. \bibinfo{pages}{19--30}.
\newblock


\bibitem[Shore et~al\mbox{.}(2022)]%
        {shore2022technology}
\bibfield{author}{\bibinfo{person}{Linda Shore}, \bibinfo{person}{Adam de Eyto}, {and} \bibinfo{person}{Leonard O’Sullivan}.} \bibinfo{year}{2022}\natexlab{}.
\newblock \showarticletitle{Technology acceptance and perceptions of robotic assistive devices by older adults--implications for exoskeleton design}.
\newblock \bibinfo{journal}{\emph{Disability and rehabilitation: assistive technology}} \bibinfo{volume}{17}, \bibinfo{number}{7} (\bibinfo{year}{2022}), \bibinfo{pages}{782--790}.
\newblock


\bibitem[Sisbot et~al\mbox{.}(2007)]%
        {sisbot2007human}
\bibfield{author}{\bibinfo{person}{Emrah~Akin Sisbot}, \bibinfo{person}{Luis~F Marin-Urias}, \bibinfo{person}{Rachid Alami}, {and} \bibinfo{person}{Thierry Simeon}.} \bibinfo{year}{2007}\natexlab{}.
\newblock \showarticletitle{A human aware mobile robot motion planner}.
\newblock \bibinfo{journal}{\emph{IEEE Transactions on Robotics}} \bibinfo{volume}{23}, \bibinfo{number}{5} (\bibinfo{year}{2007}), \bibinfo{pages}{874--883}.
\newblock


\bibitem[Slater(2009)]%
        {slater2009place}
\bibfield{author}{\bibinfo{person}{Mel Slater}.} \bibinfo{year}{2009}\natexlab{}.
\newblock \showarticletitle{Place Illusion and Plausibility in Virtual Reality}.
\newblock \bibinfo{journal}{\emph{Philosophical Transactions of the Royal Society B}} \bibinfo{volume}{364}, \bibinfo{number}{1535} (\bibinfo{year}{2009}), \bibinfo{pages}{3549--3557}.
\newblock
\href{https://doi.org/10.1098/rstb.2009.0138}{doi:\nolinkurl{10.1098/rstb.2009.0138}}


\bibitem[Strabala et~al\mbox{.}(2013)]%
        {strabala2013toward}
\bibfield{author}{\bibinfo{person}{Andrew~C. Strabala}, \bibinfo{person}{Min~Kyung Lee}, \bibinfo{person}{Anca~D. Dragan}, \bibinfo{person}{Siddhartha~S. Srinivasa}, \bibinfo{person}{Jodi Forlizzi}, {and} \bibinfo{person}{Maya Cakmak}.} \bibinfo{year}{2013}\natexlab{}.
\newblock \showarticletitle{Toward Seamless Human--Robot Handovers}.
\newblock \bibinfo{journal}{\emph{Journal of Human-Robot Interaction}} \bibinfo{volume}{2}, \bibinfo{number}{1} (\bibinfo{year}{2013}).
\newblock
\href{https://doi.org/10.5898/JHRI.2.1.Strabala}{doi:\nolinkurl{10.5898/JHRI.2.1.Strabala}}


\bibitem[Takahashi et~al\mbox{.}(2023)]%
        {takahashi2023feasibility}
\bibfield{author}{\bibinfo{person}{Kazushi Takahashi}, \bibinfo{person}{Masafumi Mizukami}, \bibinfo{person}{Hiroki Watanabe}, \bibinfo{person}{Mayumi~Matsuda Kuroda}, \bibinfo{person}{Yukiyo Shimizu}, \bibinfo{person}{Takashi Nakajima}, \bibinfo{person}{Hirotaka Mutsuzaki}, \bibinfo{person}{Hiroshi Kamada}, \bibinfo{person}{Kayo Tokeji}, \bibinfo{person}{Yasushi Hada}, {et~al\mbox{.}}} \bibinfo{year}{2023}\natexlab{}.
\newblock \showarticletitle{Feasibility and safety study of wearable cyborg Hybrid Assistive Limb for pediatric patients with cerebral palsy and spinal cord disorders}.
\newblock \bibinfo{journal}{\emph{Frontiers in Neurology}}  \bibinfo{volume}{14} (\bibinfo{year}{2023}), \bibinfo{pages}{1255620}.
\newblock


\bibitem[Takayama and Pantofaru(2009)]%
        {takayama2009influences}
\bibfield{author}{\bibinfo{person}{Leila Takayama} {and} \bibinfo{person}{Caroline Pantofaru}.} \bibinfo{year}{2009}\natexlab{}.
\newblock \showarticletitle{Influences on Proxemic Behaviors in Human--Robot Interaction}. In \bibinfo{booktitle}{\emph{Proceedings of the IEEE/RSJ International Conference on Intelligent Robots and Systems (IROS)}}. \bibinfo{pages}{5495--5502}.
\newblock
\href{https://doi.org/10.1109/IROS.2009.5354145}{doi:\nolinkurl{10.1109/IROS.2009.5354145}}


\bibitem[Tong et~al\mbox{.}(2023)]%
        {tong2023fully}
\bibfield{author}{\bibinfo{person}{Adele Tong}, \bibinfo{person}{Praneeth Perera}, \bibinfo{person}{Zhanna Sarsenbayeva}, \bibinfo{person}{Alistair McEwan}, \bibinfo{person}{Anjula~C De~Silva}, {and} \bibinfo{person}{Anusha Withana}.} \bibinfo{year}{2023}\natexlab{}.
\newblock \showarticletitle{Fully 3D-printed dry EEG electrodes}.
\newblock \bibinfo{journal}{\emph{Sensors}} \bibinfo{volume}{23}, \bibinfo{number}{11} (\bibinfo{year}{2023}), \bibinfo{pages}{5175}.
\newblock


\bibitem[Tracy(2010)]%
        {tracy2010qualitative}
\bibfield{author}{\bibinfo{person}{Sarah~J. Tracy}.} \bibinfo{year}{2010}\natexlab{}.
\newblock \showarticletitle{Qualitative quality: Eight \`{big-tent} criteria for excellent qualitative research}.
\newblock \bibinfo{journal}{\emph{Qualitative Inquiry}} \bibinfo{volume}{16}, \bibinfo{number}{10} (\bibinfo{year}{2010}), \bibinfo{pages}{837--851}.
\newblock
\href{https://doi.org/10.1177/1077800410383121}{doi:\nolinkurl{10.1177/1077800410383121}}


\bibitem[Ullman and Malle(2019)]%
        {ullman2019mdmt}
\bibfield{author}{\bibinfo{person}{Daniel Ullman} {and} \bibinfo{person}{Bertram~F Malle}.} \bibinfo{year}{2019}\natexlab{}.
\newblock \bibinfo{title}{MDMT: Multi-dimensional measure of trust}.
\newblock


\bibitem[Umezawa et~al\mbox{.}(2022)]%
        {umezawa2022bodily}
\bibfield{author}{\bibinfo{person}{Kohei Umezawa}, \bibinfo{person}{Yuta Suzuki}, \bibinfo{person}{Gowrishankar Ganesh}, {and} \bibinfo{person}{Yoichi Miyawaki}.} \bibinfo{year}{2022}\natexlab{}.
\newblock \showarticletitle{Bodily ownership of an independent supernumerary limb: an exploratory study}.
\newblock \bibinfo{journal}{\emph{Scientific reports}} \bibinfo{volume}{12}, \bibinfo{number}{1} (\bibinfo{year}{2022}), \bibinfo{pages}{2339}.
\newblock


\bibitem[Van~Lewen et~al\mbox{.}(2024)]%
        {van2024capacitive}
\bibfield{author}{\bibinfo{person}{Daniel Van~Lewen}, \bibinfo{person}{Catherine Wang}, \bibinfo{person}{Hun~Chan Lee}, \bibinfo{person}{Anand Devaiah}, \bibinfo{person}{Urvashi Upadhyay}, {and} \bibinfo{person}{Sheila Russo}.} \bibinfo{year}{2024}\natexlab{}.
\newblock \showarticletitle{Capacitive origami sensing modules for measuring force in a neurosurgical, soft robotic retractor}. In \bibinfo{booktitle}{\emph{2024 IEEE International Conference on Robotics and Automation (ICRA)}}. IEEE, \bibinfo{pages}{5302--5308}.
\newblock


\bibitem[Vatsal and Hoffman(2017)]%
        {vatsal2017wearing}
\bibfield{author}{\bibinfo{person}{Vighnesh Vatsal} {and} \bibinfo{person}{Guy Hoffman}.} \bibinfo{year}{2017}\natexlab{}.
\newblock \showarticletitle{Wearing your arm on your sleeve: Studying usage contexts for a wearable robotic forearm}. In \bibinfo{booktitle}{\emph{2017 26th IEEE International Symposium on Robot and Human Interactive Communication (RO-MAN)}}. IEEE, \bibinfo{pages}{974--980}.
\newblock


\bibitem[Vatsal and Hoffman(2018)]%
        {vatsal2018design}
\bibfield{author}{\bibinfo{person}{Vighnesh Vatsal} {and} \bibinfo{person}{Guy Hoffman}.} \bibinfo{year}{2018}\natexlab{}.
\newblock \showarticletitle{Design and analysis of a wearable robotic forearm}. In \bibinfo{booktitle}{\emph{2018 IEEE International Conference on Robotics and Automation (ICRA)}}. IEEE, \bibinfo{pages}{5489--5496}.
\newblock


\bibitem[Viteckova et~al\mbox{.}(2013)]%
        {viteckova2013wearable}
\bibfield{author}{\bibinfo{person}{Slavka Viteckova}, \bibinfo{person}{Patrik Kutilek}, {and} \bibinfo{person}{Marcel Jirina}.} \bibinfo{year}{2013}\natexlab{}.
\newblock \showarticletitle{Wearable lower limb robotics: A review}.
\newblock \bibinfo{journal}{\emph{Biocybernetics and biomedical engineering}} \bibinfo{volume}{33}, \bibinfo{number}{2} (\bibinfo{year}{2013}), \bibinfo{pages}{96--105}.
\newblock


\bibitem[Walters et~al\mbox{.}(2005)]%
        {walters2005influence}
\bibfield{author}{\bibinfo{person}{Michael~L. Walters}, \bibinfo{person}{Dag~Sverre Syrdal}, \bibinfo{person}{Kerstin Dautenhahn}, \bibinfo{person}{Ren{\'e} te Boekhorst}, {and} \bibinfo{person}{Kheng~Lee Koay}.} \bibinfo{year}{2005}\natexlab{}.
\newblock \showarticletitle{The Influence of Subjects’ Personality Traits on Personal Spatial Zones in a Human–Robot Interaction Experiment}. In \bibinfo{booktitle}{\emph{IEEE International Workshop on Robot and Human Interactive Communication (RO-MAN)}}. \bibinfo{pages}{347--352}.
\newblock
\href{https://doi.org/10.1109/ROMAN.2005.1513802}{doi:\nolinkurl{10.1109/ROMAN.2005.1513802}}


\bibitem[Wang et~al\mbox{.}(2025)]%
        {wang2025vr}
\bibfield{author}{\bibinfo{person}{Xueyang Wang}, \bibinfo{person}{Kewen Peng}, \bibinfo{person}{Chonghao Hao}, \bibinfo{person}{Wendi Yu}, \bibinfo{person}{Xin Yi}, {and} \bibinfo{person}{Hewu Li}.} \bibinfo{year}{2025}\natexlab{}.
\newblock \showarticletitle{VR Whispering: A Multisensory Approach for Private Conversations in Social Virtual Reality}.
\newblock \bibinfo{journal}{\emph{IEEE Transactions on Visualization and Computer Graphics}} (\bibinfo{year}{2025}).
\newblock


\bibitem[Wickens(2008)]%
        {Wickens2008MultipleResources}
\bibfield{author}{\bibinfo{person}{Christopher~D. Wickens}.} \bibinfo{year}{2008}\natexlab{}.
\newblock \showarticletitle{Multiple Resources and Mental Workload}.
\newblock \bibinfo{journal}{\emph{Human Factors}} \bibinfo{volume}{50}, \bibinfo{number}{3} (\bibinfo{year}{2008}), \bibinfo{pages}{449--455}.
\newblock


\bibitem[Wobbrock et~al\mbox{.}(2009)]%
        {wobbrock2009user}
\bibfield{author}{\bibinfo{person}{Jacob~O Wobbrock}, \bibinfo{person}{Meredith~Ringel Morris}, {and} \bibinfo{person}{Andrew~D Wilson}.} \bibinfo{year}{2009}\natexlab{}.
\newblock \showarticletitle{User-defined gestures for surface computing}. In \bibinfo{booktitle}{\emph{Proceedings of the SIGCHI conference on human factors in computing systems}}. \bibinfo{pages}{1083--1092}.
\newblock


\bibitem[Wo{\'z}niak et~al\mbox{.}(2024)]%
        {wozniak2024influence}
\bibfield{author}{\bibinfo{person}{Mateusz Wo{\'z}niak}, \bibinfo{person}{Davide De~Tommaso}, \bibinfo{person}{Guenther Knoblich}, \bibinfo{person}{Agnieszka Wykowska}, {and} \bibinfo{person}{Via~Enrico Melen}.} \bibinfo{year}{2024}\natexlab{}.
\newblock \showarticletitle{The influence of robot autonomy on sense of control and trust towards a robot}.
\newblock  (\bibinfo{year}{2024}).
\newblock


\bibitem[Wu et~al\mbox{.}(2025)]%
        {wu2025headturner}
\bibfield{author}{\bibinfo{person}{En-Huei Wu}, \bibinfo{person}{Po-Yun Cheng}, \bibinfo{person}{Che-Wei Hsu}, \bibinfo{person}{Cheng~Hsin Han}, \bibinfo{person}{Pei~Chen Lee}, \bibinfo{person}{Chia-An Fan}, \bibinfo{person}{Yu~Chia Kuo}, \bibinfo{person}{Kai-Jing Hu}, \bibinfo{person}{Yu Chen}, {and} \bibinfo{person}{Mike~Y Chen}.} \bibinfo{year}{2025}\natexlab{}.
\newblock \showarticletitle{HeadTurner: Enhancing Viewing Range and Comfort of using Virtual and Mixed-Reality Headsets while Lying Down via Assisted Shoulder and Head Actuation}. In \bibinfo{booktitle}{\emph{Proceedings of the 2025 CHI Conference on Human Factors in Computing Systems}}. \bibinfo{pages}{1--16}.
\newblock


\bibitem[Yu et~al\mbox{.}(2025)]%
        {yu2025designing}
\bibfield{author}{\bibinfo{person}{Jiakun Yu}, \bibinfo{person}{Hasindu Kariyawasam}, \bibinfo{person}{Shuying Wu}, \bibinfo{person}{Sriram Subramanian}, {and} \bibinfo{person}{Anusha Withana}.} \bibinfo{year}{2025}\natexlab{}.
\newblock \showarticletitle{Designing Multi-DoF Epidermal Bend Sensors Using Flexible Resistive Traces}.
\newblock \bibinfo{journal}{\emph{IEEE Sensors Journal}} (\bibinfo{year}{2025}).
\newblock


\bibitem[Yu et~al\mbox{.}(2024a)]%
        {yu2024irontex}
\bibfield{author}{\bibinfo{person}{Jiakun Yu}, \bibinfo{person}{Supun Kuruppu}, \bibinfo{person}{Biyon Fernando}, \bibinfo{person}{Praneeth~Bimsara Perera}, \bibinfo{person}{Yuta Sugiura}, \bibinfo{person}{Sriram Subramanian}, {and} \bibinfo{person}{Anusha Withana}.} \bibinfo{year}{2024}\natexlab{a}.
\newblock \showarticletitle{IrOnTex: Using Ironable 3D Printed Objects to Fabricate and Prototype Customizable Interactive Textiles}.
\newblock \bibinfo{journal}{\emph{Proceedings of the ACM on Interactive, Mobile, Wearable and Ubiquitous Technologies}} \bibinfo{volume}{8}, \bibinfo{number}{3} (\bibinfo{year}{2024}), \bibinfo{pages}{1--26}.
\newblock


\bibitem[Yu et~al\mbox{.}(2024b)]%
        {yu2024fabricating}
\bibfield{author}{\bibinfo{person}{Jiakun Yu}, \bibinfo{person}{Praneeth~Bimsara Perera}, \bibinfo{person}{Rahal~Viddusha Perera}, \bibinfo{person}{Mohammad~Mirkhalaf Valashani}, {and} \bibinfo{person}{Anusha Withana}.} \bibinfo{year}{2024}\natexlab{b}.
\newblock \showarticletitle{Fabricating customizable 3-D printed pressure sensors by tuning infill characteristics}.
\newblock \bibinfo{journal}{\emph{IEEE Sensors Journal}} \bibinfo{volume}{24}, \bibinfo{number}{6} (\bibinfo{year}{2024}), \bibinfo{pages}{7604--7613}.
\newblock


\bibitem[Yu et~al\mbox{.}(2023)]%
        {yu2023drivingvibe}
\bibfield{author}{\bibinfo{person}{Neng-Hao Yu}, \bibinfo{person}{Shih-Yu Ma}, \bibinfo{person}{Cong-Min Lin}, \bibinfo{person}{Chi-Aan Fan}, \bibinfo{person}{Luca~E Taglialatela}, \bibinfo{person}{Tsai-Yuan Huang}, \bibinfo{person}{Carolyn Yu}, \bibinfo{person}{Yun-Ting Cheng}, \bibinfo{person}{Ya-Chi Liao}, {and} \bibinfo{person}{Mike~Y Chen}.} \bibinfo{year}{2023}\natexlab{}.
\newblock \showarticletitle{DrivingVibe: Enhancing VR driving experience using inertia-based vibrotactile feedback around the head}.
\newblock \bibinfo{journal}{\emph{Proceedings of the ACM on Human-Computer Interaction}} \bibinfo{volume}{7}, \bibinfo{number}{MHCI} (\bibinfo{year}{2023}), \bibinfo{pages}{1--22}.
\newblock


\bibitem[Zeagler(2017)]%
        {zeagler2017wear}
\bibfield{author}{\bibinfo{person}{Clint Zeagler}.} \bibinfo{year}{2017}\natexlab{}.
\newblock \showarticletitle{Where to wear it: functional, technical, and social considerations in on-body location for wearable technology 20 years of designing for wearability}. In \bibinfo{booktitle}{\emph{Proceedings of the 2017 ACM International Symposium on Wearable Computers}}. \bibinfo{pages}{150--157}.
\newblock


\bibitem[Zhou et~al\mbox{.}(2024a)]%
        {zhou2024coplayingvr}
\bibfield{author}{\bibinfo{person}{Hongyu Zhou}, \bibinfo{person}{Treshan Ayesh}, \bibinfo{person}{Chenyu Fan}, \bibinfo{person}{Zhanna Sarsenbayeva}, {and} \bibinfo{person}{Anusha Withana}.} \bibinfo{year}{2024}\natexlab{a}.
\newblock \showarticletitle{CoplayingVR: Understanding User Experience in Shared Control in Virtual Reality}.
\newblock \bibinfo{journal}{\emph{Proceedings of the ACM on Interactive, Mobile, Wearable and Ubiquitous Technologies}} \bibinfo{volume}{8}, \bibinfo{number}{3} (\bibinfo{year}{2024}), \bibinfo{pages}{1--25}.
\newblock


\bibitem[Zhou et~al\mbox{.}(2025a)]%
        {zhou2025survey}
\bibfield{author}{\bibinfo{person}{Hongyu Zhou}, \bibinfo{person}{Yihao Dong}, \bibinfo{person}{Masahiko Inami}, \bibinfo{person}{Zhanna Sarsenbayeva}, {and} \bibinfo{person}{Anusha Withana}.} \bibinfo{year}{2025}\natexlab{a}.
\newblock \showarticletitle{A Survey on Methodological Approaches to Collaborative Embodiment in Virtual Reality}.
\newblock \bibinfo{journal}{\emph{arXiv preprint arXiv:2507.18877}} (\bibinfo{year}{2025}).
\newblock


\bibitem[Zhou et~al\mbox{.}(2026)]%
        {Zhou2026OneBodyTwoMinds}
\bibfield{author}{\bibinfo{person}{Hongyu Zhou}, \bibinfo{person}{Xincheng Huang}, \bibinfo{person}{Winston Wijaya}, \bibinfo{person}{Yi~Fei Cheng}, \bibinfo{person}{David Lindlbauer}, \bibinfo{person}{Eduardo Velloso}, \bibinfo{person}{Andrea Bianchi}, \bibinfo{person}{Zhanna Sarsenbayeva}, {and} \bibinfo{person}{Anusha Withana}.} \bibinfo{year}{2026}\natexlab{}.
\newblock \showarticletitle{One Body, Two Minds: Alternating VR Perspective During Remote Teleoperation of Supernumerary Limbs}. In \bibinfo{booktitle}{\emph{Proceedings of the 2026 CHI Conference on Human Factors in Computing Systems (CHI '26)}} (Barcelona, Spain). \bibinfo{publisher}{Association for Computing Machinery}, \bibinfo{address}{New York, NY, USA}.
\newblock
\showISBNx{979-8-4007-2278-3/2026/04}
\href{https://doi.org/10.1145/3772318.3791433}{doi:\nolinkurl{10.1145/3772318.3791433}}


\bibitem[Zhou et~al\mbox{.}(2025b)]%
        {zhou2025juggling}
\bibfield{author}{\bibinfo{person}{Hongyu Zhou}, \bibinfo{person}{Tom Kip}, \bibinfo{person}{Yihao Dong}, \bibinfo{person}{Andrea Bianchi}, \bibinfo{person}{Zhanna Sarsenbayeva}, {and} \bibinfo{person}{Anusha Withana}.} \bibinfo{year}{2025}\natexlab{b}.
\newblock \showarticletitle{Juggling Extra Limbs: Identifying Control Strategies for Supernumerary Multi-Arms in Virtual Reality}. In \bibinfo{booktitle}{\emph{Proceedings of the 2025 CHI Conference on Human Factors in Computing Systems}}. \bibinfo{pages}{1--16}.
\newblock


\bibitem[Zhou et~al\mbox{.}(2024b)]%
        {zhou2024pairplayvr}
\bibfield{author}{\bibinfo{person}{Hongyu Zhou}, \bibinfo{person}{Pamuditha Somarathne}, \bibinfo{person}{Treshan~Ayesh Peirispulle}, \bibinfo{person}{Chenyu Fan}, \bibinfo{person}{Zhanna Sarsenbayeva}, {and} \bibinfo{person}{Anusha Withana}.} \bibinfo{year}{2024}\natexlab{b}.
\newblock \showarticletitle{PairPlayVR: Shared Hand Control for Virtual Games}. In \bibinfo{booktitle}{\emph{Proceedings of the Augmented Humans International Conference 2024}}. \bibinfo{pages}{311--314}.
\newblock


\end{thebibliography}

\appendix

\section{Embodiment Questionnaire}\label{Embodiment}

\subsection{Questionnaire Adaptation and Validation}\label{app:adaptation}

\textit{Adaptation principles.} We aligned with the original constructs but replaced references to an artificial finger/six-fingered hand with a physical SRL, modifying component-level wording as needed. We retained the “during the task” temporal framing to preserve phase alignment, and used a 7-point, same-direction Likert scale to facilitate comparison. The after-effect item was adapted from hand width to upper-body width/arm-span to reflect the physical scale of multi-arm setups. This adaptation followed prior approaches for embodiment questionnaires in SRL research~\cite{umezawa2022bodily}.

\textit{Validation and reliability.} We ran brief think-aloud sessions (n=10) to confirm the clarity of phrases such as “looked normal” and “driving the SRL” under mixed-autonomy/PDR. Minor wording adjustments were made without altering construct intent. To check internal consistency for our two-item constructs, we computed inter–item Spearman correlations ($\rho$) and Spearman–Brown coefficients (SB)~\cite{eisinga2013reliability,hauke2011comparison}. We restricted consistency analysis to Q1\&Q5 (Agency) and Q4\&Q6 (Ownership), which jointly represent core psychological constructs of embodiment. Other items (e.g., Q2, Q3, Q7) were excluded due to construct divergence or categorical format. Results for Agency (Q1\&Q5) and Ownership (Q4\&Q6) under each autonomy condition are shown below.

\begin{table}[h]
\centering
\caption{Inter–item $\rho$ and Spearman–Brown (SB) for two–item constructs by condition.}
\begin{tabular}{lccccc}
\toprule
\multirow{2}{*}{\textbf{Construct}} & \multicolumn{2}{c}{\textbf{Inter–item $\rho$}} & & \multicolumn{2}{c}{\textbf{Spearman--Brown}} \\
\cmidrule(lr){2-3} \cmidrule(lr){5-6}
& Baseline & PDR & & Baseline & PDR \\
\midrule
Agency (Q1,Q5) & 0.46 & 0.54 & & 0.63 & 0.70 \\
Ownership (Q4,Q6) & 0.42 & 0.49 & & 0.59 & 0.66 \\
\bottomrule
\end{tabular}
\par\smallskip\noindent\scriptsize SB formula: $\,\mathrm{SB}=\tfrac{2\rho}{1+\rho}\,$.
\end{table}

\subsection{Questionnaire Items}

The questionnaire was administered immediately after each experimental phase. Participants rated seven items in English on a 7-point Likert scale ranging from 1 (Strongly disagree) to 7 (Strongly agree). The items were adapted from prior work on supernumerary-limb embodiment~\cite{umezawa2022bodily}, modified to reflect a physical supernumerary robotic limb (SRL) rather than a supernumerary finger.

\pagebreak
\noindent\textbf{Items (7-point Likert unless noted):}
\begin{enumerate}
  \item Q1: During the task, the SRL moved when I intended it to move.
  \item Q2: During the task, the augmented body with the extra arm began to look normal to me.
  \item Q3: During the task, operating the SRL was as easy as operating my own arm.
  \item Q4: During the task, I felt that the SRL was part of my body.
  \item Q5: During the task, I felt that I was driving the SRL's movements.
  \item Q6: During the task, I felt that the SRL's end-effector/hand was my hand.
  \item Q7: After removing the SRL, my sense of upper-body width / arm-span felt:
        \emph{narrower} \,/\, \emph{wider}. \hfill (categorical choice. For Q7, ``narrowed'' is assigned to 1 and ``widened'' is assigned to 7.)
\end{enumerate}

\section{Operator Action Protocol}
\label{app:operator}

\subsection{Scope \& Provenance}
This protocol standardizes the Wizard-of-Oz (WoZ) operator’s actions to emulate the high-autonomy control strategy reported for supernumerary multi-arms in VR (commands, demonstration, delegation, labeling). The goal is to ensure consistent execution across participants during early-stage elicitation, without introducing novel timing models or cue modalities. The operator remains out of view and does not converse with participants during tasks.

\subsection{High-autonomy Mode}
We structure each interaction around the following four steps; the operator acts only within the listed behaviors.

\textit{Commands}
\begin{itemize}\itemsep 0pt
\item Execute the full sequence or complex actions based on the command’s intent (e.g., directly move a labeled object to a specified position).
\item No additional confirmations, prompts, or timing cues are provided by the operator.
\end{itemize}

\textit{Demonstration}
\begin{itemize}\itemsep 0pt
\item If the participant provides a pose/path demonstration, replicate the demonstrated end-effector pose/path once, at a steady speed.
\item No corrective coaching or multi-step tutoring; accept a single demonstration attempt.
\end{itemize}
\pagebreak

\textit{Delegation}
\begin{itemize}\itemsep 0pt
\item When the participant indicates delegation of a sub-goal (e.g., align, grasp, handover), complete the sub-goal autonomously once it is contextually clear.
\item Do not add pre-cues, audio, or extra pauses beyond a single brief stop before action onset.
\end{itemize}

\textit{Labeling}
\begin{itemize}\itemsep 0pt
\item If the participant uses simple labels (e.g., my left side, table), treat them as temporary references for the current trial.
\item Do not invent new labels or query semantics; mirror only what is expressed by the participant or the script.
\end{itemize}

\subsection{Neutral Movement Constraints}
To avoid biasing elicited preferences, the operator follows neutral defaults:
\begin{itemize}\itemsep 0pt
\item \textbf{Path:} use a lateral approach by default; avoid direct frontal entries unless the task strictly requires them.
\item \textbf{Clearance:} maintain a safe visible clearance (head/face kept clearly outside near range; forearm/hand interactions only when task-defined). No hover near the face.
\item \textbf{Speed:} steady, single-band speed suitable for near-body work; no purposeful acceleration bursts or rhythm patterns.
\item \textbf{No added cues:} no audio, vibro, or light cues are introduced by the operator.
\end{itemize}

\subsection{Consistency Checklist (per trial)}
For each trial, the operator verifies:
\begin{enumerate}\itemsep 0pt
\item Four-step order respected (commands $\rightarrow$ demonstration $\rightarrow$ delegation $\rightarrow$ labeling, as applicable).
\item Neutral path used; no frontal approach unless task requires it.
\item No unsolicited prompts/cues; no empty hover in front of the face.
\item If delegation occurred, sub-goal was completed once without micro-tuning.
\end{enumerate}

\subsection{Deviations \& Exclusions}
Any deviation (e.g., accidental frontal entry, unintended pause) is logged with reason; such trials are marked for exclusion or immediate redo according to the preregistered rules.

\end{document}